\documentclass[12pt]{article}
\usepackage{mathtools}
\usepackage{diagbox}
\usepackage{savesym}
\usepackage[nosort]{cite}
\usepackage{wasysym}
\usepackage{amscd}
\usepackage{graphicx}
\usepackage{pifont}
\usepackage{float}
\usepackage{subcaption}
\usepackage[all]{xy}
\usepackage{multicol}
\usepackage{multirow}
\usepackage{amsfonts}
\usepackage{picture}
\usepackage{amssymb}
\savesymbol{iint}
\savesymbol{iiint}
\usepackage{amsmath}
\restoresymbol{TXF}{iint}
\restoresymbol{TXF}{iiint}
\usepackage{color}
\usepackage{style}
\usepackage{hyperref}
\usepackage{tikz}
\usetikzlibrary{calc,decorations.pathmorphing,arrows,decorations.markings,cd}
\tikzset{snake it/.style={decorate, decoration=snake}}
\usetikzlibrary{shapes}
\usetikzlibrary{patterns}
\usetikzlibrary{arrows,positioning,decorations.pathmorphing,
  decorations.markings, matrix, decorations.text}

\hypersetup{colorlinks=true}
\hypersetup{linkcolor=black}
\hypersetup{citecolor=black}
\hypersetup{urlcolor=black}
\usepackage{setspace}
\usepackage{multirow}
\usepackage{wrapfig}
\usepackage{verbatim}
\usepackage{dsfont}
\usepackage[vcentermath]{youngtab}

\numberwithin{equation}{section}
\usepackage{float}
\restylefloat{table}
\allowdisplaybreaks
\usepackage{accents}

\def\tilde{\widetilde}

\def\hat{\widehat}

\def\bar{\overline}

\def\1{{\mathds 1}}

\usepackage{color,soul}

\begin{document}

\date{\today}

\institution{Caltech}{\centerline{${}^{1}$Walter Burke Institute for Theoretical Physics,}}
\institution{Caltech2}{\centerline{California Institute of Technology, Pasadena, CA 91125, USA}}

\title{Symmetry-Enriched Quantum Spin Liquids in $(3+1)d$}

\authors{Po-Shen Hsin\worksat{\Caltech}\footnote{e-mail: {\tt phsin@caltech.edu}}
and Alex Turzillo\worksat{\Caltech}\footnote{e-mail: {\tt aturzillo@theory.caltech.edu}}}

\abstract{\noindent 
We use the intrinsic one-form and two-form global symmetries of (3+1)$d$ bosonic field theories to classify quantum phases enriched by ordinary ($0$-form) global symmetry. Different symmetry-enriched phases correspond to different ways of coupling the theory to the background gauge field of the ordinary symmetry. 
The input of the classification is the higher-form symmetries and a permutation action of the $0$-form symmetry on the lines and surfaces of the theory.
From these data we classify the couplings to the background gauge field by the 0-form symmetry defects constructed from the higher-form symmetry defects.
For trivial two-form symmetry the classification coincides with the classification for symmetry fractionalizations in $(2+1)d$.
We also provide a systematic method to obtain the symmetry protected topological phases that can be absorbed by the coupling, and we give the relative 't Hooft anomaly for different couplings.
We discuss several examples including the gapless pure $U(1)$ gauge theory and the gapped Abelian finite group gauge theory. As an application, we discover a tension with a conjectured  duality in $(3+1)d$ for $SU(2)$ gauge theory with two adjoint Weyl fermions.

}

\preprint{CALT-TH-2019-014}

\maketitle

\setcounter{tocdepth}{3}
\tableofcontents

\section{Introduction}

In this note we investigate the problem of classifying different ways of coupling a conformal field theory (in the broad sense including topological quantum field theory) in $(3+1)d$ to the background gauge field of an ordinary global symmetry. Such theories are said to be enriched by the global symmetry.
They can arise as the low energy effective descriptions of microscopic lattice models.
Since such theories are fixed points of the renormalization group (RG) flow, no operators are confined or decoupled\footnote{
For instance, if the theory is not conformal, the background gauge field that couples to the UV theory can decouple along the RG flow.
}, and thus any
two theories with two different couplings to the background field cannot be continuously connected to each other without breaking the global symmetry. They belong to different symmetry-enriched phases of the theory.
Symmetry-enriched phases generalize the notion of symmetry-protected topological (SPT) phases (see {\it e.g.} \cite{Chen:2011pg,Kapustin:2014tfa,Freed:2016rqq})\footnote{
Similar to SPT phases, symmetry-enriched phases can have non-trivial boundary states (see {\it e.g.} \cite{Wang:2016cto}).
}.
While an SPT phase has trivial dynamics and only depends on the background gauge field, a general symmetry-enriched phase can be described by a non-trivial coupling between a dynamical system and the background gauge field together with a decoupled SPT phase. When the theory is gapped, the symmetry-enriched phases are known as symmetry-enriched topological (SET) phases (see {\it e.g.} \cite{Maciejko:2010tx,Swingle:2010rf,Levin:2012ta,Cho:2012tx,Wan:2019oyr}).

Symmetry-enriched phases can be anomalous. This means the symmetry has an 't Hooft anomaly {\it i.e.} obstruction to gauging it. Then the system with the background gauge field cannot be defined in $(3+1)d$, but it can live on the boundary of a $(4+1)d$ bulk SPT phase that describes the anomaly. Different symmetry-enriched phases in general have different anomalies.
Since the 't Hooft anomaly is an invariant of the renormalization group flow, a symmetry-enriched phase that does not have the same anomaly as the microscopic model cannot be the only physics at the low energy\footnote{
An example is the ``symmetry-enforced gaplessness'' discussed in \cite{Wang:2014lca},  where a theory in $(2+1)d$ with a mixed anomaly between an $SU(2)$ symmetry and the time-reversal symmetry cannot flow to a gapped low energy theory that preserves the symmetries in the UV.
}.
This has applications to the infrared dualities in field theory, and the physics on domain walls or interfaces (see {\it e.g.} \cite{Gaiotto:2017yup,Komargodski:2017smk,Gaiotto:2017tne,Hsin:2018vcg}).

When the symmetry is continuous, the background gauge field can couple to a conserved current in the effective field theory. On the other hand, for theories that have non-local operators, there are additional ways to couple the system to the background gauge field. In fact, there may not even be a local conserved current, such as in a gapped SET phase.
In this note we will focus on how the presence of non-local operators affects the possible symmetry-enriched phases, where the symmetry can be continuous or discrete.

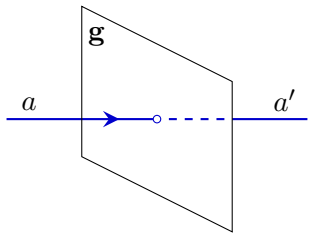
\begin{figure}[t]
\centering
\begin{tikzpicture}
\draw [thick, blue!80!black] (-1,.5) to (1,.5);
\draw [thick, blue!80!black] (2,.5) to (3,.5);
\draw (0,0) to (0,2) to (2,1) to (2,-1) to cycle;
\draw [thick, blue!80!black, dashed, dash phase=1.5, 
decoration = {markings, mark=at position -1.5 with {\arrow[scale=1.5]{stealth}}}, postaction=decorate] (1,.5) to (2,.5);
\draw [blue!80!black, fill=white] (1,.5) circle [radius=.05];
\node at (.2,1.6) {\small ${\bf g}$}; \node at (-.7,.7) {\small $a$}; \node at (2.7, .75) {\small $a'$};
\end{tikzpicture}
\caption{When a line (surface) operator of type $a$ crosses a codimension-one symmetry defect of type ${\bf g}\in G$, it leaves as a line (surface) operator of a different type $a'$. If the line (surface) operator generates a higher-form global symmetry, 
this represents the action of 0-form symmetry $G$ on the higher-form symmetries by automorphisms $a'=\rho_{\bf g}(a)$.
\label{fig: action on higher symmetry}}
\end{figure}

The classification of symmetry-enriched phases in $(2+1)d$ topological quantum field theory (TQFT) has been investigated in {\it e.g.} \cite{EtingofNOM2009,Essin:2013rca,Barkeshli:2014cna,Teo:2015xla,Tarantino:2015}. 
For the ordinary global symmetry $G$ that permutes the types of line operators (anyons) in a fixed way, the classification is
\begin{equation}\label{eqn:3dclassification}
H^2_\rho(BG,{\cal A})~,
\end{equation}
where ${\cal A}$ is the set of Abelian anyons, or the one-form global symmetry \cite{Gaiotto:2014kfa}, and $\rho$ is the action of the $G$ symmetry on the Abelian anyons (see figure \ref{fig: action on higher symmetry} with line operators).
In this note we will propose a generalization of this classification to $(3+1)d$, given in (\ref{eqn:classification}).

In general, a theory in $(3+1)d$ has line operators (particle excitations) and surface operators (string excitations)\footnote{
In this note the dimension spanned by an operator or a defect refers to the dimension in spacetime as opposed to the dimension in space. We will work in Euclidean signature and use the terms operator and defect interchangeably.
}.
Any non-trivial operator has at least one non-trivial correlation function.
Since lines cannot braid with other lines in $(3+1)d$, this implies that every topological line operator must have non-trivial braiding with some surface operator.
On the other hand, a non-trivial topological surface operator may have trivial braiding with all line operators if it braids non-trivially with other surface operators\footnote{
In $(3+1)d$ spacetime, two topological surface operators have contact interactions, while three topological surface operators can have non-trivial
triple-linking (``three-loop braiding'') correlation function \cite{Carter:2003TLK,Wang:2014xba,Jiang:2014ksa,Chen:2015gma,Tiwari:2016zru,Putrov:2016qdo}.
}. 

The non-local operators that are topological and invertible (obey group-law fusion rules) will play an important role in the discussion.
Such operators generate higher-form global symmetries \cite{Gaiotto:2014kfa}, and thus they are symmetry operators.
The symmetry line operators generate a two-form symmetry ${\cal B}$ that transforms the surface operators.
The two-form symmetry must transform at least a surface operator, otherwise the symmetry line operator would not have a non-trivial correlation function.
The symmetry surface operators generate a one-form symmetry ${\cal A}$ that transform line operators.
However, there are symmetry surface operators that generate a one-form symmetry that does not transform any line operator.
In condensed matter systems, higher-form global symmetry can emerge at low energy\footnote{
For applications of higher-form symmetry to condensed matter physics, see {\it e.g.} \cite{Hastings:2005xm,Wen:2018zux}.
}.
Different higher-form symmetries can also mix to form higher-groups, which means the backgrounds for the higher-form symmetries obey modified cocycle conditions that depend on the backgrounds of lower-form symmetries (see {\it e.g.} \cite{Kapustin:2013uxa,Cordova:2018cvg,Benini:2018reh}).\footnote{
In this note we will assume the theory does not have a non-trivial three-form global symmetry, which would be generated by a non-trivial local operator \cite{Gaiotto:2014kfa} that has scaling dimension zero for the correlation functions to be topological.}

The ordinary global symmetry (0-form symmetry) is generated by codimension-one symmetry volume defects \cite{Gaiotto:2014kfa}.
Thus different ways of coupling the theory to the background gauge field of 0-form symmetry $G$ correspond to different codimension-one symmetry $G$ defects that can be defined in the theory.

If a symmetry defect acts on line or surface operators as a one-form or two-form symmetry (namely, it transforms the operators by one-dimensional representations) then it is a one-form or two-form symmetry defect instead of 0-form symmetry defect.

In the following we will summarize several mechanisms for constructing these 0-form symmetry defects (they will be discussed in detail in section \ref{sec:classification}).
\begin{itemize}
\item[(1)] The first mechanism for defining 0-form symmetry defects is by an action on the local operators and by a permutation on the set of line and surface operators. This is the generalization of 0-form symmetry in $(2+1)d$ that permutes the types of anyons. For instance, the charge conjugation symmetry in $U(1)$ gauge theory permutes the Wilson line of charge $Q$ to the line of charge $-Q$ (see figure \ref{fig: action on higher symmetry}).
\end{itemize}
We use the higher-form symmetry defects to construct codimension-one symmetry defects, which roughly speaking correspond to ``direct products'' of higher-form symmetry defects. Some of these defects exhibit the following new features:\footnote{
There are no similar features in $(2+1)d$ TQFT without deconfined topological point operators.
}
\begin{enumerate}
\item[(2)]
There are 0-form symmetry defects that do not permute the types of non-local operators. 
Instead they dress the non-local operators that pass through them with redundant symmetry defects that modify only the contact correlation functions of the operators. In particular, they modify the junctions of open non-local operators (see figure \ref{fig:slidesurfacejunction} in section \ref{sec:permutation}).

\item[(3)]
There are $\mathbb{Z}_2$ 0-form symmetry defects that neither permute the types of non-local operators nor dress the non-local operators with redundant symmetry defects. 
Instead, they modify the line intersections between surface operators and the 0-form symmetry defects: when the orientation on the 0-form symmetry defect is reversed, there are additional symmetry line operators inserted at the line intersections (see figure \ref{fig:threegroupjunction} in section \ref{sec:permutation}).
\end{enumerate}
Finally, we have:
\begin{itemize}
\item[(4)]
The last mechanism modifies the 0-form symmetry defects by inserting higher-form symmetry defects at the junctions where three or more 0-form symmetry defects meet (see figure \ref{fig:3-junction} and figure \ref{fig:4-junction} in section \ref{sec:bgmodifyjunction}).
This modification can be detected in correlation functions involving junctions of multiple 0-form symmetry defects, 
as opposed to the symmetry defects produced by the first two mechanisms which can be detected in correlation functions involving only a single 0-form symmetry defect.
The modifications on the junctions must satisfy certain consistency conditions that can be described by higher-form symmetries (in general, a three-group global symmetry).

\end{itemize}

In this note we will focus on the classification of symmetry-enriched phases with a fixed action of the 0-form symmetry on local operators and a fixed permutation on the species of non-local operators.
Namely, we will focus on the above mechanisms except for the first one. 
The list is also not meant to be complete, but it reproduces and generalizes many results in the literature (see section \ref{sec:comparelit} for a comparison).

If the theory does not have a non-trivial two-form symmetry, then we show the classification of symmetry-enriched phases is the same as in $(2+1)d$, by $H^2_\rho(BG,{\cal A})$.
If the theory has two-form symmetry ${\cal B}$, then the classification is modified to be\footnote{
The two-form symmetry ${\cal B}$ also contributes additional one-form symmetry $H^2({\cal B},U(1))\subset {\cal A}$, which yields additional couplings by $H^2_\rho(G,{\cal A})$. See section \ref{sec:classification} for details.
}
\begin{equation}\label{eqn:classification}
(\eta_2,\nu_3,\xi)\in
H^2_\rho(BG,{\cal A})\times C^3(BG,{\cal B})\times H^1_\sigma(BG,H^3({\cal B},U(1))')~,
\end{equation}
where $H^3({\cal B},U(1)))'$ is the quotient of $H^3({\cal B},U(1))$ by the subgroup generated by its antisymmetric elements, and $\rho,\sigma$ are the $G$-actions induced by permuting the generators of the higher-form symmetries.
There is a constraint that specifies $\delta_\sigma\nu_3$ (differential acting on $\nu_3$ twisted by the $G$-action $\sigma$) in terms of $\eta_2,\xi$, and the equivalence relation on $\eta_2\in H^2_\rho(BG,{\cal A})$ induces a equivalence relation on $\nu_3$ by the constraint.
We determine how the constraint on $\delta_\sigma\nu_3$ depends on $\xi$. We discuss how to obtain the general constraint on $\delta_\sigma \nu_3$ by correlation functions.

The classification (\ref{eqn:classification}) of couplings does not include SPT phases of $G$ symmetry.
This is in the same spirit of the classification (\ref{eqn:3dclassification}) for symmetry fractionalizations in $(2+1)d$, where two symmetry fractionalizations that differ by an SPT phase are equivalent symmetry fractionalizations.
In fact, the classification of inequivalent SPT phases for $G$ symmetry depends on the coupling in (\ref{eqn:classification}). 
We provide a systematic method for obtaining the inequivalent SPT phases for each element in the classification (\ref{eqn:classification}). More on this later in section \ref{sec:couplingsSPT}.

The classification (\ref{eqn:classification}) in general includes theories with both anomalous and anomaly free $G$ symmetry.
We provide a systematic method of obtaining the relative anomaly for $G$ symmetry among theories with different couplings in (\ref{eqn:classification}) using the anomaly of higher-form symmetries.
The different couplings in (\ref{eqn:classification}) correspond to 0-form symmetry $G$ with different 't Hooft anomalies in general.
The 't Hooft anomaly of 0-form symmetry in $(3+1)d$ bosonic theory is classified by $H^5(BG,U(1))$, and it is given by the phase difference that appears between different ways of fusing five 0-form symmetry defects.
When changing the way of fusing 0-form symmetry defects, the higher-form symmetry defects inserted at the junctions can cross each other and induce additional 't Hooft anomalies of the 0-form symmetry from the 't Hooft anomalies of the higher-form symmetries.
In section \ref{sec:classification} we will discuss how the 't Hooft anomaly of 0-form symmetry depends on the couplings (\ref{eqn:classification}).

In this note we will focus on unitary internal global symmetry. We will also assume the theory is bosonic. In particular, all local operators must be bosonic. This means the theory does not require a spin (or more generally pin$^\pm$) structure.

\subsection{SPT phases trivialized by symmetry-enriched phases}
\label{sec:couplingsSPT}

In addition to the couplings (\ref{eqn:classification}) to the background gauge field, the symmetry-enriched phases also include the SPT phases of 0-form symmetry. The SPT phases for 0-form symmetry $G$ are classified in \cite{Chen:2011pg,Kapustin:2014tfa,Freed:2016rqq}.
We will show that some of the SPT phases become trivial when the background gauge field couples to the dynamics.
Examples of this phenomena were discussed in \cite{Wang:2016cto,Zou:2017ppq} using free $U(1)$ gauge theory in $(3+1)d$.
The existence of such SPT phases has dynamical consequence: it implies the system cannot be trivially gapped, even if the 0-form symmetry does not have an 't Hooft anomaly.

We show that such SPT phases of 0-form symmetry trivialized by the dynamics can be obtained from the 't Hooft anomaly of hidden anomalous higher-form symmetries.
In these examples, the SPT phases that are trivialized by the dynamics are reduced to lower-dimensional SPT phases on non-local operators.
For instance, in $(1+1)d$ $\mathbb{Z}_2$ gauge theory where the dynamical gauge field $a$ couples to the backgrounds $X,Y$ of $\mathbb{Z}_2\times\mathbb{Z}_2$ ordinary global symmetry as $\pi\int_{2d}aY$, the $(1+1)d$ SPT phase $\pi\int_{2d}XY$ can be removed by the $\mathbb{Z}_2$ one-form symmetry $a\rightarrow a+X$ \cite{Kapustin:2014gua} that gives rise to the $(0+1)d$ SPT phase $\pi\oint_{1d} X$ on the Wilson line $\pi\oint a$.

More generally, if the higher-form symmetry has anomaly described by the SPT phase in the $(1+D)$ dimensional bulk
\begin{equation}
\omega(B_2,B_3,\cdots B_n)=\omega(\{B_i\})
\end{equation}
where $B_i$ are the $i$-form gauge field for $(i-1)$-form symmetry ${\cal A}_i$, and we denote
the anomaly inflow to the $D$ dimensional boundary $M_D$ under the transformation $B_i\rightarrow B_i+\delta\lambda_{i-1}$ by
\begin{equation}\label{eqn:inflowgeneral}
\omega(\{B_i+\delta \lambda_{i-1}\})=\omega(\{B_i\})+ \int_{M_D} {\cal A}(\{B_i\},\{\lambda_i\})~.
\end{equation}
Then for the symmetry-enriched phase corresponds to $B_i=X^*\eta_i$ with $\eta_i\in H^i_{\rho_i}(BG,{\cal A}_{i-1})$ and $X$ being the background $G$ gauge field, the SPT phases that are ``eaten'' by the symmetry-enriched phases are classified by
\begin{equation}
\mu_{i-1}\in H^{i-1}_{\rho_{i-1}}(BG,{\cal A}_{i-1})~.
\end{equation}
The trivialized SPT phase corresponding to each $\{\mu_{i-1}\}\in \prod_i H^{i-1}_{\rho_{i-1}}(BG,{\cal A}_{i-1})$ is
\begin{equation}
{\cal A}(\{\eta_i\},\{\mu_{i-1}\})\in H^{D}(BG,U(1))~,
\end{equation}
and they are obtained by the anomaly inflow relation (\ref{eqn:inflowgeneral}) with the higher-form symmetry transformations $B_i\rightarrow B_i+\delta\lambda_{i-1}$ using the parameters $\lambda_{i-1}=X^*\mu_{i-1}$.

\subsection{One-form symmetry that acts trivially on all lines in TQFT}

In the application to the finite Abelian topological gauge theory in $(3+1)d$ we discover that there exists symmetry surface operators that braid trivially with all line operators, but they have non-trivial triple linking correlation functions with two other surface operators that have non-trivial holonomy of flat gauge connection.
This means that the one-form symmetry generated by these symmetry surface operators acts trivially on all line operators.

An example is $\mathbb{Z}_2\times\mathbb{Z}_2$ gauge theory without any Dijkgraaf-Witten term.
It can be described by the following BF theory
\begin{equation}
\frac{2}{2\pi}a^{(1)}db^{(1)}+\frac{2}{2\pi}a^{(2)}db^{(2)}~,
\end{equation}
where $a^{(i)}$ are $U(1)$ gauge fields and $b^{(i)}$ are $U(1)$ two-form gauge fields.
The theory has $\mathbb{Z}_2$ subgroup one-form symmetry generated by the following symmetry surface operator
\begin{equation}\label{eqn:projsurf}
\exp\left(\frac{2i}{2\pi}\oint a^{(1)}a^{(2)}\right)~,
\end{equation}
which describes a Dijkgraaf-Witten $\mathbb{Z}_2\times\mathbb{Z}_2$ gauge theory \cite{Dijkgraaf:1989pz} on the $(1+1)d$ surface, classified by $H^2(\mathbb{Z}_2\times\mathbb{Z}_2,U(1))=\mathbb{Z}_2$. The surface operator (\ref{eqn:projsurf}) braids trivially with the Wilson line operators $e^{i\oint a^{(i)}}$, and thus the one-form symmetry it generates acts trivially on all lines in the theory.
Nevertheless, the operator (\ref{eqn:projsurf}) is non-trivial as it has the following triple linking correlation function computed in section \ref{sec:correlationfunctionZn}:
\begin{equation}\label{eqn:tlkcorrel}
\left\langle
\exp\left(i\oint_{\Sigma}b^{(1)}\right)\exp\left(i\oint_{\Sigma'}b^{(2)}\right)
\exp\left(i{2\over 2\pi}\oint_{\Sigma''} a^{(1)}a^{(2)}\right)
\right\rangle
=(-1)^{\text{Tlk}(\Sigma,\Sigma',\Sigma'')}~ ,
\end{equation}
where $\Sigma,\Sigma',\Sigma''$ are three closed surfaces, and Tlk denotes the triple linking number \cite{Carter:2003TLK}.
As we will discuss in section \ref{sec:Abelianfinitegauge}, the one-form symmetry generated by the symmetry surface operator $\oint a^{(1)}a^{(2)}$ participates in a three-group global symmetry.

In general, the triple linking correlation functions are called three-loop braiding in the condensed matter literature \cite{Wang:2014xba,Jiang:2014ksa,Chen:2015gma,Tiwari:2016zru,Putrov:2016qdo}. However, the previous literature discussed a different braiding process where every surface operator involved braids non-trivially with at least one line operator, in contrast to the process (\ref{eqn:tlkcorrel}) where the surface operator (\ref{eqn:projsurf}) braids trivially with all lines.

We remark that the surface operator (\ref{eqn:projsurf}) factorizes on the torus, and thus the state it gives rise to on the torus can also be obtained from lines.
Nevertheless, such surface operators give rise to new symmetry-enriched phases via the one-form symmetry they generate.

\subsection{Comparison with literature}
\label{sec:comparelit}

In the following we would like to make some comparisons with two existing methods in the literature to study symmetry-enriched phases in $(3+1)d$.

One method begins with an SPT phase of an ordinary symmetry and then promotes the gauge field of a subgroup to be dynamical (see {\it e.g.} \cite{Mesaros:2012yd,Hung:2012nf,Guo:2017xex}).
This construction relies on having a microscopic gauge theory description for the effective theory\footnote{
See \cite{Lan:2018vjb} for an argument in favor of the existence of a gauge theory description for any bosonic $(3+1)d$ TQFT.
On the other hand, there are conformal field theories that do not have a Lagrangian description that preserves the global symmetry.
}, and there can be multiple microscopic descriptions for the same low energy theory.
This is unlike our approach which does not require a microscopic model.
In addition, any symmetry-enriched phase obtained from gauging an SPT phase with dynamical gauge field cannot have an 't Hooft anomaly for the remaining 0-form symmetry in the original SPT phase\footnote{There is a generalization that starts with a certain almost trivial theory instead of an SPT phase and gauges a subgroup symmetry with dynamical gauge field. This construction can produce a subset of anomalous SET phases, see {\it e.g.} \cite{Wang:2017loc,Wan:2018djl,Kobayashi:2019lep} and the review in section 2.6 of \cite{Tachikawa:2017gyf}.}.
On the contrary, the symmetry-enriched phases discussed in our approach can have 't Hooft anomalies, 
which have additional applications such as 't Hooft anomaly matching.
Furthermore, the approach of studying symmetry-enriched phases by gauging SPT phases does not distinguish how the global symmetry permutes the non-local operators.
On the other hand, in our approach we study the classification for fixed permutation, and thus we can identify the different mechanisms involved in the symmetry-enriched phases.
In section \ref{sec:comparegaugingSPT} we will make this comparison for the example of a finite group $H$ gauge theory with unitary finite group $G$ 0-form symmetry\footnote{
The finite groups $G,H$ can be non-Abelian.
}, by comparing the symmetry-enriched phases classified in our approach and the symmetry-enriched phases obtained by gauging the symmetry $H$ of the SPT phases with $G\times H$ symmetry.

Another method is to enumerate the anomalous quantum numbers of particle and string excitations under the 0-form symmetry. Different  quantum numbers correspond to different ways the 0-form symmetry ``acts'' on the non-local operators (see {\it e.g.} \cite{Cheng:2015kvt,Wang:2016cto,Zou:2017ppq}).
This is incorporated and generalized in the last mechanism that we discussed above.
If the higher-form symmetry defects inserted at the junctions of the 0-form symmetry defects generate a faithful higher-form symmetry, such insertions can be detected by correlation functions with the corresponding charged objects.
Such correlation functions can be interpreted as the line or surface operators carrying an 't Hooft anomaly of the 0-form symmetry on their worldvolumes. For instance, when the charged object that detects the insertions is a line operator (particles), this means that different insertions on the junction give the particle excitation different projective representations of the 0-form symmetry\footnote{
This is similar to the symmetry fractionalization classified by one-form symmetry in $(2+1)d$ (see {\it e.g.} \cite{EtingofNOM2009,Barkeshli:2014cna,Benini:2018reh}).
}.
Similarly, the modification of junctions by inserting symmetry line operators changes the 't Hooft anomaly of 0-form symmetry on the worldvolume of surface operators that are charged under the two-form symmetry.

On the other hand, there can be symmetry surface operators inserted at the junctions that do not generate a faithful one-form symmetry {\it i.e.} they have trivial braiding with all line operators. These modifications cannot be detected by the anomalous quantum number on the particle excitations and string excitations\footnote{
As we will discuss in section \ref{sec:classification} and \ref{sec:Abelianfinitegauge}, there are one-form symmetry defects that have non-trivial correlation functions only with at least two surface operators of another type.
}, but they nevertheless represent new 0-form symmetry defects.

\subsection{Summary of models}

We discuss several examples to illustrate the methods.
A gapless example is the pure $U(1)$ gauge theory in $(3+1)d$ with time-reversal symmetry,
\begin{equation}\label{eqn:U1action}
-\frac{1}{4e^2}F\wedge\star F + \frac{\theta}{2(2\pi)^2}F\wedge F~,
\end{equation}
where $F$ is the $U(1)$ field strength and the time-reversal symmetry requires $\theta\in \pi\mathbb{Z}$. The theory has line operators, labelled by the electric and magnetic charges $(q_\textbf{e},q_\textbf{m})$, and they
 transform under ${\cal A}=U(1)\times U(1)$ one-form global symmetry.
On the other hand, the theory has trivial two-form symmetry.
We classify different ways to couple the theory to the background of the time-reversal symmetry (more precisely, the time-reversing Lorentz symmetry $O(4)$), where the time-reversal symmetry ${\cal T}$ permutes the line operators as\footnote{
Such time-reversal symmetry does not commute with the electric $U(1)$ gauge rotation and is denoted by $U(1)_\text{gauge}\rtimes \mathbb{Z}_2^{\cal T}$.
}
\begin{equation}\label{eqn:Tparticle}
{\cal T}(q_\textbf{e})=q_\textbf{e},\qquad {\cal T}(q_\textbf{m})=-q_\textbf{m}~.
\end{equation}
The classification (\ref{eqn:classification}) in this case reduces to $H^2_{\cal T}(BO(4),U(1)\times U(1))$,
and it reproduces the classification of symmetry-enriched phases in \cite{Wang:2016cto}.

A gapped example is the Abelian finite group gauge theory in $(3+1)d$ with trivial Dijkgraaf-Witten action \cite{Dijkgraaf:1989pz}.
We discuss the $\mathbb{Z}_2$ gauge theory with time-reversal symmetry, and we show only a subset of the couplings can be obtained by Higgsing the time-reversal symmetric $U(1)$ theory.
We also reproduce the results in \cite{Chen:2016pec} for $\mathbb{Z}_2$ gauge theory.

As an application, we find the $\mathbb{Z}_2$ gauge theory coupled to the background gauge fields of 0-form and one-form symmetries with the 't Hooft anomaly as required to be the candidate topological sector in the non-supersymmetric $(3+1)d$ duality proposed in \cite{Cordova:2018acb,Bi:2018xvr}. The duality conjectures that $SU(2)$ gauge theory with one massless adjoint Dirac fermion flows at low energy to a free massless Dirac fermion with a decoupled $\mathbb{Z}_2$ gauge theory.
The theory in the UV has $\mathbb{Z}_8$ 0-form symmetry under which the fermions have charge one, and $\mathbb{Z}_2$ one-form symmetry that transforms the $SU(2)$ fundamental Wilson line. The symmetries have a mixed anomaly (see {\it e.g.}\cite{Anber:2018tcj,Cordova:2018acb}).
By scanning the different ways to couple the low energy theory to the background gauge field of 0-form symmetry we find that within the list of couplings we discussed, for the low energy theory to preserve the $\mathbb{Z}_8$ 0-form symmetry the $\mathbb{Z}_2$ one-form symmetry in the UV would have to be spontaneously broken\footnote{
A broken one-form symmetry means that the line operators transformed under the symmetry are deconfined at low energies \cite{Gaiotto:2014kfa}.
This is in contrast to what is proposed in \cite{Bi:2018xvr}, and the standard lore that adjoint QCD with small number of flavors confines (see {\it e.g.} \cite{Shifman:2013yca} and the references therein).
In the context of symmetry-enriched phases, a theory enriched by a one-form symmetry only means that it couples to the backgrounds of the symmetry, where the symmetry may or may not be broken depending on the coupling.
}. Our finding is consistent with \cite{Cordova:2019bsd}, which proves that such UV 't Hooft anomaly cannot be realized by symmetry-preserving gapped theories.
The $\mathbb{Z}_2$ gauge field is the corresponding Goldstone boson for the broken one-form symmetry. 
We also describe a general method to construct a TQFT of the Goldstone modes that matches any given 't Hooft anomaly of spontaneously broken finite group higher-form symmetries.

In another scenario \cite{Cordova:2018acb}, the $\mathbb{Z}_8$ 0-form symmetry is spontaneously broken to $\mathbb{Z}_4$ and the theory has a domain wall interpolating between the two vacua. The low energy theory in the vacua is proposed in \cite{Cordova:2018acb} to be a massless Dirac fermion with a decoupled $\mathbb{Z}_2$ gauge theory.
We propose a $\mathbb{Z}_2$ gauge theory coupled to background fields that realizes the anomaly in the UV, and we discuss several possible theories on the domain wall that match the anomaly.

The note is organized as follows.
In section \ref{sec:classification} we discuss different couplings to backgrounds of the 0-form symmetry using the higher-form symmetries.
In section \ref{sec:Abelianfinitegauge} we discuss the example of Abelian finite group gauge theory.
In section \ref{sec:adjointQCD4} we discuss the application to adjoint QCD$_4$ with two flavors.
In section \ref{sec:U(1)timereversal} we discuss $U(1)$ gauge theory with time-reversal symmetry.
In section \ref{sec:appendixSPT} we discuss how gauging higher-form symmetries in different symmetry-enriched phases can lead to different SPT phases. In particular, we show the $(3+1)d$ Gu-Wen phase \cite{Gu:2012ib} can be obtained by gauging the two-form symmetry of the symmetry-enriched $\mathbb{Z}_2$ gauge theory that has fermionic particle together with magnetic string that carries anomalous quantum number.
In section \ref{sec:moreexamples} we discuss several other examples.

There are several appendices. In appendix \ref{sec:appendixmath} we summarize some mathematical facts about cochains and higher cup products.
In appendix \ref{sec:ZnunitaryZn} we discuss the dimensional reduction to $(2+1)d$ for different symmetry-enriched phases of $\mathbb{Z}_n$ gauge theory.
In appendix \ref{sec:appendixUoneSO3} we reproduce the classification in \cite{Zou:2017ppq} for $U(1)$ gauge theory with $SO(3)$ and time-reversal symmetry.

\section{Classification of symmetry-enriched phases from higher-form symmetry}
\label{sec:classification}

This section classifies the symmetry enrichments of a theory as its couplings to a background field for the symmetry, i.e. as representations of the symmetry by codimension-one symmetry operators in the theory. We proceed by discussing several mechanisms for constructing these operators: permutation of non-local operators, forming products of higher codimension symmetry operators, and modifying defect junctions, discussed in sections 2.1, 2.2, and 2.3, respectively. Couplings related to the product mechanism have an invariant $\xi\in H^1_\sigma(BG,H^3(\mathcal{B},U(1))')$, while the junction modification mechanism yields invariants $(\eta_2,\nu_3)\in H^2_\rho(BG,\mathcal{A})\times C^3(BG,\mathcal{B})$. These three invariants must satisfy a constraint; this is the subject of section 2.4. We continue in section 2.5 with a description of the relative 't Hooft anomaly in terms of the three invariants. Section 2.6 studies the phenomenon of SPT absorption and its connection to the anomaly. Finally, we compare our methods with the procedure of building symmetry enriched phases by gauging SPT phases, in section 2.7.

\subsection{Symmetry defects from permuting non-local operators}

In this section we will discuss codimension-one symmetry defects that permute the species of non-local operators.
When a line (surface) operator pierces the codimension-one symmetry defect, it becomes another type of line (surface) operator (see figure \ref{fig: action on higher symmetry}). The local operators are also transformed when passing through the codimension-one defect.
Such codimension-one symmetry defects are determined by the detailed dynamics of the theory, and we will not attempt to classify them here.
The complete set of such codimension-one symmetry defects generates a 0-form symmetry ${\cal S}$ that we will call the intrinsic 0-form symmetry. Then the theory can couple to the $G$ gauge field by a homomorphism
\begin{equation}\label{eqn:permutationsymmetryf}
f\in \text{Hom}(G,{\cal S})~.
\end{equation}
The coupling corresponds to depositing on $g\in G$ symmetry defects the $f(g)\in {\cal S}$ symmetry defect that permutes the non-local operators.

The homomorphism $f$ satisfies constraints that arise from the requirement that coupling the system to the background gauge field for the 0-form symmetry does not activate independent background gauge fields for the higher-form symmetries. This restricts the possible couplings to the background of 0-form symmetry to those that do not combine with the one-form symmetry to be a two-group symmetry and do not combine with the two-form symmetry to be a three-group symmetry (for a discussion about higher-group symmetry see {\it e.g.} \cite{Kapustin:2013uxa,Cordova:2018cvg,Benini:2018reh}).
Denote the backgrounds for the one-form and two-form symmetries by $B_2, B_3$, and let $\rho, \sigma$ denote the permutation actions of $G$ on $\mathcal{A}, \mathcal{B}$ that arise from the action $f$ on line and surface operators. The higher group structure is described by the constraint
\begin{equation}\label{eqn:constraintf}
\delta_{X^*\rho}B_2=X^*\Theta_3,\qquad\delta_{X^*\sigma}B_3=X^*\Theta_4~,
\end{equation}
where $X^*$ denotes the pullback and $\delta_{X^*\rho}, \delta_{X^*\sigma}$ the twisted coboundary operations. The twisted classes $\Theta_3\in H^3_\rho(BG,{\cal A})$ and $\Theta_4\in H^4_\sigma(BG,{\cal B})$ arise from $f$.
The constraint on $f$ means that $\Theta_3$ and $\Theta_4$ vanish in cohomology, {\it i.e.} $\Theta_3=\delta_\rho\eta_2$ and $\Theta_4=\delta_\sigma\nu_3$ for $\eta_2\in C^2(BG,{\cal A})$ and $\nu_3\in C^3(BG,{\cal B})$. 
Thus the theory can couple to only the background $X$, with fixed backgrounds of the higher-form symmetries $B_2=X^*\eta_2$ and $B_3=X^*\nu_3$.

The group cocycles $\Theta_3$ and $\Theta_4$ in the constraint (\ref{eqn:constraintf}) can be detected by correlation functions.
Since the background field is Poincar\'e dual to the locus where the generating symmetry defects are inserted, 
the first equation in (\ref{eqn:constraintf}) says the intersection of four 0-form symmetry defect $g_1,g_2,g_3,g_1g_2g_3$ emits
the two-form symmetry surface defect $\Theta_3(g_1,g_2,g_3)$\footnote{This uses the property that the Poincar\'e dual of $\delta B_2$ is the boundary of the Poncar\'e dual of $B_2$.}.
Namely, different ways of fusing $g_1,g_2,g_3$ defects differ by additional symmetry surface defect $\Theta_3(g_1,g_2,g_3)$, and the latter can be detected by correlation functions.
Similarly $\Theta_4$ can be detected by the correlation functions for different ways of fusing four 0-form symmetry defects. 
$\Theta_3,\Theta_4$ are the generalizations of the $H^3$ obstruction to symmetry fractionalization discussed in \cite{EtingofNOM2009,Barkeshli:2014cna,Barkeshli:2017rzd}. In particular, they do not represent an 't Hooft anomaly of the 0-form symmetry, as emphasized in \cite{Cordova:2018cvg,Benini:2018reh}.

\subsection{Symmetry defects from higher-codimension symmetry defects}
\label{sec:permutation}

In this section we will construct symmetry defects from the generators of the higher-form symmetries.
For instance, when two symmetry defects are mutually local (do not have a non-trivial braiding correlation function), then taking their products give new symmetry defects of lower codimensions.
We will first give an example using $\mathbb{Z}_n$ gauge theory, and then discuss the generalization.

\subsubsection{$\mathbb{Z}_n$ gauge theory}

The bosonic $\mathbb{Z}_n$ gauge theory in $(3+1)d$ can be described by a $\mathbb{Z}_n$ one-cochain $u$ and a $\mathbb{Z}_n$ two-cochain $v_2$ with the action
\begin{equation}\label{eqn:Zngaugetheory}
\frac{2\pi}{n}\int u\delta v_2~.
\end{equation}
The equations of motion for $u,v_2$ impose that they are $\mathbb{Z}_n$ cocycles.

The theory has invertible topological line and surface operators
\begin{equation}\label{eqn:lineandsurfaceZn}
U=\exp\left({2\pi i\over n}\oint u\right),\qquad V=\exp\left({2\pi i\over n}\oint v_2\right)~.
\end{equation}
They generate a $\mathbb{Z}_n$ two-form symmetry and a $\mathbb{Z}_n$ one-form symmetry, respectively.
The theory couples to the backgrounds $B_2,B_3$ for the one-form and two-form symmetry by $\frac{2\pi}{n}\int uB_3$ and $\frac{2\pi}{n}\int B_2v_2$.
A direct computation (see {\it e.g.} \cite{Putrov:2016qdo}) shows that the line $U$ and the surface $V$ have a $\mathbb{Z}_n$ correlation function that depends on their linking number: for $U,V$ supported on the line $\gamma$ and surface $\Sigma$, it is
\begin{equation}\label{eqn:Znbraidingphase}
\langle U_\gamma V_\Sigma\rangle=\exp\left(-\frac{2\pi i}{n}\text{link}(\gamma,\Sigma)\right)~.
\end{equation}
This implies that the surface $V$ is charged under the two-form symmetry generated by $U$. Likewise,
the line $U$ is charged under the one-form symmetry generated by $V$. This means that the one-form and two-form symmetries have a mixed 't Hooft anomaly.

In addition, the theory has the following codimension-one symmetry defect from the $\mathbb{Z}_n$ one-form gauge field
\begin{equation}\label{eqn:0-formdefect2-form}
{\cal U}=\exp\left({2\pi i\over n}\oint u(\delta \tilde u/n) \right),\quad {\cal U}^n=1~,
\end{equation}
where the integrand equals $u\text{Bock}(u)$, with the Bockstein homomorphism for the short exact sequence $1\rightarrow \mathbb{Z}_n\rightarrow\mathbb{Z}_{n^2}\rightarrow \mathbb{Z}_n\rightarrow 1$. The tilde denotes a lift of $u$ to a $\mathbb{Z}_{n^2}$ cochain, and different lifts only changes the integrand by exact cocycle, while the integral is independent of the lift. Thus we will drop the tilde notation from now on.
The worldvolume of the defect supports a non-trivial $(2+1)d$ Dijkgraaf-Witten theory for $\mathbb{Z}_n$ gauge group \cite{Dijkgraaf:1989pz}.

The theory also has the following codimension-one symmetry defect from the $\mathbb{Z}_n$ two-form gauge field 
\begin{equation}\label{eqn:0-formdefect1-form}
{\cal V}=\exp\left({2\pi i\over n}\oint \delta \tilde v_2/n\right),\quad {\cal V}^n=1~,
\end{equation}
where the integrand equals $\text{Bock}(v_2)$. 
This codimension-one defect is a ``fractional surface operator'' that depends on the three-dimensional bounding manifold.
At the three-junctions of this codimension-one symmetry defect ${\cal V}^{q_1},{\cal V}^{q_2},{\cal V}^{-[q_1+q_2]_n}$ where $[q]_n=q\text{ mod }n$,
there is a symmetry defect $V^{\left(q_1+q_2-[q_1+q_2]_n\right)/n}$ for the one-form symmetry. Thus coupling the theory to background field using this codimension-one symmetry defect is the same as modifying the three-junction by generators of the one-form symmetry. This belongs to a special case in section \ref{sec:bgmodifyjunction} and we will not count it as an independent coupling.

In addition to the codimension-one defects discussed above, there are also codimension-two symmetry defects from the two-form symmetry, generated by
\begin{equation}\label{eqn:codimensiontwodefecttwoform}
{\cal W}=\exp\left({2\pi i\over n}\oint \delta u/n\right),\quad {\cal W}^n=1~.
\end{equation}
Similar to ${\cal V}$, this is a ``fractional symmetry line defect'' that is specified by the symmetry line operator appears at the three-junction of ${\cal W}$.
As demonstrated in section \ref{sec:Abelianfinitegauge},
such defects have only contact correlation functions {\it i.e.} they are redundant symmetry defects.
Therefore we will not count them as non-trivial surface operators. 
Similar to the defect (\ref{eqn:0-formdefect1-form}), they can form junctions that support non-trivial line operators.

If the gauge group is $\mathbb{Z}_{n_1}\times\mathbb{Z}_{n_2}$ instead of $\mathbb{Z}_n$, with $\mathbb{Z}_{n_1}$ and $\mathbb{Z}_{n_2}$ cochains $(u^1,v^1_2),(u^2,v^2_2)$, then there are additional codimension-one symmetry defects
\begin{equation}
{\cal U}^{12}=\exp\left(
{2\pi i\over n_1n_2}\oint u^1\delta u^2\right)~,
\end{equation}
and $\oint u^1 v_2^2$, $\oint u^2 v^1_2$. These are valid because $\oint u^1$ is mutually local with $\oint \text{Bock}(u^2),\oint v_2^2$.

Similarly, there is an additional symmetry surface defect
\begin{equation}
{\cal W}^{12}=\exp\left({2\pi i\over\gcd(n_1,n_2)}\oint u^1 u^2\right)~.
\end{equation}
Unlike $\oint \text{Bock}(u^1)$ or $\oint \text{Bock}(u^2)$, the symmetry defect ${\cal W}^{12}$ has non-trivial correlation functions. For instance, it can link with two other surface operators $\oint v_2^1,\oint v_2^2$, and we compute the correlation functions in section \ref{sec:Abelianfinitegauge}. Thus the symmetry surface defect ${\cal W}^{12}$ is a non-trivial operator in the theory.

\subsubsection{Generalization}

Here we will generalize the above construction to any theory with finite two-form symmetry.

Our argument uses the property that the two-form symmetry in $(3+1)d$ does not have an 't Hooft anomaly by itself (while there can be mixed anomalies with other symmetries\footnote{
An example is the mixed anomaly between $\mathbb{Z}_2$ two-form symmetry and the bosonic Lorentz symmetry, with the 5$d$ SPT \cite{Gaiotto:2015zta,Kapustin:2017jrc}
$
\pi\int_{5d} Sq^2(B_3)=\pi\int_{5d}(w_2+w_1^2)B_3,
$
where $B_3$ is the background for the two-form symmetry, and $w_1,w_2$ are the first and second Stiefel-Whitney classes for the manifold.}).
This follows from the fact that there are no non-trivial correlation functions that involve only the symmetry line operators.
Another way to see this is that one cannot write down a corresponding 5$d$ SPT phase.
This property of two-form symmetry enables us to gauge the two-form symmetry and obtain a new theory without two-form symmetry.

Any finite Abelian group is isomorphic to a product of cyclic groups ${\cal B}=\prod_i \mathbb{Z}_{n_i}$. Let us first consider the case where the two-form symmetry is ${\cal B}=\mathbb{Z}_n$.

We gauge the two-form symmetry in the theory by introducing a dynamical three-form gauge field $b_3$. The resulting theory has an emergent $\mathbb{Z}_n$ 0-form symmetry generated by $\exp\left({2\pi i\over n}\oint b_3\right)$, which we could then gauge to recover the original theory. Denote the gauge field for this $\mathbb{Z}_n$ 0-form gauge symmetry by $u$. It couples as a Lagrangian multiplier $\frac{2\pi}{n}\int ub_3$.
Thus we find the exact duality
\begin{equation}\label{eqn:dualitytwoform}
\text{Theory A}\quad\longleftrightarrow\quad \text{Theory B}={\text{Theory A}\times (\mathbb{Z}_n\text{ gauge theory})\over \mathbb{Z}_n^{(2)} }~,
\end{equation}
where the quotient denotes gauging the diagonal $\mathbb{Z}_n$ two-form symmetry with dynamical gauge field $b_3$, and the Wilson lines of the $\mathbb{Z}_n$ gauge theory are generated by $\oint u$. The duality (\ref{eqn:dualitytwoform}) is nothing but a $\mathbb{Z}_n$ discrete Fourier transform and its inverse.

The advantage of the duality (\ref{eqn:dualitytwoform}) is that in the dual Theory B the two-form symmetry of the original theory is now generated by the $\mathbb{Z}_n$ Wilson line $U=\exp\left({2\pi i\over n}\oint u\right)$, and we can use the $\mathbb{Z}_n$ gauge field $u$ to construct the symmetry defects ${\cal U,W}$ in (\ref{eqn:0-formdefect2-form}),(\ref{eqn:codimensiontwodefecttwoform}). 

The generalization to arbitrary discrete two-form symmetry ${\cal B}=\prod \mathbb{Z}_{n_i}$ is straightforward. 
The 0-form symmetry defects can be constructed using the $\mathbb{Z}_{n_i}$ gauge fields in the dual description. On their worldvolumes, the defects have $(2+1)d$ Dijkgraaf-Witten theories \cite{Dijkgraaf:1989pz} with gauge group ${\cal B}$, and thus they are classified by $H^3({\cal B},U(1))$.

Similarly, there are symmetry surface operators of the form $\oint u u'$ that arise from the two-form symmetry. These generate a subgroup $H^2({\cal B},U(1))\subset {\cal A}$ of the one-form symmetry.
Below we will argue that this subgroup one-form symmetry does not transform any line operators, but the surface operators (of the type $\oint uu'$) that generate this one-form symmetry have non-trivial correlation functions with other surface operators.
On the other hand, the surface operators $\oint\text{Bock}(u)$ are trivial in $H^2({\cal B},U(1))$ and thus they are excluded in this classification. This is consistent with the discussion below where we find that they have only contact correlation functions {\it i.e.} they are redundant symmetry defects.

\subsubsection{Correlation functions}
\label{sec:correlationgeneral}

We can use the duality (\ref{eqn:dualitytwoform}) to compute the correlation functions of the above symmetry defects built from the two-form symmetry. 

First, we need to understand the operators in the dual theory B.
The line operators are all possible tensor products of the lines in theory A and the Wilson lines in the $\mathbb{Z}_n$ gauge theory, subject to the identification using the generators of the diagonal two-form gauge symmetry.
On the other hand, the surface operators must be invariant under the diagonal gauge two-form symmetry.
Since the surface operator $V$ is charged under the two-form symmetry in the $\mathbb{Z}_n$ gauge theory, it must be paired with another surface in theory A that has the opposite two-form charge to make the tensor product of the surface operators gauge invariant.

The above consideration implies that the symmetry defects built from the two-form symmetry (denote schematically by ${\cal U}$ of codimension-one and ${\cal W}$ of codimension-two) have trivial correlation functions with the operators from the theory A.
Moreover, since the symmetry defects ${\cal U,W}$ also have trivial correlation functions with lines in the ${\cal B}$ gauge theory, as demonstrated in section \ref{sec:correlationfunctionZn}, these symmetry defects have trivial correlation functions with all line operators.
On the other hand, their correlation functions with the surface operators only depend on the two-form charges of the surface operators and can be computed by replacing each surface operator ${\cal O}^{\{q_i\}}_\text{surface}$ of two-form symmetry charges $\{q_i\}\in {\cal B}=\prod_i\mathbb{Z}_{n_i}$ with the surface operator in ${\cal B}$ gauge theory of the same two-form charge:\footnote{
The correlation function (\ref{eqn:substitutionZn}) is normalized by that without ${\cal U},{\cal W}$ insertions to describe the interactions between ${\cal U},{\cal W}$ and other operators.
}
\begin{equation}\label{eqn:substitutionZn}
\large\langle \cdots {\cal O}^{\{q_i\}}_\text{surface}\cdots {\cal U}\cdots {\cal W}\cdots\large\rangle=
\large\langle \cdots \left(\prod_{i} (V_i)^{q_i}\right)\cdots {\cal U}\cdots {\cal W}\cdots\large\rangle_\text{${\cal B}$ gauge theory}~,
\end{equation}
where $V_i$ is the basic surface operator in the $\mathbb{Z}_{n_i}$ gauge theory charged under the two-form symmetry.
This replacement is carried out for each operator in the correlation function.
The correlation functions in the ${\cal B}$ gauge theory are discussed in section \ref{sec:correlationfunctionZn}.

In particular, from (\ref{eqn:substitutionZn}) and the correlation functions in the ${\cal B}$ gauge theory we find that the symmetry surface defects $\oint \text{Bock}(u^i)$ have only contact correlation functions, and therefore they should not be included in the list of non-trivial operators.

It is also interesting to look at the correlation functions of the non-trivial symmetry surface operators constructed from symmetry line operators, classified by $H^2({\cal B},U(1))$ for two-form symmetry ${\cal B}$.
From the computation in section \ref{sec:correlationfunctionZn} we find this symmetry surface operator has non-trivial triple-linking correlation function with two other surface operators, and thus this operator is non-trivial.
On the other hand, it has trivial correlation functions with any line operators. 
Therefore, any theory in $(3+1)d$ with a finite two-form symmetry ${\cal B}$ with non-trivial $H^2({\cal B},U(1))$ has a subgroup one-form symmetry $H^2({\cal B},U(1))$ generated by non-trivial surface operators, but nevertheless this one-form symmetry does not act on any line operators.\footnote{
In $(2+1)d$ fermionic theories there is also $\mathbb{Z}_2$ one-form symmetry generated by the transparent fermion line $\psi$ that does not transform any line operators. However, the theory here can be bosonic or fermionic.
}

\subsubsection{Action on non-local operators}
\label{sec:permutationactionopr}

Consider the codimension-one symmetry defect given by the ``direct product'' $L\times L'$ of a symmetry line defect and a symmetry surface defect that are mutually local with each other ({\it i.e.} they have trivial mutual braiding). 
This is the codimension-one symmetry defect with the feature that when it wraps $\Sigma_2\times S^1$, the symmetry line and symmetry surface operators $L,L'$ can wrap $S^1$ and $\Sigma_2$.
When an operator that is charged under the one-form or two-form symmetries generated by the symmetry defects $L,L'$ pierces the codimension-one symmetry defect, the intersection is not well-defined. Therefore the operator is modified after passing through the symmetry defect (with an exception to be discussed later).

For any theory with two-form symmetry ${\cal B}$, we can use the duality (\ref{eqn:dualitytwoform}) to construct the following codimension-one symmetry defects using the ${\cal B}$ gauge field.
They fall into the following three categories. As we will show below, the defects of the first two categories permute the non-local operators, while those of the third category does not.

The first category uses the subgroup two-form symmetry $\mathbb{Z}_{n_1}\times \mathbb{Z}_{n_2}\times\mathbb{Z}_{n_3}\subset {\cal B}$ gives the following codimension-one symmetry defect of order $\ell=\gcd(n_1,n_2,n_3)$:
\begin{equation}\label{eqn:defecttwoformpermute}
{\cal U}_1=\exp\left({2\pi i\over \ell}\oint u^1 u^2 u^3\right)~,
\end{equation}
where $u^i$ is the $\mathbb{Z}_{n_i}$ gauge field.
The theory can couple to background gauge field $X$ of $\mathbb{Z}_\ell$ 0-form symmetry using this symmetry defect:
\begin{equation}
\int\left(\sum_{i=1}^3 \frac{2\pi}{n_i}u^i\delta v_2^i\right)+\frac{2\pi}{\ell}\int u^1u^2u^3X~.
\end{equation}
The equation of motion for $v_2^i$ constrains $u^i$ to be a $\mathbb{Z}_{n_i}$ cocycle, while the equation of motion for $u^i$ gives
\begin{equation}
\delta v_2^1+{n_1\over \ell}u^2u^3X=0\text{ mod }n_1~,
\end{equation}
and two similar equations given by the cyclic permutations of 1,2,3.
Under a background transformation $X\rightarrow X-\delta\lambda$ changes $v_2^1$ as
\begin{equation}\label{eqn:examplepermuteabeliancubic}
v_2^1\;\longrightarrow\; v_2^1+{n_1\over \ell}u^2u^3\lambda~,
\end{equation}
and similarly it changes $v_2^2,v_2^3$ as above with cyclic permutations.
The 0-form global symmetry transformation corresponds to constant $\lambda\in \mathbb{Z}_\ell$.
This means that the codimension-one symmetry defect (\ref{eqn:defecttwoformpermute}) permutes the surface operators by multiplying with additional codimension-two symmetry defects ${\cal W}$ built from the two-form symmetry:
\begin{equation}\label{eqn:permutationsymma1a2a3}
V_1=\exp\left({2\pi i\over n_1}\oint v_2^1\right)\;\longrightarrow\; V_1\,{\cal W}_{23}^\lambda,\qquad
{\cal W}_{23}=\exp\left({2\pi i\over \ell}\oint u^2u^3\right)~.
\end{equation}
It is easy to verify that the permutation is indeed a symmetry of the correlation functions.
More generally, using the duality (\ref{eqn:dualitytwoform}) the permutation symmetry acts the same way for all surface operators with the same two-form charge as $V_1$.

The second category uses the subgroup one-form symmetry $\mathbb{Z}_{n_1}\times\mathbb{Z}_{n_2}$ to construct the following codimension-one defect of order $\gcd(n_1,n_2)$
\begin{equation}
{\cal U}_2=\exp\left({2\pi i\over \gcd(n_1,n_2)}\oint u^1v_2^2\right)~.
\end{equation}
By a similar computation as above, we find this defect permutes the types of both the line and surface operators: under a $\mathbb{Z}_{\gcd(n_1,n_2)}$ global symmetry with parameter $\lambda$,
\begin{align}
\oint u^2\rightarrow \oint u^2-\frac{n_2}{\gcd(n_1,n_2)}\oint u^1\lambda,\qquad
\oint v_2^1\rightarrow \oint v_2^1+\frac{n_1}{\gcd(n_1,n_2)}\oint v_2^2\lambda~.
\end{align}

\begin{figure}[t]
  \centering
    \includegraphics[width=0.35\textwidth]{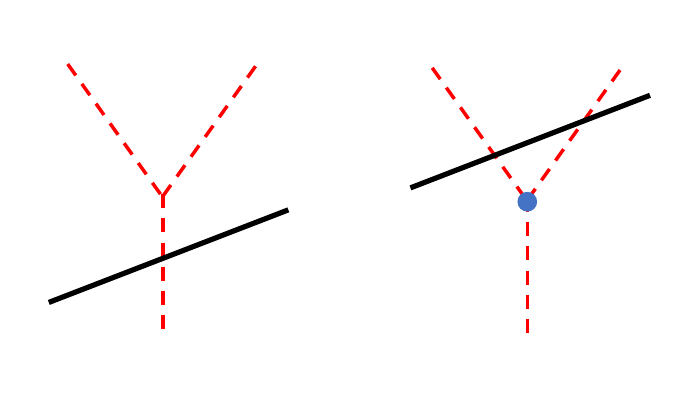}
      \caption{Sliding the 0-form symmetry defect (\ref{eqn:symmetrydefectnotpermute}) (denoted by the solid black lines) across the junction of three surface operators (denoted by the red dashed lines) produces an additional symmetry line defect at the junction (denoted by the blue dot). The 0-form symmetry defect acts on the surface operators on the lower side of the solid black line in the figure as in (\ref{eqn:transformthreegroupdefect}).
}\label{fig:slidesurfacejunction}
\end{figure}

The third category uses the subgroup two-form symmetry $\mathbb{Z}_n\subset {\cal B}$ or $\mathbb{Z}_{n_1}\times\mathbb{Z}_{n_2}\subset {\cal B}$. They correspond to the elements in $H^3({\cal B},U(1))$ that are not completely antisymmetric (roughly speaking, a Chern-Simons-like cocycle).
As we will show below, they give rise to symmetry defects that only modify the surface operators by a redundant symmetry defect.
From the subgroup of two-form symmetry $\mathbb{Z}_{n_1}\times\mathbb{Z}_{n_2}\subset {\cal B}$ one can construct the following codimension-one symmetry defect of order $\ell'=\gcd(n_1,n_2)$\footnote{
The image of the Bockstein homomorphism has order $\ell'$: since $\gcd(n_1,n_2)=\alpha_2 n_1+\alpha_1 n_2$ for some integers $\alpha_1,\alpha_2$, $\ell'\text{Bock}(u^2)=n_1(\alpha_2\text{Bock}(u^2))+\delta(\alpha_1 u^2)$ is trivial in the cohomology with $\mathbb{Z}_{n_1}$ coefficient.
}:
\begin{equation}\label{eqn:symmetrydefectnotpermute}
{\cal U}'=\exp\left({2\pi i\over n_1n_2}\oint u^1\delta u^2\right)~,
\end{equation}
where $u^i$ are $\mathbb{Z}_{n_i}$ gauge fields, and $\delta u^2/n_2\text{ mod }n_1=\text{Bock}(u^2)$ uses the Bockstein homomorphism of the short exact sequence $1\rightarrow\mathbb{Z}_{n_1}\rightarrow\mathbb{Z}_{n_1n_2}\rightarrow\mathbb{Z}_{n_2}\rightarrow 1$.
By a similar computation as before one finds that the codimension-one symmetry defect ${\cal U}'$ implements the following transformation with constant $\lambda'\in\mathbb{Z}_{\ell'}$:
\begin{equation}\label{eqn:transformthreegroupdefect}
V_1=\exp\left({2\pi i\over n_1}\oint v_2^1\right)\;\longrightarrow\; V_1\; {\cal W}_2^{\lambda'},\quad
{\cal W}_2=\left({2\pi i \over \ell'}\oint\frac{\delta u^2}{n_2}\right)~.
\end{equation}
Therefore the surface operator is modified by multiplication with a codimension-two symmetry defect that is not a non-trivial operator.
Since these codimension-one symmetry defects do not implement non-trivial permutations on the non-local operators, they give rise to new mechanisms of coupling the theory to the background field of the 0-form symmetry.
The redundant symmetry defect ${\cal W}_2$ has the symmetry line defect $\oint u^2$ at its three-junctions, and thus the transformation (\ref{eqn:transformthreegroupdefect}) modifies the three-junctions of the surface operators $\oint v_2^1$ (where three open surfaces meet at a line) by inserting additional symmetry line defects (see figure \ref{fig:slidesurfacejunction}).

For general ${\cal B}=\prod_{i}\mathbb{Z}_{n_i}$ two-form symmetry, the defect (\ref{eqn:symmetrydefectnotpermute}) can be parametrized as
\begin{equation}\label{eqn:generalsurfacenotpermute}
\prod_i\exp\left(\alpha^i \frac{2\pi i}{n_i}\oint u^i{\delta u^i\over n_i}\right)
\prod_{i>j}\exp\left(\alpha^{ij}{2\pi i\over n_i}\oint u^i{\delta u^j\over n_j}\right)~
\end{equation}
with integers $\alpha^i\in\mathbb{Z}_{n_i}$, $\alpha^{ij}\in \mathbb{Z}_{n_{ij}}$.
The 0-form symmetry defect acts on the surface operators as follows: ($\lambda^i,\lambda^{ij}$ are numbers valued in $\mathbb{Z}_{n_i}$ and $\mathbb{Z}_{n_{ij}}$)
\begin{equation}
\oint v_2^i\longrightarrow
\oint  v_2^i+2\alpha^i\oint \frac{\delta u^i}{n_i}\lambda^i+\sum_{j\neq i}\alpha^{ij}\frac{n_i}{n_{ij}}\oint \frac{\delta u^j}{n_j}\lambda^{ij}
~.
\end{equation}

We remark that in general not every dressing that is compatible with the fusion rules is a symmetry of the theory. For instance, in $\mathbb{Z}_{n}\times\mathbb{Z}_{n}$ gauge theory, the transformation that
changes the surface operator $\oint v_2^1\rightarrow \oint v_2^1+\oint \text{Bock}(u^2)$ but leaves invariant $\oint v_2^2$ does not preserve the correlation function between $\oint v_2^2$ and the three-junction of surfaces $\oint v_2^1$.
On the other hand, in $\mathbb{Z}_n$ gauge theory any dressing of surface operators by redundant symmetry defects that is compatible with the fusion rules of surface operators corresponds to a 0-form symmetry defect.

\begin{figure}[t]
  \centering
    \includegraphics[width=0.2\textwidth]{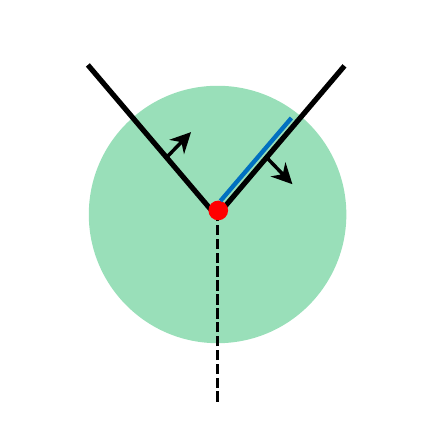}
      \caption{The junction of two $\mathbb{Z}_2$ 0-form symmetry defects (\ref{eqn:exception}) 
      (indicated by the black solid line) 
      fusing into the trivial defect (indicated by the dashed line). The 0-form symmetry defects meet at the codimension-two red surface.
      The black arrows indicate their orientations.
Intersecting the junction with the green surface $\oint v_2$ (that fills the plane in the figure) emits an additional symmetry line defect $\oint u$ (indicated by the blue line) on one of the two branches.}
\label{fig:threegroupjunction}
\end{figure}

So far we have described the 0-form symmetry defects constructed as ``direct products'' that act on non-local operators by permuting the types of non-local operators or dressing them with redundant symmetry defects.
The exception occurs for the following special case of (\ref{eqn:generalsurfacenotpermute}) given by the $\mathbb{Z}_2$ 0-form symmetry defects constructed from every subgroup $\mathbb{Z}_n\subset {\cal B}$ two-form symmetry for even $n$:
\begin{equation}\label{eqn:exception}
{\cal U}''=\exp\left(\pi i\oint u\frac{\delta u}{n}\right)~.
\end{equation}
Inserting such a defect by turning on background $X$ for the $\mathbb{Z}_2$ 0-form symmetry changes the equation of motion of $u$ into
\begin{equation}\label{eqn:eomuv}
\delta v_2 = u\text{Bock}(X)~.
\end{equation}
This means that $\oint v_2$ is invariant under a background gauge transformation $X\rightarrow X-d\lambda$.
Thus one finds that such defect does not permute the types of non-local operators, and it also does not dress the surface operators with redundant symmetry defects.
From the equation of motion (\ref{eqn:eomuv}), under $X\rightarrow X+2Y$ for integer one-cochain $Y$, the surface operator $\oint v_2$ transforms as
\begin{equation}
\oint v_2\rightarrow \oint (v_2+uY)~.
\end{equation}
In particular, changing the orientation $X\rightarrow -X$ inserts an additional symmetry line defect $\oint u$ at the intersection of the surface operator with the 0-form symmetry defect\footnote{
The defect ${\cal U}''$ that intersects with a surface operator $\oint_\Sigma v_2$ is no longer $\mathbb{Z}_2$ valued: ${\cal U}''^n=\exp(\pi i \oint u\text{PD}(\Sigma))$.
}.
Another way to see the effect of the $\mathbb{Z}_2$ 0-form symmetry defect is by intersecting the surface operator $\oint v_2$ with the three-junction of the 0-form symmetry defects, which creates an additional symmetry line operator emitted from the intersection (see figure \ref{fig:threegroupjunction})\footnote{
When the surface operator generates a one-form symmetry, this describes a three-group symmetry.
}.

To summarize, the defects in the first two categories belong to the defects constructed by permutation.
On the other hand, the defects from the third category and the exceptions are new 0-form symmetry defects. They are classified by the elements in
\begin{equation}\label{eqn:newdefectnotpermute}
H^3({\cal B},U(1))'=H^3({\cal B},U(1))/H^3({\cal B},U(1))_A~,
\end{equation}
where ${\cal B}$ is the two-form symmetry, $H^3({\cal B},U(1))_A$ is the subgroup generated by the antisymmetric elements in $H^3({\cal B},U(1))$.
For general ${\cal B}=\prod_i\mathbb{Z}_{n_i}$ the groups are
$H^3({\cal B},U(1))_A=\prod_{i,j,k|i>j>k}\mathbb{Z}_{\gcd(n_i,n_j,n_k)}$ and
$H^3({\cal B},U(1))'=\prod_{i,j|i\geq j}\mathbb{Z}_{\gcd(n_i,n_j)}$, and $H^3({\cal B},U(1))$ is isomorphic to their direct product
$H^3({\cal B},U(1))_A\times H^3({\cal B},U(1))'$.

The new defects in (\ref{eqn:newdefectnotpermute}) give rise to additional ways to couple the theory to the background of 0-form symmetry $G$, as classified by
\begin{equation}\label{eqn:xicoupling}
H^1_\sigma(BG,H^3({\cal B},U(1))')~,
\end{equation}
where $\sigma$ is the action induced by the permutation of $G$ on the symmetry line operators $\oint u$.

\begin{figure}[t]
  \centering
      \includegraphics[width=0.3\textwidth]{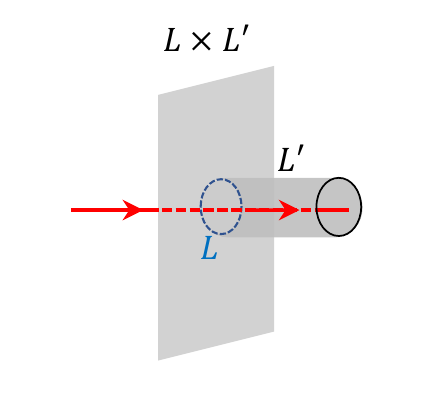}
  \caption{The red operator pierces the codimension-one symmetry defect $L\times L'$ where $L$ supports the symmetry defect not local with respect to the red operator. Then by shrinking the circumference of the cylinder we find the red operator leaves the codimension-one symmetry defect with (a power of) additional symmetry defect $L'$ attached to it. }\label{fig:permsymm}
\end{figure}

We can give an intuitive picture for the modification of the operators passing through the codimension-one symmetry defect (excluding the exception (\ref{eqn:exception}) where the operators are not modified).
Since the operator coming toward the codimension-one symmetry defect cannot intersect the defect, it deforms the nearby region on the codimension-one defect into a long cigar that encompasses the incoming operator.
Then as the operator passes through the codimension-one symmetry defect, the cigar becomes an infinitely-long thin cylinder enclosing the incoming operator. The circumference of the cylinder wraps the operator with the constituent ($L$ or $L'$) of the defect that is not local with the incoming operator, while the longitudinal direction of the cylinder has the other constituent symmetry defect that is local with respect to the incoming operator (see figure \ref{fig:permsymm}).
Then by shrinking the circumference of the cylinder to be zero we find the ``outgoing operator'' that leaves the codimension-one symmetry defect is given by the incoming operator attached with the constituent symmetry defect which is local with respect to the incoming operator. 
(More precisely, due to the braiding between the constituent defect on the circumference and the incoming operator, the shrinking produces a power of the longitudinal defect that depends on the braiding.) This reproduces the computation earlier.

\subsection{Coupling by modifying the defect junctions}\label{sec:bgmodifyjunction}

After the action of the 0-form symmetry $G$ on the operators is fixed, and we choose a coupling in (\ref{eqn:xicoupling}), there are additional ways to couple the theory to the background $G$ gauge field given by modifying the 0-form symmetry defect junctions with different insertions of symmetry operators that generate higher-form symmetries\footnote{
Since the 0-form symmetry defects are topological and invertible, the insertions must also be operators that are topological and invertible {\it i.e.} higher-form symmetry defects.}.

Since the insertion of a symmetry operator is equivalent to turning on the background gauge field for the corresponding symmetry (given by the Poincar\'e dual of the cycle where the operator is inserted),
modifying the junctions of 0-form symmetry defects by insertions of symmetry operators that generate higher-form symmetries is equivalent turning on different backgrounds of higher-form symmetries that are specified by the 0-form symmetry background. 
In this language, we can derive the consequences of such modifications from the properties of higher-form symmetries as discussed in \cite{Gaiotto:2014kfa}.

In section \ref{sec:modifyjunctions} we will demonstrate explicitly how the different fixed backgrounds of higher-form symmetries are related to different modifications on the junctions of 0-form symmetry defects. 
The discussion is similar to the classification of symmetry fractionalizations in $(2+1)d$ discussed in \cite{Barkeshli:2014cna}.

\subsubsection{Coupling by fixed higher-form symmetry backgrounds}

The 0-form symmetry, one-form symmetry and two-form symmetry in general form a three-group symmetry.
This means that there are constraints between their background gauge fields $X,B_2,B_3$ in the form of modified cocycle conditions\footnote{
An example where the higher-form symmetries are $\mathbb{Z}_n$ is
\begin{equation}
\delta_\rho B_2=X^*\Theta_3,\qquad
\delta_\sigma B_3 = B_2 X^*\Theta_2+X^*\Theta_4~,
\end{equation}
where $\Theta_3\in H^3_\rho(BG,\mathbb{Z}_n)$, $\Theta_i\in H^i_\sigma(BG,\mathbb{Z}_n)$ for $i=2,4$.
The case of interest here is $\Theta_3=0,\Theta_4=0$, thus it is consistent with setting $B_2=0,B_3=0$ with non-trivial $X$.
}
.
Since in the end we would like to couple the theory only to the 0-form symmetry background $X$, we will restrict to the subgroups of ${\cal A},{\cal B}$ such that we can turn off $B_2=0,B_3=0$ with non-trivial $X$ in the constraints. 
In particular, the background for the one-form symmetry must be a twisted cocycle $\delta_\rho B_2=0$, where $\rho$ denotes the action of 0-form symmetry by permutation.

The constraints between $B_2,B_3,X$ depends on the choice of couplings classified by $H^1_\sigma(BG, H^3({\cal B}),U(1))'$.
This will be discussed in section \ref{sec:constraint}.

Once the constraints are determined, the theory can couple to the background $X$ of the 0-form symmetry $G$ in different ways by turning on fixed backgrounds $B_2,B_3$ for each coupling\footnote{
The theory can couple to the backgrounds $B_2,B_3$ for the higher-form symmetries (not specified by $X$) in the following way: 
we replace $B_2,B_3$ by $B_2+X^*\eta_2$ and $B_3+X^*\nu_3$ instead of by $X^*\eta_2,X^*\nu_3$ as in (\ref{eqn:4dbackgrounds}).
See appendix \ref{sec:appendixSPT} for an application of this method of coupling the theory to both the 0-form and higher-form symmetry backgrounds.
}
\begin{equation}\label{eqn:4dbackgrounds}
B_2=X^*\eta_2,\quad B_3=X^* \nu_3~,
\end{equation}
where $\eta_2\in H_\rho^2(BG,{\cal A})$ is a twisted group two-cocycle, and $\nu_3\in C^3(BG,{\cal B})$ satisfies constraints induced from the constraints on the backgrounds $B_2,B_3,X$. 
The coupling parameter $\eta_2$ is defined in the group cohomology, since an exact cocycle corresponds to a background one-form gauge transformation and thus decouples from the theory.
Similarly, the background gauge transformations of $X,B_2,B_3$ also induce gauge transformations on $\nu_3$, making it into a torsor. 
The constraints and the gauge transformations will be discussed in more details in section \ref{sec:constraintidentifications}.

In addition to the gauge transformations, the couplings $(\eta_2,\nu_3)$ are subject to the identification from the remaining intrinsic  0-form symmetry after turning on the background gauge field of symmetry $G$:
\begin{equation}\label{eqn:identificationsymmetryset}
N(f(G),{\cal S})/f(G)~,
\end{equation}
where $f:G\rightarrow {\cal S}$ specifies how $G$ symmetry permutes the non-local operator as in (\ref{eqn:permutationsymmetryf}), and
$N(f(G),{\cal S})$ is the normalizer of $f(G)$ in ${\cal S}$.
Every element in (\ref{eqn:identificationsymmetryset}) generates an action on the higher-form symmetry backgrounds (modulo $G$ actions) by permuting their symmetry generators and thus identifies the couplings $(\eta_2,\nu_3)$ in (\ref{eqn:4dbackgrounds}) related by the action. In particular, if $f$ is surjective then there is no non-trivial identification from (\ref{eqn:identificationsymmetryset}).

After imposing the constraint and identifications on $\eta_2,\nu_3$, the resulting parameters $(\eta_2,\nu_3)$ represent different couplings to the 0-form symmetry background $G$.

As an example, consider the special case where the 3-group symmetry is a trivial product of the 0-form symmetry, one-form and two-form symmetries. We allow the 0-form symmetry to act on the higher-form symmetry by permutations $\rho,\sigma$ as before.
Then the backgrounds of the higher-form symmetries satisfy the twisted cocycle conditions
\begin{equation}
\delta_\rho B_2=0,\qquad \delta_\sigma B_3=0~.
\end{equation}
This implies that the parameters $(\eta_2,\nu_3)$ takes value in
\begin{equation}\label{eqn:4dbackgroundsordinary}
(\eta_2,\nu_3)\in H^2_\rho(BG,{\cal A})\times H^3_\sigma(BG,{\cal B})~
\end{equation}
up to the identification by the intrinsic permutation 0-form symmetry (\ref{eqn:identificationsymmetryset}).

Let us explore the consequences of different coupling parameters $(\eta_2,\nu_3)$ in (\ref{eqn:4dbackgrounds}) using the properties of higher-form global symmetry.

In the presence of the backgrounds $B_2,B_3$ for the one-form and two-form symmetries, the line and surface operators that are charged under the symmetries are attached to open surfaces and open volumes that carry fluxes of the classical fields $\int B_2,\int B_3$, with coefficients specified by the one-form and two-form charges of the line and surface operators.
Thus for different backgrounds $B_2,B_3$ in (\ref{eqn:4dbackgrounds}), the attached fluxes change the anomaly of 0-form symmetry $G$ on the worldvolume of line and surface operators by anomaly inflow. 

As an example, consider the 0-form symmetry $G=\mathbb{Z}_2\times \mathbb{Z}_2$ that does not act on the higher-form symmetries, 
and thus $B_2$ obeys the standard cocycle condition. For simplicity take the one-form symmetry to be $\mathbb{Z}_2$. Denote the backgrounds for the 0-form symmetry by the $\mathbb{Z}_2$ gauge fields $X,X'$.
Then the fixed background $B_2=XX'$ implies the line operators charged under the one-form symmetry have an anomaly for the $\mathbb{Z}_2\times\mathbb{Z}_2$ 0-form symmetry on the worldline, described by the SPT phase $\pi\int XX'$.
In other words, the particles on such line operators transform as the projective representation of the 0-form symmetry where the two $\mathbb{Z}_2$s do not commute on the Hilbert space for the $(0+1)d$ quantum mechanics.

Another consequence for different coupling parameters $(\eta_2,\nu_3)$ is that they give rise to different selection rules on  amplitudes.
Consider the path integral on a compact spacetime with a non-local operator charged under the higher-form symmetry and wrapping a non-trivial cycle.
In the absence of backgrounds $B_2,B_3$, the path integral must vanish due to the higher-form global symmetry \cite{Gaiotto:2014kfa}.
If we turn on backgrounds $B_2,B_3$, and if there is a non-trivial mixed anomaly between the one-form and two-form symmetries, then the  higher-form symmetry backgrounds modify the selection rules by inserting extra symmetry defects that themselves carry higher-form symmetry charges.
Thus different couplings $(\eta_2,\nu_3)$ that produce different backgrounds $B_2,B_3$ as in (\ref{eqn:4dbackgrounds}) lead to different selection rules on the amplitude. We will give an explicit example using $\mathbb{Z}_n$ gauge theory in section \ref{sec:selectionrule}.

\subsubsection{Modification on the defect junctions}
\label{sec:modifyjunctions}

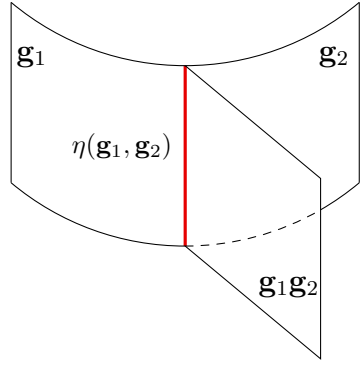
\begin{figure}[t]
\centering
\begin{tikzpicture}[scale=.6]
\draw [very thick, red!90!black] (0,-2) to (0,2); % vertical red line
\draw (0,2) to (3,-.5) to (3,-4.5) to (0,-2); % planes
\draw ([shift=(-130:6)] 0,8) arc (-130:-50:6);
\draw ([shift=(-130:6)] 0,4) arc (-130:-90:6); \draw [dashed, line cap=round] ([shift=(-90:6)] 0,4) arc (-90:-60:6); \draw ([shift=(-60:6)] 0,4) arc (-60:-50:6);
\draw ([shift=(-130:6)] 0,4) to ++(0,4); \draw ([shift=(-50:6)] 0,4) to ++(0,4);
\node at (-3.4,2.2) {${\bf g}_1$}; \node at (3.3,2.2) {${\bf g}_2$}; \node at (2.3,-2.9) {${\bf g}_1{\bf g}_2$}; % labels
\node at (-1.4,.2) {\footnotesize $\eta({\bf g}_1, {\bf g}_2)$};
\end{tikzpicture}
\caption{A junction (in red) where three codimension-one 0-form symmetry defects of type ${\bf g}_1,{\bf g}_2,{\bf g}_1{\bf g}_2\in G$ meet in codimension two, with one additional dimension suppressed in the figure.
The junction can be modified by inserting codimension-two symmetry defects of one-form symmetry $\eta({\bf g}_1,{\bf g}_2)\in{\cal A}$.
\label{fig:3-junction}}
\end{figure}

Three $G$ defects meet at a codimension-two junction (see figure \ref{fig:3-junction}) specified by $g_1,g_2\in G$. 
A line $x$ encircling the three-junction produces the phase $\hat \eta_{x}(g_1,g_2)$: 
\begin{equation}\label{eqn:eta}
\raisebox{-4.5em}{\begin{tikzpicture}
\draw [thick, densely dashed, decoration = {markings, mark=at position .3 with {\arrow[scale=1.5]{stealth}}}, postaction=decorate] (0,-1.5) node[below] {${\bf g}_1{\bf g}_2$} to (0,0);
\draw [thick, densely dashed, decoration = {markings, mark=at position .9 with {\arrow[scale=1.5]{stealth}}}, postaction=decorate] (0,0) to (-1,1.5) node[above] {${\bf g}_1$};
\draw [thick, densely dashed, decoration = {markings, mark=at position .9 with {\arrow[scale=1.5]{stealth}}}, postaction=decorate] (0,0) to (1,1.5) node[above] {${\bf g}_2$};
\draw [thick, decoration = {markings, mark=at position .49 with {\arrowreversed[scale=1.5,rotate=5]{stealth}}}, postaction=decorate] (0,.1) circle [radius = .7];
\node at (-1,0) {$x$}; \node at (0,1.2) {$\sigma_{{\bf {\bar g}}_1}x$}; \node at (1,-.8) {$\sigma_{{\bf{\bar g}}_2 {\bf {\bar g}}_1}x$};
\filldraw[fill=red,draw=red] (0,0) circle [radius=.06];
\end{tikzpicture}}
= \; \hat\eta_x({\bf g}_1, {\bf g}_2) \;
\raisebox{-4.5em}{\begin{tikzpicture}
\draw [thick, densely dashed, decoration = {markings, mark=at position .5 with {\arrow[scale=1.5]{stealth}}}, postaction=decorate] (0,-1.5) node[below] {${\bf g}_1{\bf g}_2$} to (0,0);
\draw [thick, densely dashed, decoration = {markings, mark=at position .7 with {\arrow[scale=1.5]{stealth}}}, postaction=decorate] (0,0) to (-1,1.5) node[above] {${\bf g}_1$};
\draw [thick, densely dashed, decoration = {markings, mark=at position .7 with {\arrow[scale=1.5]{stealth}}}, postaction=decorate] (0,0) to (1,1.5) node[above] {${\bf g}_2$};
\draw [thick, decoration = {markings, mark=at position .48 with {\arrowreversed[scale=1.5,rotate=8]{stealth}}}, postaction=decorate] (-.9,-.5) circle [radius = .5];
\node at (-1.7,-.55) {$x$};
\filldraw[fill=red,draw=red] (0,0) circle [radius=.06];
\end{tikzpicture}}
\end{equation}
where dashed lines denote the codimension-one 0-form symmetry $G$ defects (see figure \ref{fig:3-junction}). The three 0-form symmetry defects meet at a codimension-two junction denoted by the red point (this is the red line in figure \ref{fig:3-junction}), which braids with the line operator.

The subgroup of the 0-form symmetry that does not permute the line operator $x$ (namely, the stabilizer subgroup $G_x\subset G$) is a 0-form symmetry in the $(0+1)d$ wordline of $x$.
The phase $\hat\eta_x$ in (\ref{eqn:eta}) implies the Hilbert space on the worldline is in a projective representation of the global symmetry $G_x$, and thus the phase $\hat\eta_x$ describes the anomaly of the global symmetry $G_x$ on the $(0+1)d$ worldline of $x$ \cite{Benini:2018reh}.

A consistency condition for the phase $\hat\eta$ is as follows. Consider the four-junction of 0-form symmetry defects in figure \ref{fig:4-junction}, and sliding a loop $x$ from the bottom to the top of the junction. Since the intersection of the four-junction has codimension-three, it has trivial braiding with the loop and thus the phases produced from the bottom and top three-junctions must agree:
\begin{equation}\label{eqn:constrainteta}
\hat\eta_{\sigma_{{\bar{\bf g}}_1}(x)}({\bf g}_2,{\bf g}_3)\hat\eta_x({\bf g}_1,{\bf g}_2{\bf g}_3)
=
\hat\eta_x({\bf g}_1,{\bf g}_2)\hat\eta_x({\bf g}_1{\bf g}_2,{\bf g}_3)~.
\end{equation}
In particular, for 0-form symmetry defects in the stabilizer subgroup $G_x$, this is consistent with $\hat\eta_x$ being the anomaly on the $(0+1)d$ worldvolume of the line $x$, which takes values in $H^2(BG_x,U(1))$, and (\ref{eqn:constrainteta}) coincides with the cocycle condition.

\begin{figure}[t]
\centering
\begin{tikzpicture}
\draw [thick, red!90!black] (0,-2) to (0,2); % vertical line
\draw [thick, red!90!black] (0,0) arc (125:155.3:4.5); % wavy line
\draw [thick, red!90!black, dashed, line cap=round] (0,0) arc (-55:-31.7:3.8); 
\draw [thick, red!90!black] (1.5,2.19) arc (-14:-31.5:3.8);
\draw (0,2) to (3,-.5) to (3,-4.5) to (0,-2); % plane in front
\draw ([shift=(-130:6)] 0,8) arc (-130:-50:6);
\draw ([shift=(-130:6)] 0,4) arc (-130:-90:6); \draw [dashed, line cap=round] ([shift=(-90:6)] 0,4) arc (-90:-60:6); \draw ([shift=(-60:6)] 0,4) arc (-60:-50:6);
\draw ([shift=(-130:6)] 0,4) to ++(0,4); \draw ([shift=(-50:6)] 0,4) to ++(0,4);
\draw [dashed, dash phase = -2, line cap=round] (-1.5,-1.81) to ++(-1,1.8); \draw (1.5,2.19) to ++(-1,1.8); % plane in back
\draw [dashed, line cap=round] (-2.5,-0.01) to ([shift=(53.1:2.61)] -2.5,-0.01);
\draw [line cap=round] ([shift=(53.1:2.61)] -2.5,-0.01) to ([shift=(53.1:5)] -2.5,-0.01);
\node at (-3.5,2.5) {${\bf g}_1$}; \node at (.5,3.3) {${\bf g}_2$}; \node at (3.4,2.5) {${\bf g}_3$}; \node at (2.4,-3.3) {${\bf g}_1{\bf g}_2{\bf g}_3$}; % labels
\node at (-.7,-1.5) {\small ${\bf g}_1{\bf g}_2$}; \node at (0.9,1.75) {\small ${\bf g}_2{\bf g}_3$};
\node at (-1.3,.2) {\footnotesize $\zeta({\bf g}_1, {\bf g}_2,{\bf g}_3)$};
\filldraw[fill=blue,draw=blue] (0,0) circle [radius=.05];
\end{tikzpicture}
\caption{A junction where four codimension-one 0-form symmetry defects of type ${\bf g}_1$, ${\bf g}_2$, ${\bf g}_3$, ${\bf g}_1{\bf g}_2{\bf g}_3\in G$ meet in codimension three, with one additional dimension suppressed in the figure.
The junctions of three codimension-one defects are in red, and their intersection is the blue point. The codimension-three intersection can be modified by inserting codimension-three symmetry defects of two-form symmetry $\nu({\bf g}_1, {\bf g}_2,{\bf g}_3)\in {\cal B}$. 
Note the four-junction considered here does not have additional defects emanating from the codimension-three intersection, in contrast to the four-junction that described two-group symmetry discussed in \cite{Benini:2018reh}. In particular, a line operator encircling the upper-half of the junction produces the same phase as the line encircling the lower-half, since a line has trivial braiding with the codimension-three intersection.
\label{fig:4-junction}}
\end{figure}

The three-junction can be modified by inserting a codimension-two symmetry defect of the one-form symmetry\footnote{
If the symmetry surface defects are redundant one-form symmetry defects, the insertion is equivalent to inserting symmetry line defects on higher-codimension junctions and thus we do not need to count such cases twice.}
$\eta(\textbf{g}_1,\textbf{g}_2)\in{\cal A}$, see figure \ref{fig:3-junction}.
Since line operators braid with the defects of the one-form symmetry (the braiding gives the one-form charges of the line operators),
the modification of the junction changes the phase $\hat \eta_x$ as 
\begin{equation}\label{eqn:changeeta}
\hat\eta_x({\mathbf g}_1,{\mathbf g}_2)\quad\longrightarrow\quad \hat\eta_x({\mathbf g}_1,{\mathbf g}_2)\cdot {\cal M}_{x,\;\eta({\mathbf g}_1,{\mathbf g}_2)}~,\quad \eta({\mathbf g}_1,{\mathbf g}_2)\in {\cal A}~,
\end{equation}
where ${\cal M}_{x,\eta({\mathbf g}_1,{\mathbf g}_2)}$ denotes the eigenvalue of $x$ under the one-form symmetry element $\eta({\mathbf g}_1,{\mathbf g}_2)$ by braiding.
To satisfy the constraint (\ref{eqn:constrainteta}), $\eta$ has to be a group cocycle twisted by the action $\rho$ (note ${\cal M}_{\sigma_{\bf g}(x),\;\rho_{\bf g}(y)}={\cal M}_{x,\;y}$).
This makes ${\hat \eta}_x\in H^2(BG,U(1))$ into a $H^2_\rho(BG,{\cal A})$ torsor.
Thus different couplings $\hat\eta_x$ are classified by $\eta\in H^2_\rho(BG,{\cal A})$. The modification of the junction is equivalent to changing the background of the one-form symmetry (\ref{eqn:4dbackgrounds}).
The modification of the phase $\hat \eta_x$ in (\ref{eqn:changeeta}) implies that the worldline of $x$ has an anomaly of the global symmetry $G_x$ that depends on $\eta$. This is consistent with (\ref{eqn:4dbackgrounds}), since changing the background $B_2$ for the one-form symmetry attaches the line operator to additional surface $\int \Delta B_2=\int X^*\eta$ that describes the $(1+1)d$ SPT phase for the global symmetry on its boundary worldline.

Similarly, consider the four-junction of $G$ defects meeting at a codimension-three line (see figure \ref{fig:4-junction}), specified by $g_1,g_2,g_3\in G$. A surface operator $y$ encircling the four-junction produces the phase $\hat \nu_y(g_1,g_2,g_3)$:
\begin{equation}\label{eqn:zeta}
\raisebox{-4.5em}{\begin{tikzpicture}
\filldraw [fill=lightgray,draw=lightgray, postaction=decorate] (0,.1) circle [radius = .7];
\node at (-1,0) {$y$}; 
\draw [thick, densely dashed, decoration = {markings, mark=at position .3 with {\arrow[scale=1.5]{stealth}}}, postaction=decorate] (0,-1.5) node[below] {${\bf g}_1{\bf g}_2$} to (0,0);
\draw [thick, densely dashed, decoration = {markings, mark=at position .9 with {\arrow[scale=1.5]{stealth}}}, postaction=decorate] (0,0) to (-1,1.5) node[above] {${\bf g}_1$};
\draw [thick, densely dashed, decoration = {markings, mark=at position .9 with {\arrow[scale=1.5]{stealth}}}, postaction=decorate] (0,0) to (1,1.5) node[above] {${\bf g}_2$};
\draw [thick, densely dashed, decoration = {markings, mark=at position .7 with {\arrow[scale=1.5]{stealth}}}, postaction=decorate] (0,0) to (1.7,0.2) node[above] {${\bf g}_3$};
\filldraw[fill=blue,draw=blue] (0,0) circle [radius=.06];
\end{tikzpicture}}
\;= \quad \hat\nu_y({\bf g}_1, {\bf g}_2,{\bf g}_3) \;
\raisebox{-4.5em}{\begin{tikzpicture}
\draw [thick, densely dashed, decoration = {markings, mark=at position .5 with {\arrow[scale=1.5]{stealth}}}, postaction=decorate] (0,-1.5) node[below] {${\bf g}_1{\bf g}_2{\bf g}_3$} to (0,0);
\draw [thick, densely dashed, decoration = {markings, mark=at position .7 with {\arrow[scale=1.5]{stealth}}}, postaction=decorate] (0,0) to (-1,1.5) node[above] {${\bf g}_1$};
\draw [thick, densely dashed, decoration = {markings, mark=at position .7 with {\arrow[scale=1.5]{stealth}}}, postaction=decorate] (0,0) to (1,1.5) node[above] {${\bf g}_2$};
\draw [thick, densely dashed, decoration = {markings, mark=at position .7 with {\arrow[scale=1.5]{stealth}}}, postaction=decorate] (0,0) to (1.7,0.2) node[above] {${\bf g}_3$};
\filldraw [fill=lightgray,draw=lightgray, postaction=decorate] (-.9,-.5) circle [radius = .5];
\node at (-1.7,-.55) {$y$};
\filldraw[fill=blue,draw=blue] (0,0) circle [radius=.06];
\end{tikzpicture}}
\end{equation}
where dashed lines denote the codimension-one 0-form symmetry $G$ defects (see figure \ref{fig:4-junction}). 

The subgroup of the 0-form symmetry that does not permute the surface operator $y$ (namely, the stablizer subgroup $G_y\subset G$) is a 0-form symmetry in the $(1+1)d$ wordvolume of $y$.
Consider sliding a surface operator $y$ from the bottom to the top of the four-junction (\ref{eqn:zeta}).
The phase in (\ref{eqn:zeta}) implies that the $F$-move of 0-form symmetry defect on the worldvolume produces a phase $\hat\nu_y$:
\begin{equation}
\label{eqn:2danomaly}
(\text{Surface worldvolume}):\;
\raisebox{-4em}{\begin{tikzpicture}
\draw [thick] (0,-.8) node[below] {${\bf g}_1{\bf g}_2{\bf g}_3$} to (0,0);
\draw [thick] (0,0) to (-1.2,1.6) node[above] {${\bf g}_1$};
\draw [thick] (0,0) to (1.2,1.6) node[above] {${\bf g}_3$};
\draw [thick] (-.4,.533) to (0,1) to (0,1.6) node[above] {${\bf g}_2$};
\end{tikzpicture}}
\; = \; \hat\nu_y({\bf g}_1,{\bf g}_2,{\bf g}_3)\;\;
\raisebox{-4em}{\begin{tikzpicture}
\draw [thick] (0,-.8) node[below] {${\bf g}_1{\bf g}_2{\bf g}_3$} to (0,0);
\draw [thick] (0,0) to (-1.2,1.6) node[above] {${\bf g}_1$};
\draw [thick] (0,0) to (1.2,1.6) node[above] {${\bf g}_3$};
\draw [thick] (.4,.533) to (0,1) to (0,1.6) node[above] {${\bf g}_2$};
\end{tikzpicture}} ~,
\end{equation}
where the solid line denotes the $G_y$ symmetry defects. Thus the phase $\hat\nu_y$ describes the anomaly of the global symmetry $G_y$ on the $(1+1)d$ worldvolume of $y$ \cite{Benini:2018reh}.

When there 0-form and higher-form symmetries do not mix into a three-group, one can derive an identity similar to (\ref{eqn:constrainteta}) to show the phase $\hat\nu$ is a cocycle.

The four-junction can be modified by inserting a codimension-three symmetry defect of the two-form symmetry $\nu(\textbf{g}_1,\textbf{g}_2,\textbf{g}_3)\in {\cal B}$, see figure \ref{fig:4-junction}. 
Since surface operators braid with the symmetry defects of the two-form symmetry (the braiding gives the two-form charge of the surface operators),
the modification changes the phase $\hat\nu_y$ as
\begin{equation}\label{eqn:changezeta}
\hat\nu_y(g_1,g_2,g_3)\quad\longrightarrow\quad \hat\nu_y(g_1,g_2,g_3)\cdot {\cal M}_{\nu(g_1,g_2,g_3),\;y}~,\quad \nu(g_1,g_2,g_3)\in {\cal B}~.
\end{equation}
where ${\cal M}_{\nu(g_1,g_2,g_3),y}$ is the eigenvalue of $y$ under the two-form symmetry element $\nu(g_1,g_2,g_3)$ by braiding.
When the symmetries do not form a three-group, the constraint on $\hat\nu$ implies $\nu\in H^3_\sigma(BG,{\cal B})$, and thus $\hat\nu_y\in H^3(BG,U(1))$ is a $H^3_\sigma(BG,{\cal B})$ torsor.
The modification of the phase $\hat \nu_y$ in (\ref{eqn:changezeta}) implies the worldvolume of $y$ has an anomaly of the global symmetry $G_y$ that depends on $\nu$. This is consistent with changing the background $B_3$ by $X^*\nu$ in (\ref{eqn:4dbackgrounds}).

\subsection{Constraint and equivalence relation}\label{sec:constraintidentifications}
\label{sec:constraint}

Consider the theory coupled to the backgrounds $B_2,B_3,X$ with the constraint
\begin{equation}\label{eqn:originalthreegroup}
\delta_\sigma B_3=\Xi(B_2,X)~.
\end{equation}
We want to study how the cocycle $\Xi$ depends on
the coupling $\xi\in H^1_\sigma(BG,H^3({\cal B},U(1))')$ in (\ref{eqn:xicoupling}), which corresponds to codimension-one symmetry defects constructed from the two-form symmetry that have trivial permutation action.

First, we need to identify the one-form symmetry that can be described by the ${\cal B}$ gauge theory in the duality (\ref{eqn:dualitytwoform}).
The full one-form symmetry ${\cal A}$ is the group extension of an Abelian group ${\cal A}'$ by $H^2({\cal B},U(1))$, where the subgroup $H^2({\cal B},U(1))$ is generated by the symmetry surface operators of the type $\oint uu'$ in the ${\cal B}$ gauge theory as discussed in section \ref{sec:permutation}. 
Let $r_{\cal A}':{\cal A}\rightarrow {\cal A}'={\cal A}/H^2({\cal B},U(1))$ be the quotient map.
The background $B_2$ of the one-form symmetry ${\cal A}$ can be expressed in terms of the two-form background $r_{\cal A}'(B_2)$ valued in ${\cal A}'$ and the background $r_{\cal A}(B_2)$,\footnote{
The notation $r_{\cal A}$ here does not refer to a homomorphism unless the group extension of ${\cal A}'$ by $H^2({\cal B},U(1))$ splits.
} for the subgroup one-form symmetry $H^2({\cal B},U(1))$ with the constraint $\delta r_{\cal A}(B_2)=\text{Bock}(r_{\cal A}'(B_2))$, where Bock is the Bockstein homomorphism for the short exact sequence $1\rightarrow H^2({\cal B},U(1))\rightarrow {\cal A}\rightarrow {\cal A}'\rightarrow 1$.

The ${\cal B}$ gauge theory also has one-form symmetry generated by the symmetry surfaces $\oint v_2$. 
However, due to the two-form gauge symmetry in the duality (\ref{eqn:dualitytwoform}), every surface $\oint v_2$ must pair with another surface operator which may not be a symmetry surface, thus the surface operators $\oint v_2$ do not correspond to a subgroup of the one-form symmetry.

Denote by ${\cal A}_{\cal B}$ the subgroup of the one-form symmetry that consists of symmetry surface operators that are not charged under the two-form symmetry. Namely, they are the symmetry surface operators that have trivial braiding with all symmetry line operators. Then denote the quotient map
\begin{equation}\label{eqn:oneformmappi}
\pi:\quad {\cal A}\longrightarrow {\cal A}/{\cal A}_{\cal B}~.
\end{equation}
The quotient $\pi({\cal A})$ is the one-form symmetry corresponding to $\oint v_2$. Since $H^2({\cal B},U(1))\subset {\cal A}_{\cal B}$, there is a subgroup ${\cal A}_{\cal B}'\subset {\cal A}'$ with the quotient map $\pi':{\cal A}'\rightarrow {\cal A}'/{\cal A}_{\cal B}'\cong {\cal A}/{\cal A}_{\cal B}$. 

We will also need the property that the one-form symmetry and two-form symmetry have a bilinear function $\langle\cdot,\cdot\rangle: {\cal A}\times{\cal B}\rightarrow U(1)$ given by the braiding between the symmetry surface operators and symmetry line operators. 
From the bilinear function one can define a linear map
\begin{equation}
{\cal M}_{\cal AB}:\quad {\cal A}\longrightarrow \text{Hom}({\cal B},U(1))={\hat {\cal B}}\cong {\cal B}
\end{equation}
by ${\cal M}_{\cal AB}(\alpha)=\langle \alpha,\cdot\rangle$ for each $\alpha\in {\cal A}$.
In particular, ker ${\cal M}_{\cal AB}\cong{\cal A}_{\cal B}$ and $\text{im }{\cal M}_{\cal AB}\cong {\cal A}/{\cal A}_{\cal B}= \pi({\cal A})$.
Thus ${\cal M}_{\cal AB}$ defines another linear map
\begin{equation}
{\cal M}_{\cal AB}'\quad \pi({\cal A})\longrightarrow {\cal M}_{\cal AB}({\cal A})\xhookrightarrow{} \text{Hom}({\cal B},U(1))={\hat {\cal B}}\cong {\cal B}~.
\end{equation}
In the following, we will use the same symbol $\pi$ for ${\cal M}'_{\cal AB}\circ \pi$ as a short hand notation.
For simplicity, we will take a basis in ${\cal B}=\prod_i\mathbb{Z}_{n_i}$, with $H^2({\cal B},U(1))\cong \prod_{i>j}\mathbb{Z}_{\gcd(n_i,n_j)}$.

Using the duality (\ref{eqn:dualitytwoform}), the theory can couple to the backgrounds $B_2,B_3,X$ using the ${\cal B}$ gauge theory with (\ref{eqn:Abeliantheorycoupling}) for $B_2^i=\pi(B_2)^i=\pi'(r_{\cal A}'(B_2))^i$, $C_2^{ij}=r_{\cal A}(B_2)^{ij},(X^{ij})=X^*(\xi^{ij})$:\footnote{
If ${\cal A}\neq H^2({\cal B},U(1))\times {\cal A}'$, then $\delta C_2^{ij}=\text{Bock}(r_{\cal A}'(B_2))\neq 0$ and the ${\cal B}$ gauge field has a bulk dependence that depends on $r_{\cal A}'(B_2)$ (see (\ref{eqn:bulkAbeliangaugedepn})), which must cancel the bulk dependence from the other sector in the theory B description in (\ref{eqn:dualitytwoform}) for the coupling to the background $r_{\cal A}'(B_2)$ to be consistent.
}
\begin{align}\label{eqn:couplingextra}
&\sum_{i\geq j}\frac{2\pi}{n_{ij}n_j}\int \left(u^i\delta_\sigma u^j-u^i\pi(B_2)^j-u^j \pi(B_2)^i
+\pi(B_2)^i\cup_1\delta_\sigma u^j \right)X^*(\xi^{ij})\cr
&\qquad+\sum_{i>j}\frac{2\pi}{n_{ij}}\int u^iu^jr_{\cal A}(B_2)^{ij}+\sum_i\frac{2\pi}{n_i}\int u^i B_3^i~,
\end{align}
with $\delta u^i=\pi(B_2)^i$ mod $n_i$.
In particular, the theory depends on $B_3$, $r_{\cal A}(B_2)$ and $\xi$ only by the coupling (\ref{eqn:couplingextra}), while the ${\cal B}$ gauge theory couples to $r_{\cal A}'(B_2)$ only through the projection $\pi(r_{\cal A}'(B_2))=\pi(B_2)$.
Thus without loss of generality, we can replace $B_2$ in (\ref{eqn:originalthreegroup}) by $r_{\cal A}'(B_2)$, and in the following we will discuss the consequence of turning on the couplings in (\ref{eqn:couplingextra}) with non-trivial $\xi$ and $r_{\cal A}(B_2)$.

The original three-group symmetry (\ref{eqn:originalthreegroup}) and (\ref{eqn:bulkAbeliangaugedepn}) together imply the following bulk dependence on the ${\cal B}$ gauge field $u^i$ in the presence of non-trivial $\xi$ and $r_{\cal A}(B_2)$:
\begin{align}\label{eqn:totalbulkdepn}
&\sum_i\frac{2\pi}{n_i}\int_{5d} 
	u^i\left(\Xi(r_{\cal A}'(B_2),X)^i+
	2{\pi(\delta_\rho B_2)^i\over n_i}X^*\xi^i-\pi(B_2)^iX^*{\delta_\sigma \xi^i\over n_i}\right.\cr
&\qquad\;
	+\sum_{j>i}\left(
		{n_i\over n_{ij}}{\pi(\delta_\rho B_2)^j\over n_j}X^*\xi^{ji}-\pi(B_2)^jX^*{\delta_\sigma \xi^{ji}\over n_{ij}}
		+{n_i\over n_{ij}} \pi(B_2)^jr_{\cal A}(B_2)^{ji}\right)
	\cr
	&\qquad\;\left.
	+\sum_{j<i}\left( 
		{n_i\over n_{ij}}{\pi(\delta_\rho B_2)^j\over n_j}X^*\xi^{ij}-{n_i\over n_{ij}} \pi(B_2)^jr_{\cal A}(B_2)^{ij}\right)
	-\delta_\sigma B_3^i\right)	~,
\end{align}
where we used the isomorphism $\text{Hom}({\cal B},U(1))={\hat{\cal B}}\cong {\cal B}$ on $\Xi(r_{\cal A}'(B_2),X)\in{\cal B}$.
Therefore, to cancel the bulk dependence of $u^i$, the two-form symmetry background $B_3$ satisfies the following constraint:
\begin{align}\label{eqn:constraintmodify}
&\delta_\sigma B_3^i
=\Xi(r_{\cal A}'(B_2),X)^i+
2{\pi(\delta_\rho B_2)^i\over n_i}X^*(\xi^i)-\pi(B_2)^iX^*{\delta_\sigma \xi^i\over n_i}\cr
&\qquad\qquad
+\sum_{j>i}\left(
	{n_i\over n_{ij}}{\pi(\delta_\rho B_2)^j\over n_j}X^*(\xi^{ji}) 
	+ {n_i\over n_{ij}} \pi(B_2)^jr_{\cal A}(B_2)^{ji}-\pi(B_2)^jX^*{\delta_\sigma \xi^{ji}\over n_{ij}}\right)
\cr
&\qquad\qquad	+\sum_{j<i}\left( {n_i\over n_{ij}}{\pi(\delta_\rho B_2)^j\over n_j}X^*(\xi^{ij})
- {n_i\over n_{ij}} \pi(B_2)^jr_{\cal A}(B_2)^{ij}\right)
\text{ mod }n_i~,
\end{align}
For a fixed $G$ symmetry action, if the three-group symmetry constraint between $B_2,B_3,X$ for some coupling $\xi$ is known, then this gives $\Xi$ and thus the constraint for general couplings $\xi$ is given by (\ref{eqn:constraintmodify}).

Using (\ref{eqn:threegroupgeneralgaugetransf}) with $\lambda_1^i=\pi(\lambda_1)^i$ and $\lambda_1^{ij}=r_{\cal A}(\lambda_1)^{ij}$,\footnote{
$r_{\cal A}(\lambda_1)$ is defined similarly to $r_{\cal A}(B_2)$.
} we can find the transformation of $B_3^i$ under the background gauge transforms $B_3\rightarrow B_3+\delta_\sigma\lambda_2$ and $B_2\rightarrow B_2+\delta_\rho \lambda_1$:
\begin{align}\label{eqn:backgroundgaguetransfgeneral}
&B_3^i\longrightarrow B_3^i
+\delta_\sigma\lambda_2
+\Delta(r_{\cal A}'(B_2),X,r_{\cal A}'(\lambda_1))
-\pi(\lambda_1)^iX^*{\delta_\sigma \xi^i\over n_i}
\cr
&\quad\quad+\sum_{j>i}\left({n_i\over n_{ij}} 
		(\pi(\lambda_1)^j r_{\cal A}(B_2)^{ji} + \pi(B_2)^jr_{\cal A}(\lambda_1)^{ji}
		+\pi(\lambda_1)^j r_{\cal A}(\delta_\rho\lambda_1)^{ji})
-\pi(\lambda_1)^jX^*{\delta_\sigma \xi^{ji}\over n_{ij}} \right)
\cr
&\quad\quad
-\sum_{j<i}\left({n_i\over n_{ij}} (\pi(\lambda_1)^jr_{\cal A}(B_2)^{ij}
+\pi(B_2)^jr_{\cal A}(\lambda_1)^{ij}+\pi(\lambda_1)^jr_{\cal A}(\delta_\rho \lambda_1)^{ij})\right)
\text{ mod }n_i~,
\end{align}
where $\delta\Delta(r_{\cal A}'(B_2),X,r_{\cal A}'(\lambda_1))
=\Xi(r_{\cal A}'(B_2)+r_{\cal A}'(\delta_\rho \lambda_1),X)-\Xi(r_{\cal A}'(B_2),X)$.

In particular, in the absence of the background $B_2$ of one-form symmetry, 
the one-form global symmetry transformation parametrized by cocycles $\lambda_1=\rho_1$ with $\delta_\rho\rho_1=0$
 transforms the background $B_3$ as
\begin{equation}
B_3^i\longrightarrow B_3^i-\sum_{j\geq i}\pi(\rho_1)^j X^*(\delta_\sigma \xi^{ji}/n_{ji})~.
\end{equation}

$\Xi$ is a twisted cocycle whose structure can be classified in principle, but we will not do it here.
The different structures in $\Xi$ can be detected by correlation functions.
As an example, if ${\cal A}'={\cal B}=\mathbb{Z}_2$, then the structure 
$\Xi(r_{\cal A}'(B_2),X)=r_{\cal A}'(B_2)X^*\zeta_2$ with $\zeta\in H^2(BG,\mathbb{Z}_2)$ describes a junction where three codimension-one 0-form symmetry defects meet at a codimension-two surface and the surface intersects the symmetry surface defect for ${\cal A}'$ at a point, which emits a symmetry line defect\footnote{
This uses the fact that the Poincar\'e dual of $\delta B_3$ describes the boundary of the Poincar\'e dual of $B_3$.
}.
The symmetry line defect can then be detected by correlation functions.

The different couplings to the background $X$ of 0-form symmetry $G$ are then specified by $\xi$, and the couplings $(\eta_2,\nu_3)\in H^2_\rho(BG,{\cal A})\times C^3(BG,{\cal B})$ by the fixed higher-form symmetry backgrounds
\begin{equation}\label{eqn:substitutionordinary}
B_2=X^*\eta_2,\quad B_3=X^*\nu_3~
\end{equation}
with $(\eta_2,\nu_3)$ satisfy additional constraints depending on $\xi$ as induced by (\ref{eqn:constraintmodify}) 
\begin{align}\label{eqn:constraintcouplinggeneral}
&\delta_\sigma \nu_3^i=
\Xi_G(r_{\cal A}'(\eta_2))
+2{\pi(\delta_\rho\eta_2)^i\over n_i}\xi^i-\pi(\eta_2)^i{\delta_\sigma \xi^i\over n_i}\cr
&\qquad\qquad
+\sum_{j>i}\left({n_i\over n_{ij}}{\pi(\delta_\rho\eta_2)^j\over n_j}\xi^{ji}
-\pi(\eta_2)^j{\delta_\sigma \xi^{ji}\over n_{ij}}+\pi(\eta_2)^j r_{\cal A}(\eta_2)^{ji}\right)
\cr
&\qquad\qquad	+\sum_{j<i}\left( {n_i\over n_{ij}}{\pi(\delta_\rho \eta_2)^j\over n_j}\xi^{ij}-\pi(\eta)_2^jr_{\cal A}(\eta_2)^{ij}\right)
\text{ mod }n_i~,
\end{align}
where $\Xi(r_{\cal A}(X^*\eta_2),X)=X^*\Xi_G(r_{\cal A}'(\eta_2))$ with $\Xi_G\in H^4_\sigma(BG,{\cal B})$.

The couplings $(\eta_2,\nu_3)$ have equivalence relations that depend on $\xi$, given by the background gauge transformations (\ref{eqn:backgroundgaguetransfgeneral}) with $\lambda_1=X^*s_1,\lambda_2=X^*s_2$ for $s_1\in C^1(BG,{\cal A})$ and $s_2\in C^2(BG,{\cal B})$.
Namely, $(\eta_2,\nu_3)$ is equivalent to $(\eta_2',\nu_3')$ with
\begin{align}\label{eqn:backgroundgaguetransfgeneralcoupling}
&\eta_2'=\eta_2+\delta_\rho s_1\cr
&\nu_3'^i= \nu_3^i+\delta_\sigma s_2
+\Delta_G(r_{\cal A}'(\eta_2),r_{\cal A}'(s_1))
-\pi(s_1)^i{\delta_\sigma \xi^i\over n_i}
\cr
&\quad
+\sum_{j>i}\left(
{n_i\over n_{ij}} 
		(\pi(s_1)^j r_{\cal A}(\eta_2)^{ji} + \pi(\eta_2)^jr_{\cal A}(s_1)^{ji}+\pi(s_1)^jr_{\cal A}(\delta_\rho s_1)^{ji})		
		-\pi(s_1)^j{\delta_\sigma \xi^{ji}\over n_{ij}} \right)\cr
&\quad
-\sum_{j<i}\left({n_i\over n_{ij}} (\pi(s_1)^jr_{\cal A}(\eta_2)^{ij}+\pi(\eta_2)^jr_{\cal A}(s_1)^{ij}
+\pi(s_1)^j r_{\cal A}(\delta_\rho s_1)^{ij})\right)
\text{ mod }n_i~,
\end{align}
where $X^*\Delta_G(r_{\cal A}'(\eta_2),r_{\cal A}'(s_1))=\Delta(r_{\cal A}'(X^*\eta_2),X,r_{\cal A}'(X^*s_1))$.

If at some coupling $\xi=0$ the constraint on $\delta_\sigma B_3$ is known, then $\Xi$ (and thus $\Xi_G$) can be determined and the couplings to background of 0-form symmetry $G$ are classified by
\begin{equation}\label{eqn:classificationfinal}
(\eta_2,\nu_3,\xi)\in 
H^2_\rho(BG,{\cal A})\times C^3(BG,{\cal B})\times H^1_\sigma(BG,H^3({\cal B},U(1))')
\end{equation}
subject to the constraint (\ref{eqn:constraintcouplinggeneral}) and the identification from the corresponding gauge transformation.

We will give examples with trivial and non-trivial $\Xi$.
An example of trivial $\Xi$ is the Abelian gauge theory with zero Dijkgraaf-Witten action \cite{Dijkgraaf:1989pz} to be discussed in  section \ref{sec:Abelianfinitegauge}. An example of non-trivial $\Xi$ is the $O(2)$ gauge theory to be discussed in section \ref{sec:Otwogaugetheory}.

When the 0-form symmetry $G$ does not permute the non-local operators or act on local operators, $\Xi(B_2,X)=\Xi(B_2)$ only depends on the background for the one-form symmetry\footnote{
If there are other 0-form symmetry defects not included in those we have discussed, then $\Xi$ can also depend the background for these 0-form symmetries. In this note we do not consider such backgrounds.
}. Then (\ref{eqn:originalthreegroup}) becomes
\begin{equation}\label{eqn:originalthreegroupspecial}
\delta B_3=\Xi(B_2)=q(B_2,B_2)
\end{equation} 
where $q$ is a bilinear function on ${\cal A}\times{\cal A}$ that takes values in ${\cal B}$.
This constraint describes the following junction: two symmetry surface defects meet at a point that emits a symmetry line defect.
The symmetry line defect can then be detected by correlation functions.
The constraint (\ref{eqn:originalthreegroupspecial}) implies that it is inconsistent to gauge only the one-form symmetry but not the two-form symmetry. This is analogous to the $H^3$ obstruction discussed in \cite{EtingofNOM2009,Barkeshli:2014cna}.

\subsection{Anomalies for different couplings}\label{sec:anomcoulpling}

First we need to understand how the 't Hooft anomaly for $B_2,B_3,X$ depends on $B_3$.
If we gauge the two-form symmetry with dynamical gauge field, and then gauge the dual 0-form symmetry with ${\cal B}$ gauge field $u^i$, then the background $B_3$ only couples through $u^i$. This is just a restatement of the duality (\ref{eqn:dualitytwoform}).
Thus the 't Hooft anomaly involving $B_3$ only comes from the modification on $\delta u^i$, which equals $\pi(B_2)^i$.

On the other hand, after gauging the two-form symmetry with dynamical field there can be an 't Hooft anomaly that depends on $r_{\cal A}'(B_2)$ and $X$, which we will denote by $S^0(r_{\cal A}'(B_2),X)$.
Together with the 't Hooft anomaly (\ref{eqn:threegroupAbeliananom}) in the ${\cal B}$ gauge theory, we find the anomaly is given by
\begin{align}\label{eqn:anomalyp}
&S_\text{anom}=
S^0(r_{\cal A}'(B_2),X)
+\sum_i\frac{2\pi}{n_i}\int_{5d}\pi(B_2)^i B_3^i
\cr
&-\sum_{i>j}\frac{2\pi}{n_{ij}n_j}\int_{5d}\left\{ 
	(\pi(B_2)^i\pi(B_2)^j
	-\pi(\delta_\rho B_2)^i\cup_1 \pi(B_2)^j)X^*\xi^{ij}
-(\pi(B_2)^i\cup_1 \pi(B_2)^j)X^*\delta_\sigma \xi^{ij}\right\}\cr
&+\sum_i\frac{2\pi}{n_i^2}\int_{5d}
(\pi(B_2)^i\cup_1 \pi(B_2)^i)X^*\delta_\sigma\xi^{i}
-
(\pi(B_2)^i\pi(B_2)^i-\pi(\delta_\rho B_2)^i\cup_1 \pi(B_2)^i)X^*\xi^{i}
~ ,\quad
\end{align}
where the part that depends on $\xi$ shows that the coupling $\xi$ produces an additional 't Hooft anomaly.

The 't Hooft anomaly for different couplings from modification of the 0-form symmetry defect junction is given by substituting the fixed backgrounds of the higher-form symmetries (\ref{eqn:4dbackgrounds}) into the 't Hooft anomaly of the higher-form and 0-form symmetries \cite{Benini:2018reh}.
The 't Hooft anomaly of 0-form symmetry $G$ for different $(\eta_2,\nu_3)$ depends on the coupling $\xi\in H^1_\sigma(BG,H^3({\cal B},U(1))')$ by
\begin{align}\label{eqn:anomalcouplinggeneral}
&S_\text{anom}=\int_{5d}X^*(\omega_5^0+\omega_5^1)~,\qquad \text{where }\int_{5d}X^*\omega_5^0=S^0(r_{\cal A}'(X^*\eta_2),X)
~,\cr
&\omega_5^1=\sum_i\frac{2\pi}{n_i}
\left\{
\pi(\eta_2)^i \nu_3^i
-\frac{1}{n_i}\left( 
	(\pi(\eta_2)^i\pi(\eta_2)^i-\pi(\delta_\rho\eta_2)^i\cup_1 \pi(\eta_2)^i)\xi^{i}
	-(\pi(\eta_2)^i\cup_1 \pi(\eta_2)^i)\delta_\sigma \xi^{i}\right)
\right\}
\cr
&\qquad-\sum_{i>j}\frac{2\pi}{n_{ij}n_j}\left\{ (\pi(\eta_2)^i\pi(\eta_2)^j- \pi(\delta_\rho\eta_2)^i\cup_1 \pi(\eta_2)^j)\xi^{ij}
	-(\pi(\eta_2)^i\cup_1 \pi(\eta_2)^j)\delta_\sigma \xi^{ij}\right\}~.
\end{align}

\subsection{SPT phases trivialized by coupling to the dynamics}\label{sec:wSPTgeneral}

A general symmetry-enriched phase consists of a coupling to the background gauge field and a decoupled SPT phase of the 0-form symmetry. Different SPT phases correspond to different local counterterms of the 0-form symmetry background gauge field. 

As we will discuss, some of the SPT phases become equivalent when there is a non-trivial coupling between the $G$ gauge field and the dynamics. 
Thus the set of distinct SPT phases depends on the coupling to the dynamics.

These trivialized SPT phases are equivalent to depositing SPT phases on the worldvolume where some non-local operators are supported. To preserve the correlation functions, this means that the worldvolume SPT phases correspond to a global symmetry that transform the non-local operators. Thus the SPT phases that can be absorbed in this way are related to higher-form global symmetries in the theory.

To produce an SPT phase from the higher-form symmetry, there must be insertion of non-local operators that are charged under the higher-form symmetry with locus specified by the background of the 0-form symmetry.
When the operators are invertible, they are symmetry defects, and the insertion corresponds to activating the backgrounds for higher-form symmetries as in (\ref{eqn:4dbackgrounds}).
Since the operator inserted is charged under the higher-form symmetry, this means that there is an 't Hooft anomaly for the higher-form symmetry.
Thus the 't Hooft anomaly of the higher-form symmetry can absorb some of the SPT phases into lower-dimensional SPT phases on the worldvolumes as the phase induced by a higher-form symmetry transformation.

In a theory that is trivially gapped {\it i.e.} gapped and has a unique vacuum, every SPT phase must be distinct. 
The trivialization of SPT phases relies on the dynamics.
If there exists a single SPT phase that is non-trivial on its own, but can be absorbed by the dynamics as above, then the theory cannot be trivially gapped. 
If there exists such an SPT phase in the UV, then it must persist in the IR, 
since otherwise the same UV theory with or without such SPT phase would flow to distinct symmetry-enriched phases in the IR and thus it leads to a contradiction. This is consistent with the fact that an 't Hooft anomaly for the higher-form symmetry implies that the theory is not trivially gapped.

We remark that an analogous, but distinct phenomenon that occurs in the $4d$ $SU(N)$ Yang-Mills theory coupled to the background gauge field for the $\mathbb{Z}_N$ one-form symmetry. 
The theory has a continuous $\theta$ parameter, and by changing $\theta$ in a multiple of $2\pi$ some of the SPT phases of the $\mathbb{Z}_N$ one-form symmetry can be cancelled \cite{Gaiotto:2017yup}\footnote{
One way to incorporate this particular example into our approach is to interpret $\theta$ as a ``background gauge field'' that transforms as $\theta\rightarrow \theta+2\pi$. Then the SPT phases that are trivialized come from the mixed anomaly between the $\mathbb{Z}_N$ one-form symmetry and this shift symmetry.
}. 
We will argue by contradiction that this implies at some value of $\theta$ the system cannot be trivially gapped. Suppose the converse were true {\it i.e.} the theory were trivially gapped at all values of $\theta$, then the SPT phase would be smoothly deformed into a different SPT phase by tuning the continuous parameter $\theta$ across multiples of $2\pi$, and thus we have a contradication. See \cite{Cordova:2019jnf,Cordova:2019uob} for an alternative argument using the anomaly in the space of continuous couplings.
The above phenomenon is different from the previous discussion that does not require a continuous parameter.

Although we focus on $(3+1)d$, the same discussion can be generalized to other spacetime dimensions.
In the following we will discuss some concrete examples.

\subsubsection{Examples}

The $(2+1)d$ Chern-Simons theory $U(1)_4$ is a non-trivial TQFT. It has $\mathbb{Z}_4$ one-form symmetry, and it can couple to the background gauge field of $\mathbb{Z}_2$ ordinary global symmetry that does not permute the anyons as follows
\begin{equation}
\frac{4}{4\pi}ada+\frac{1}{2\pi}adX~,
\end{equation}
where $a$ is the dynamical $U(1)$ gauge field, $X$ is the $\mathbb{Z}_2$ gauge field expressed as a classical $U(1)$ gauge field with holonomy $\oint X\in \pi\mathbb{Z}$. 
The coupling to $X$ uses the $\mathbb{Z}_4$ one-form symmetry background $B_2=\frac{1}{4}dX$~.
Now, the field redefinition $a\rightarrow a+X$ for the dynamical field $a$ gives
\begin{equation}
\frac{4}{4\pi}ada+\frac{1}{2\pi}adX+\frac{1}{2\pi}XdX+\left(\frac{4}{4\pi}XdX+\frac{4}{2\pi}daX\right)
=\frac{4}{4\pi}ada+\frac{1}{2\pi}adX+\frac{1}{2\pi}XdX~,
\end{equation}
where the last term on the left hand side is trivial due to the holonomy of the background $X$ only lives in $\pi\mathbb{Z}$.
Thus the field redefinition produces the SPT $\frac{1}{2\pi}XdX$, which is the non-trivial Dijkgraaf-Witten theory in $(2+1)d$ for $\mathbb{Z}_2$ gauge field \cite{Dijkgraaf:1989pz}. This SPT phase is non-trivial on its own, but when the background gauge field couples to $U(1)_4$ as above it can be removed by the field redefinition as above, which introduces a worldline SPT phase: for Wilson line of charge $Q$, the $(0+1)d$ SPT phase can be obtained from $a\rightarrow a+X$ as
\begin{equation}
\exp\left(iQ\oint a\right)\rightarrow \exp\left(iQ\oint a\right)(-1)^{Q\oint X/\pi}~.
\end{equation}
Since $X$ is a $\mathbb{Z}_2$ gauge field, the redefinition of $a$ is equivalent to a $\mathbb{Z}_2$ one-form global symmetry.
The one-form symmetry deposits the worldline SPT phase $Q\oint X$ for to the Wilson line of $U(1)$ charge $Q$.
This $\mathbb{Z}_2$ subgroup one-form symmetry has a mixed anomaly with the full $\mathbb{Z}_4$ one-form symmetry, as their generating line operators have non-trivial mutual braiding \cite{Hsin:2018vcg}. The anomaly is $\mathbb{Z}_2$ valued, and it can absorb the $\mathbb{Z}_2$ valued Dijkgraaf-Witten SPT.

Another example is QED$_3$ with one Dirac fermion of charge four\footnote{
The theory can be obtained by gauging the subgroup $\mathbb{Z}_4\subset U(1)$ magnetic 0-form symmetry in QED$_3$ with two fermions of charge one, which is conjectured to be gapless and enjoy a self-duality \cite{Xu:2015lxa,Hsin:2016blu,Benini:2017dus,Wang:2017txt}.
}
coupled to the background gauge field $X$ of $\mathbb{Z}_4$ 0-form symmetry.
The theory has $\mathbb{Z}_4$ one-form symmetry, which has an anomaly given by the SPT phase $\frac{8}{4\pi}\int_{4d}B_2B_2$ with background two-form $\mathbb{Z}_4$ gauge field $B_2$. 
The anomaly implies that under a background one-form gauge transformation $B_2\rightarrow B_2+d\lambda$, the theory changes by $\frac{8}{4\pi}\lambda d\lambda+\frac{8}{2\pi}B_2\lambda$.
The theory can couple to $\mathbb{Z}_4$ gauge field $X$ by $B_2 = \frac{1}{4}dX$. Then the global one-form symmetry transformation $\lambda=X$ produces $\frac{2}{2\pi}XdX$ ($X$ is a $\mathbb{Z}_4$ gauge field, so this is not a one-form gauge transformation). This means that adding this counterterm of $X$ does not change the theory.

An example with continuous 0-form symmetry is $\mathbb{Z}_n$ gauge theory in $(3+1)d$ coupled to a background $U(1)$ gauge field $A$ by activating the $\mathbb{Z}_n$ one-form symmetry background gauge field $B_2=\frac{1}{n}dA$.
The theory also has $\mathbb{Z}_n$ two-form symmetry, which has a mixed anomaly with the one-form symmetry.
Due to the mixed anomaly, performing a $\mathbb{Z}_n$ two-form global symmetry transformation with parameter $\mathbb{Z}_n$ two-cocycle $\lambda_2=q\frac{dA}{2\pi}$ mod $n$ for some integer $q$ shifts the theory by the action
\begin{equation}\label{eqn:sptuonezn}
\frac{n}{2\pi}\int B_2\left(\frac{q}{n}dA\right)
=\frac{q/n}{2\pi}\int dAdA~.
\end{equation}
Thus the counterterm of the $U(1)$ background gauge field $A$ with $\theta=4\pi q/n$ can be absorbed by the worldvolume SPT phase $q_\textbf{m}\oint (q/n)dA$ on the surface operator of charge $q_\textbf{m}$ under the two-form symmetry.

In the absence of time-reversal symmetry, the SPT phase (\ref{eqn:sptuonezn}) can be continuously tuned to zero since the theta parameter can be any real number.
On the other hand, if the 0-form symmetry is $U(1)\times \mathbb{Z}_2^{\cal T}$ then $\theta=\pi$ is a non-trivial SPT phase on its own. For $n=4m$ a multiple of 4, taking $q=n/4$ we conclude the $\theta=\pi$ SPT phase for the $U(1)\times\mathbb{Z}_2^{\cal T}$ symmetry becomes trivial due to the dynamics of the system.

The above examples have SPT phases living on the lines or surface operators. An example with SPT living on the domain wall is the TQFT
\begin{equation}
\frac{2}{2\pi}\int b_{d-1}d\phi~,
\end{equation}
where $\phi\sim\phi+2\pi$ is a periodic scalar field, $b_{d-1}$ is a $(d-1)$-form $U(1)$ gauge field. 
The theory has 0-form $\mathbb{Z}_2$ symmetry generated by the operator $\exp(i\oint b_{d-1})$, and $(d-1)$-form $\mathbb{Z}_2$ symmetry generated by the point operator $\exp(i\phi)$. Denote their backgrounds by $B_1,B_{d}$. 
The two symmetries have the mixed 't Hooft anomaly described by $\frac{2}{2\pi}\int B_1B_{d}$.
The theory can couple to the background gauge field $X$ for a 0-form $\mathbb{Z}_2$ symmetry by $B_1=X$ {\it i.e.} insert the symmetry generator, a domain wall, at the Poincare dual of $X$.
Then the $(d-1)$-form global symmetry transformation by the parameter $\lambda_{d-1}=\pi (X/\pi)^{d-1}$ deposits on the domain wall the SPT phase $\oint \lambda_{d-1}=\pi\oint (X/\pi)^{d-1}$ and produces the SPT phase $\pi (X/\pi)^d$. Thus this SPT phase in $d$-dimensional spacetime is equivalent to an SPT phase on the $(d-1)$-dimensional domain wall.

\subsection{Comparison with gauging SPT phases}
\label{sec:comparegaugingSPT}

We would like to make some comparisons with the method of obtaining symmetry-enriched topological phases by gauging a subgroup symmetry in an SPT phase. 
We will focus on finite group $H$ gauge theory with unitary finite group 0-form symmetry $G$.

\subsubsection{Finite group $H$ gauge theory}
\label{sec:finitegroupHgaugetheory}

We begin by reviewing some properties of the $H$ gauge theory.

The theory has Wilson lines (particle excitations), labelled by the representations of $H$. 
There are also surface operators (string excitations), labelled by the holonomies of flat connections around the (codimension-two) surface operators {\it i.e.} labelled by the conjugacy classes of $H$.

The two-form symmetry ${\cal B}$ of the theory is generated by line operators that are topological and invertible {\it i.e.} the Wilson lines in the one-dimensional representations. Thus
\begin{equation}
{\cal B}=\hat H=\text{Hom}(H,U(1))=H^1(H,U(1))~.
\end{equation}
They can be obtained by starting with $(0+1)d$ SPT phases of $H$ symmetry, and then gauging the symmetry with a dynamical gauge field.
The two-form symmetry acts on the surface operators by evaluating the conjugacy classes on the one-dimensional representations.

If the $H$ gauge theory has trivial Dijkgraaf-Witten action \cite{Dijkgraaf:1989pz}, then the theory has one-form symmetry $Z(H)$ (the center of the gauge group $H$) that acts on the Wilson lines of $H$ by evaluating the representation of the line operator on the center of the gauge group $H$.
The background gauge field of this one-form symmetry $Z(H)$ leads to selection rules on the Wilson lines and thus replaces the gauge bundle with an $H/Z(H)$ bundle. If the theory has a non-trivial Dijkgraaf-Witten action for the dynamical gauge field, then the one-form symmetry $Z(H)$ is explicitly broken to a subgroup.\footnote{
An example is $\mathbb{Z}_2^4$ gauge theory with dynamical $\mathbb{Z}_2$ gauge fields $a^i$. Without a Dijkgraaf-Witten action there is $Z(\mathbb{Z}_2^4)=\mathbb{Z}_2^4$ one-form symmetry that transforms the line operators by $a^i\rightarrow a^i+x^i$ with classical $\mathbb{Z}_2$ cocycles $x^i$ as $\exp(\pi i\oint a^i)\rightarrow \exp(\pi i \oint a^i)(-1)^{\oint x^i}$.
If there is Dijkgraaf-Witten action $\pi\int a^1a^2a^3a^4$, the theory is not invariant under the one-form symmetry $a^i\rightarrow a^i+x^i$, as the change cannot be absorbed by a non-trivial background field.
} 

The theory has another class of surface operators, labelled by $H^2(H,U(1))$. They can be obtained by starting with $(1+1)d$ SPT phases with $H$ symmetry and gauging the symmetry with a dynamical gauge field. They also corresponds to the projective representations of $H$.
Such surface operators are topological and invertible, and thus they generate a one-form symmetry.
They are mutually local with the Wilson lines of $H$ (the $H$ bundle is well-defined inside and outside such surface operators), unlike the surfaces that generate the one-form symmetry $Z(H)$.
Thus for trivial Dijkgraaf-Witten action of the $H$ gauge group the theory has one-form symmetry\footnote{
Since the one-form symmetry must be Abelian \cite{Gaiotto:2014kfa}, the elements in $Z(H),H^2(H,U(1))\subset {\cal A}$ commute with one another. Furthermore, the non-trivial elements in the subgroups $Z(H),H^2(H,U(1))$ are distinct, and thus the one-form symmetry factorizes. 
}
\begin{equation}
{\cal A}=Z(H)\times H^2(H,U(1))~.
\end{equation}
The presence of a Dijkgraaf-Witten action breaks the one-form symmetry $Z(H)$ to a subgroup but does not affect the one-form symmetry $H^2(H,U(1))$.

Since the symmetry surface defects in $H^2(H,U(1))$ have trivial braiding with the Wilson lines of one-dimensional representations that generate the two-form symmetry, there is no mixed anomaly between the two-form symmetry and the one-form symmetry $H^2(H,U(1))$.

Gauging the one-form symmetry $H^2(H,U(1))$ restricts the set of $H$ bundles that is summed over in the path integral.
This leads to an extension of the gauge group.
To see this, denote the group cocycle by $\omega_2\in H^2(H,U(1))=\prod_i\mathbb{Z}_{m_i}$, and the $H$ gauge field by $u$.
The two-form gauge field for the one-form symmetry $H^2(H,U(1))$ can be described by the pairs of $\mathbb{Z}_{m_i}$ two-cochain $b_2^i$ and one-cochain $a^i$ that couple to the theory as
\begin{equation}
\frac{2\pi}{m_i}\int b_2^i\, u^*(\omega_2)^i+\frac{2\pi}{m_i}\int b_2^i\,\delta a^i~.
\end{equation}
The equation of motion for $b_2^i$ then implies
\begin{equation}\label{eqn:extensionHB}
\delta a^i=u^*(\omega_2)^i~.
\end{equation}
Namely, the path integral only includes the $H$ gauge field $u$ such that the cohomology class of $u^*\omega_2$ is trivial.
The gauge fields $(a^i,u)$ in (\ref{eqn:extensionHB}) describe the gauge field of a new gauge group, 
given by the group extension of $H$ by $H^2(H,U(1))$ specified by $\omega_2$.\footnote{
An example is $H=\mathbb{Z}_2\times\mathbb{Z}_2$, which has the one-form symmetry $H^2(H,U(1))=\mathbb{Z}_2$.
Then gauging this one-form symmetry extends the gauge group to be the Dihedral group of order $8$.
}
For more general continuous gauge group, the corresponding one-form symmetry is $H^2(BH,U(1))$, and it is the magnetic one-form symmetry\footnote{
An example is $SO(3)$ gauge theory, where gauging the $\mathbb{Z}_2$ magnetic one-form symmetry generated by the surface operator $\oint w_2(SO(3))$ (the second Stiefel-Whitney class of the $SO(3)$ bundle) extends the gauge group to be its double covering $SU(2)$.
}.

We remark that since the one-form symmetry $H^2(H,U(1))$ can be gauged as above without activating a background for the two-form symmetry, the two symmetries do not combine into a three-group symmetry {\it i.e.} the cocycle $\Xi$ vanishes in (\ref{eqn:originalthreegroupspecial}), and the background $B_3$ for the two-form symmetry obeys the usual cocycle condition.
Moreover, since this one-form symmetry can be gauged, it does not have an anomaly on its own.

The theory has intrinsic 0-form symmetry $\text{Out}(H)$ that permutes the non-local operators. We will focus on the case that the 0-form symmetry has trivial permutation action.

\subsubsection{Comparison with gauging SPT phase}

Consider finite groups $G,H$, the bosonic SPT phases for $G\times H$ are classified by $H^4(G\times H,U(1))$, which can be decomposed using the K\"unneth formula\footnote{See \cite{Hung:2012nf} for a discussion of this form of the K\"unneth formula.} into the product
\begin{align}\label{eqn:SPTclassif}
&H^4(H,U(1))\times H^4(G,U(1))\cr
&\qquad\times H^3(G,H^1(H,U(1)))\times H^2(G,H^2(H,U(1)))\times H^1(G,H^3(H,U(1)))~.
\end{align}
Gauging the symmetry $H$ with dynamical gauge field in such an SPT yields an SET, a $G$-enriched $H$ gauge theory. We would like to compare \eqref{eqn:SPTclassif} to our classification data \eqref{eqn:classification}
\begin{equation}\label{eqn:classification2}
(\eta_2,\nu_3,\xi)=H^2_\rho(BG,{\cal A})\times C^3(BG,{\cal B})\times H^1_\sigma(BG,H^3({\cal B},U(1))')~.
\end{equation}
In the comparison with (\ref{eqn:SPTclassif}) we will take $G$ to have trivial actions $\rho,\sigma$ on the one-form and two-form symmetries.

When we gauge the SPT phases with $G\times H$ symmetry, the theory has the standard $G\times H$ gauge fields.
This means that there is no background for the one-form symmetry $Z(H)$ (or a subgroup of it).
This means we will only obtain couplings for which $\eta_2$ receives contributions from the one-form symmetry $H^2(H,U(1))\subset {\cal A}$: $\eta_2\in H^2(G,H^2(H,U(1))$.\footnote{A non-trivial $\eta_2\in H^2(G,Z(H))$ corresponds to a non-trivial group extension of $G$ by $H$.} 
This matches the $H^2$ term in the SPT gauging classification scheme (\ref{eqn:SPTclassif}). Such special couplings have $\delta\eta_3=0$, i.e. modulo the appropriate equivalence, $\eta_3\in H^3(BG,{\cal B})$. In an $H$ gauge theory, the two-form symmetry group is ${\cal B}=H^1(H,U(1))$. Therefore, the $H^3$ terms in the two classification schemes also match.

The remaining terms in the classification are not quite the same, and there are several sources to the disagreement. Gauging an SPT phase can give an SET where the $G$ symmetry permutes the non-local operators of the $H$ gauge theory.
This comes from the last term in (\ref{eqn:SPTclassif}), where $H^3(H,U(1))$ represents the 0-form symmetry domain wall defects that support $H$ gauge theories in $(2+1)d$ \cite{Dijkgraaf:1989pz}. Some of these 0-form symmetry defects permute the types of the surface operators that correspond to non-trivial conjugacy classes of $H$\footnote{
In general, the symmetry $G$ can permute non-local operators even when the symmetry group is a product $G_\text{global}\times H_\text{gauge}$
}.
Consider such a surface operator piercing the wall at a line intersection. There is a non-trivial holonomy of the $H$ connection around the line intersection on the wall that equals the holonomy around the surface operator, and thus the intersection represents a magnetic line in the $H$ gauge theory on the wall. If it is a genuine line operator\footnote{\label{foot:lineH}
The set of line operators in finite group $H$ gauge theory in $(2+1)d$ depends on the topological action classified by $H^3(H,U(1))$ \cite{Dijkgraaf:1989pz} (see {\it e.g.} \cite{Roche:1990hs,Coste:2000tq,Hu:2012wx}). For $\omega\in H^3(H,U(1))$, define (where we use the addition notation for $U(1)$) \cite{Roche:1990hs,Coste:2000tq,Hu:2012wx}
\begin{equation}
\beta_h(h_1,h_2)=\omega\left( h,h_1,h_2\right)-
\omega\left( h_1,h_1^{-1}hh_1,h_2\right)
+\omega\left( h_1,h_2,(h_1h_2)^{-1}h\, h_1h_2\right),\quad h,h_1,h_2\in H~.
\end{equation}
$\beta_h$ obeys a ``twisted cocycle condition'' with $U(1)$ coefficient \cite{Roche:1990hs,Coste:2000tq,Hu:2012wx}, and for $h_1,h_2\in C(h,H)$ (the centralizer of $h$ in $H$) it is an ordinary cocycle $\beta_h\big|_{C(h,H)}\in H^2(C(h,H),U(1))$. Then every line in the $H$ gauge theory in $(2+1)d$ is labelled by a conjugacy class $[h]$ and a projective representation for $C(h,H)$ specified by $\beta_h\big|_{C(h,H)}$. 
When $\beta_h\big|_{C(h,H)}$ is a non-trivial cocycle, the pure magnetic line is not a genuine line operator. Therefore
 we expect the $(2+1)d$ domain wall of $H$ gauge theory with topological action $\omega$ to permute the surface operators when $\beta_h\big|_{C(h,H)}$ is a non-trivial two-cocycle. For Abelian finite group $H$ (where the two-form symmetry is ${\cal B}\cong H$) this is indeed the case for $\omega\in H^3(H,U(1))_A$.
}, 
the type of the surface operator that passes through the wall is not changed. On the other hand, if it is not a genuine line operator on the wall, then the wall changes the type of the surface operator. This can be shown explicitly for Abelian $H$.
As an example, consider $H=\mathbb{Z}_2^3$ gauge theory as discussed in section \ref{sec:permutationactionopr}. The wall corresponding to the non-trivial element in $H^3(H,U(1))_A=\mathbb{Z}_2\subset H^3(H,U(1))$ permutes the surface operators that have non-trivial conjugacy classes of $H$. 
The corresponding $H$ gauge theory on the wall is
\begin{equation}
\pi\int_{3d}\left( u^1 u^2u^3+\sum_{i=1}^3 u^i\delta v^i\right)
\end{equation}
with $\mathbb{Z}_2$ one-cochains $u^i,v^i$. The equation of motion for $u^1$ gives $\delta v^1=u^2u^3$, which implies that the pure magnetic line $\oint v^1$ is not a genuine line operator as it depends on the surface that bounds the line as $\int \delta v^1=\int u^2u^3$ (the line can be made gauge invariant by dressing with a projective representation of $H$ instead of a linear representation, see also footnote \ref{foot:lineH}). Indeed, as shown in section \ref{sec:permutationactionopr}, such domain wall permutes the set of surface operators as $\oint v_2^1\rightarrow \oint (v_2^1+u^2u^3)$ (see (\ref{eqn:examplepermuteabeliancubic})).
This is one origin of the overcounting in \eqref{eqn:SPTclassif} versus \eqref{eqn:classification2}\footnote{
Here we used $H/[H,H]\cong H^1(H,U(1))=\hat H$, and the projection $H\rightarrow H/[H,H]\cong \hat H$ implies the inclusions 
$H^3(H,U(1))\supset H^3(\hat H,U(1))\supset H^3(\hat H,U(1))'$.
}, where our classification \eqref{eqn:classification2} is for a fixed permutation action\footnote{
As another example of this type of overcounting, consider the continuous gauge group $H=U(1)$. Then $H^3(H,U(1))=\mathbb{Z}$, which represents domain walls with bosonic Chern-Simons actions of the $U(1)$ gauge field  (see {\it e.g.} \cite{Ning:2019ffr}) {\it i.e.} the walls across which the $\theta$ angle of the $U(1)$ gauge theory changes by multiples of $4\pi$.
The Witten effect \cite{Witten:1979ey} implies that these domain walls permute the set of line operators: magnetic 't Hooft lines that pass through the walls become dyonic lines. Thus they belong to the 0-form symmetry defects constructed by the first mechanism. In this case the two-form symmetry ${\cal B}$ is trivial (the $U(1)$ gauge field is not flat and thus the lines are invertible but not topological) and thus $H^3({\cal B},U(1))' $ is trivial.
}. 
It will be interesting to investigate the walls in $H^3(H,U(1))$ that do not permute the types of the non-local operators and do not correspond to the walls in $H^3({\cal B},U(1))'$ (see footnote \ref{foot:lineH}), which requires non-Abelian $H$. We will leave it to the future work.

Another way in which gauging SPT phases gives a redundant counting is that distinct classical actions can become equivalent under a field redefinition of the dynamical gauge field.
For example, the SPT phase with $\mathbb{Z}_2\times\mathbb{Z}_2$ symmetry described by the $\mathbb{Z}_2\times\mathbb{Z}_2$ gauge fields $X,Y$ as $\pi\int\left( X^3Y+XY^3\right)$, becomes equivalent to $\pi\int X^3Y$ when $X$ is promoted to be a dynamical gauge field, by the field redefinition $X\rightarrow X+Y$. 
In our approach there is no such redundancy by the equivalence relation on the coupling $\nu_3$ in (\ref{eqn:backgroundgaguetransfgeneralcoupling}) (see section \ref{sec:Z2gaugetheoryunitary} for a discussion on this case).

Finally, as demonstrated in section \ref{sec:wSPTgeneral}, it is also possible that some SPT phases with $G$ symmetry classified by $H^4(G,U(1))$ in the expansion \eqref{eqn:SPTclassif} become trivial in the presence of the SET phase. 
The classification (\ref{eqn:classification2}) does not include the SPT phases for $G$ symmetry.

As a remark, we can also compare the 't Hooft anomaly obtained from the two approaches.
The symmetry-enriched phases obtained from gauging a subgroup symmetry in the SPT phases are non-anomalous.
In our method, the 't Hooft anomaly is given by (\ref{eqn:anomalcouplinggeneral}), which indeed vanishes for trivial permutation and $\eta_2$ that receives contribution from the one-form symmetry $H^2(H,U(1))$ (using the fact that the one-form symmetry $H^2(H,U(1))$ is non-anomalous as shown in section \ref{sec:finitegroupHgaugetheory}).

\section{Abelian finite group gauge theory with unitary symmetry}
\label{sec:Abelianfinitegauge}

In this section we will discuss the bosonic gauge theory in $(3+1)d$ for Abelian gauge group ${\cal B}=\prod_i \mathbb{Z}_{n_i}$, with the trivial Dijkgraaf-Witten action \cite{Dijkgraaf:1989pz}. 
We will first consider the example ${\cal B}=\mathbb{Z}_n$, and later discuss the general case.

\subsection{$\mathbb{Z}_n$ gauge theory}

As discussed earlier, the $\mathbb{Z}_n$ gauge theory (\ref{eqn:Zngaugetheory}) has $\mathbb{Z}_n$ one-form and two-form symmetries. Denote their backgrounds by $B_2,B_3$.
The $\mathbb{Z}_n$ gauge theory has 0-form symmetry that permutes the operators given by the automorphism group of $\mathbb{Z}_n$.

We will couple the theory to background $X$ of a unitary 0-form symmetry $G$.
If the 0-form symmetry acts on the higher-form symmetry, we replace the ordinary coboundary operation $\delta$ with the twisted coboundary operation $\delta_X$. 
In the following we will focus on the case that the 0-form symmetry does not act on the higher-form symmetry, while the same discussion also applies with the above replacement when the 0-form symmetry acts by an automorphism of $\mathbb{Z}_n$.

We will first investigate the constraints on the backgrounds $B_2,B_3,X$ for different couplings $\xi\in H^1(BG,H^3(\mathbb{Z}_n,U(1)))=H^1(BG,\mathbb{Z}_n)$. The non-trivial codimension-one symmetry defect from the two-form symmetry is $\exp\left({2\pi i\over n}\oint u\text{Bock}(u)\right)$ for $\mathbb{Z}_n$ gauge field $u$.
The background $B_2$ modifies the $\mathbb{Z}_n$ gauge field $u$ to be the cochain
\begin{equation}
\delta u=B_2\quad\text{mod }n~.
\end{equation}
The backgrounds $X,B_3$ couples as
\begin{equation}
{2\pi\over n} \int \tilde u{\delta \tilde u\over n}X^*\xi+{2\pi\over n}\int u B_3+I[\tilde u,X,B_2,B_3]
\end{equation}
where tilde denotes a lift of $u$ to a $\mathbb{Z}_{n^2}$ one-cochain, and the remaining terms $I$ are to be determined such that the couplings are well-defined in the presence of backgrounds $B_2,B_3$.
In particular, the theory must not depend on the lift of $u$. Changing the lift $\tilde u\rightarrow \tilde u+nc$ with $\mathbb{Z}_{n^2}$ one-cochain $c$ changes the first two terms by
\begin{equation}
{2\pi\over n}\int \left( cB_2+B_2c\right)X^*\xi={2\pi\over n}\int \left(2cB_2-B_2\cup_1\delta c\right) X^*\xi\quad\text{mod }2\pi\mathbb{Z}~,
\end{equation}
where we used (\ref{eqn:cochainrules}) and $\delta u=B_2$ mod $n$. 
To cancel the change, we can take $I={2\pi\over n^2}\int \left(-2\tilde uB_2+B_2\cup_1\delta \tilde u\right) X^*\xi$, and the theory is
\begin{equation}\label{eqn:Zngaugetheorycouplebackground}
{2\pi\over n^2} \int \left(\tilde u \delta \tilde u-2\tilde uB_2+B_2\cup_1\delta \tilde u\right) X^*\xi+{2\pi\over n}\int u B_3~.
\end{equation}
Since the theory does not depend on the lift of $u$, we will drop the tilde notation.

In addition to changing the lift, we need to impose $\mathbb{Z}_n$ gauge symmetry of the dynamical field $u$.
One way to study the gauge invariance is to look for the bulk dependence.
The theory depends on the $(4+1)d$ bulk by
\begin{align}
&\frac{2\pi}{n}\int_{5d}\delta \left\{\frac{1}{n}\left(
u\delta u -2u B_2+B_2\cup_1 \delta u\right)X^*\xi +uB_3\right\}
\cr
&\quad\;
=\frac{2\pi}{n^2}\int_{5d}
\left(\delta u\delta u -2\delta u B_2+2u\delta B_2
+\delta B_2\cup_1 \delta u -B_2\delta u+\delta u B_2\right)X^*\xi
\cr
&\qquad\;
-\frac{2\pi}{n^2}\int_{5d}\left(
u\delta u-2u B_2+B_2\cup_1\delta u\right) X^*\delta \xi
+\frac{2\pi}{n}\int_{5d}\left( B_2B_3 -u\delta B_3\right)
\cr
&\quad\;=\frac{2\pi}{n}\int_{5d}u\left(-\delta B_3^{(n)}+{1\over n}\left(2\delta B_2^{(n)} X^*\xi+B_2^{(n)}X^*\delta \xi\right)\right)
+S_\text{anom} \quad\text{mod}\ 2\pi\mathbb{Z}~,
\end{align}
where that last equality used $\delta u=B_2^{(n)}$ mod $n$, (\ref{eqn:cochainrules}), and completing the square with $(\delta u-B_2)^2=0$ mod $n^2$. The last term $S_\text{anom}$ depends only on the background fields,
\begin{equation}\label{eqn:anomalyZnZm}
S_\text{anom}=\int_{5d} 
\left(
	\frac{2\pi}{n}B_2B_3
	-\frac{2\pi}{n^2}
		\left(
		{\frak P}'(B_2)X^*\xi  +  (B_2\cup_1B_2)X^*\delta \xi
		\right) 
\right)~
\end{equation}
with ${\frak P}'(B_2)\equiv B_2\cup B_2 - \delta B_2\cup_1 B_2$, which equals the Pontryagin square (see {\it e.g.} \cite{Whitehead1949}) of $B_2$ when $n$ is even.
Thus in order for the dynamical field $u$ to live in $4d$, the backgrounds must satisfy the relation
\begin{equation}\label{eqn:3-group}
\delta B_3= 2\text{Bock}(B_2)X^*\xi + B_2X^*\text{Bock}(\xi)\quad \text{ mod }n~.
\end{equation}
The background fields satisfy (\ref{eqn:3-group}) describe a three-group symmetry that combines the $G$ 0-form, $\mathbb{Z}_n$ one-form and $\mathbb{Z}_n$ two-form symmetries.
Under a background one-form gauge transformation $B_2\rightarrow B_2+\delta\lambda_1$, the background $B_3$ transforms as
\begin{equation}
B_3\rightarrow B_3+\lambda_1 X^*\text{Bock}(\xi)~.
\end{equation}

From the constraint (\ref{eqn:3-group}) we obtain the coupling parameters $(\eta_2,\nu_3)\in H^2(BG,\mathbb{Z}_2)\times C^3(BG,\mathbb{Z}_n)$ by $B_2=X^*\eta_2,B_3=X^*\nu_3$ with the constraint
\begin{equation}\label{eqn:Znthreegroupcosntraintcoupling}
\delta \nu_3=2\text{Bock}(\eta_2)\xi+\eta_2\text{Bock}(\xi)\quad \text{mod }n~
\end{equation}
subject to the background gauge transformation of three-group symmetry
\begin{equation}\label{eqn:Znthreegorupgaugetrasfmcoupling}
\eta_2\rightarrow \eta_2+\delta s_1,\qquad
\nu_3\rightarrow \nu_3+\delta s_2+s_1 \text{Bock}(\xi)~,
\end{equation}
where we use the transformation $\lambda_1=X^*s_1,\lambda_2=X^*s_2$ with $s_1\in C^1(BG,\mathbb{Z}_n)$ and $s_2\in C^2(BG,\mathbb{Z}_n)$.

The 't Hooft anomaly of the couplings $(\eta_2,\nu_3,\xi)$ is given by the 't Hooft anomaly of the three-group symmetry, which is described by the bulk $5d$ SPT phase (\ref{eqn:anomalyZnZm}).

By substitution, we find
\begin{equation}\label{eqn:anomalyZncouplings}
S_\text{anom}=\int_{5d} X^*
\left(
	\frac{2\pi}{n}\eta_2 \nu_3
	-\frac{2\pi}{n^2}
		\left(
		{\frak P}'(\eta_2)\xi  +  (\eta_2\cup_1 \eta_2)\delta \xi
		\right) 
\right)=\int_{5d}X^*\omega_5~,
\end{equation}
where $\omega_5\in H^5(BG,U(1))$. In particular, one can verify $\omega_5$ is a five-cocycle using the constraint (\ref{eqn:Znthreegroupcosntraintcoupling}).

\subsubsection{$\mathbb{Z}_2$ gauge theory with unitary symmetry $G$}
\label{sec:Z2gaugetheoryunitary}

As an example, consider the $\mathbb{Z}_2$ gauge theory. 
The theory only has two non-trivial operators: the $\mathbb{Z}_2$ Wilson line and the basic surface operator in (\ref{eqn:lineandsurfaceZn}) with $n=2$.
Thus the intrinsic 0-form symmetry ${\cal S}$ is trivial, and the 0-form symmetry defects are constructed from the other mechanisms.
In particular, the 0-form symmetry does not act on the higher-form symmetries.
The constraint (\ref{eqn:Znthreegroupcosntraintcoupling}) and gauge transformation (\ref{eqn:Znthreegorupgaugetrasfmcoupling}) reduce to
\begin{align}\label{eqn:Z2couplingconsraint}
&\delta \nu_3=\eta_2\text{Bock}(\xi)\;\; \text{mod }2,\quad
\;(\eta_2,\nu_3)\rightarrow (\eta_2+\delta s_1,\nu_3+\delta s_2+s_1 \text{Bock}(\xi))~.
\end{align}
The couplings have the 't Hooft anomaly (\ref{eqn:anomalyZncouplings})
\begin{equation}
\omega_5=\pi\left(\eta_2 \nu_3 -\frac{1}{2}{\frak P}(\eta_2)\xi\right)\quad\text{ mod }2\pi\mathbb{Z}~.
\end{equation}
with $\frak P$ the ordinary Pontryagin square operation, and we used the property $\eta_2\cup_1\eta_2=Sq^1(\eta_2)=\text{Bock}(\eta_2)$ from (\ref{eqn:Steenrodsquarerule}) and thus the last term in (\ref{eqn:anomalyZncouplings}) is trivial for $n=2$.

For the 0-form symmetry $G=\mathbb{Z}_2$, $\xi$ is classified by $H^1(G,\mathbb{Z}_2)=\mathbb{Z}_2$:
\begin{itemize}
\item
If $\xi$ is trivial, then the couplings $(\eta_2,\nu_3)$ are classified by $H^2(G,\mathbb{Z}_2)\times H^3(G,\mathbb{Z}_2)=\mathbb{Z}_2\times\mathbb{Z}_2$. Only $(\eta_2,\nu_3)$ that are both non-trivial have an anomaly, given by
\begin{equation}
\pi\int_{5d} X^*\omega_5= \pi\int_{5d}X^5~.
\end{equation}

\item
If $\xi$ is non-trivial, then the coupling $\eta_2$ is trivial. Fix the gauge $\eta_2=0$, that leaves the transformation with $s_1$ being a cocycle. The other coupling $\nu_3$ then satisfies $\delta \nu_3=0$ and has the gauge transformation $\nu_3\rightarrow \nu_3+\delta s_2+s_1\text{Bock}(\xi)$. The gauge transformation with $s_1=\xi$ removes the non-trivial element in $H^3(G,\mathbb{Z}_2)$, and thus the only distinct coupling is $(\eta_2,\nu_3)=(0,0)$ for non-trivial $\xi$.

\end{itemize}
Thus we find four non-anomalous symmetry-enriched phases and one anomalous symmetry-enriched phase. 
This agrees with \cite{Chen:2016pec}.

Another example is the 0-form symmetry $G=\mathbb{Z}_4$. The coupling $\xi$ is classified by $H^1(\mathbb{Z}_4,\mathbb{Z}_2)=\mathbb{Z}_2$. In particular, Bock$(\xi)$ is trivial. Thus the three-group symmetry (\ref{eqn:3-group}) reduces to the trivial product of 0-form, one-form and two-form symmetries, with the 't Hooft anomaly (\ref{eqn:anomalyZnZm}):
\begin{equation}\label{eqn:Z2gaugetheoryanomaly}
S_\text{anom}=\pi \int_{5d} 
\left( B_2B_3
	-\frac{1}{2}{\frak P}(B_2)X\right)~,
\end{equation}
where the backgrounds are ordinary cocycles.
In section \ref{sec:adjointQCD4} when we discuss the example $G=\mathbb{Z}_8$, we use the above result with $X$ being a special $\mathbb{Z}_4$ gauge field that can be lifted to a $\mathbb{Z}_8$ gauge field.

\subsection{General case}

Denote the $\mathbb{Z}_{n_i}$ dynamical gauge field by $u^i$, and $n_{ij}\equiv \gcd(n_i,n_j),n_{ijk}\equiv \gcd(n_i,n_j,n_k)$.
The 0-form symmetry defects are classified by $H^3({\cal B},U(1))=\prod_{i\geq j}\mathbb{Z}_{n_{ij}}\times \prod_{i>j>k} \mathbb{Z}_{n_{ijk}}$.
We will couple the theory to ordinary $\mathbb{Z}_{n_{ij}}$ background gauge fields $X^{ij}$ and ordinary $\mathbb{Z}_{n_{ijk}}$ background gauge fields $X^{ijk}$ as follows
\begin{equation}\label{eqn:abeliancoupling}
\sum_{i\geq j}\frac{2\pi}{n_{ij}}\int u^i\text{Bock}(u^j)X^{ij}+\sum_{i>j>k}\frac{2\pi}{n_{ijk}}\int u^iu^ju^k X^{ijk}~.
\end{equation}
We will denote $X^{ii}=X^i$.
Then the theory can couple to the background $X$ of 0-form symmetry $G$ by $(X^{ij},X^{ijk})=X^*\xi$ with $\xi\in H^1(BG,H^3({\cal B},U(1)))$.

We will also turn on the backgrounds for the one-form and two-form symmetries and study the constraints between the backgrounds of different higher-form symmetries.
From the constraint one can then derive the set of couplings (\ref{eqn:4dbackgrounds}).

The theory has two-form symmetry ${\cal B}=\prod\mathbb{Z}_{n_i}$ generated by the Wilson lines $\oint u^i$, with backgrounds $B_3^i$.
The theory has one-form symmetry ${\cal A}=\prod \mathbb{Z}_{n_i}$ that acts on the line operators, with backgrounds $B_2^i$ that modify the cocycle conditions for $u^i$ into $\delta u^i=B_2^i$ mod $n_i$. 
The theory also has codimension-two symmetry defects, labelled by $H^2({\cal B},U(1))=\prod_{i>j}\mathbb{Z}_{n_{ij}}$, and we couple them to the backgrounds $C_2^{ij}$.

The second coupling in (\ref{eqn:abeliancoupling}) permutes the types of surface operators as discussed in section \ref{sec:permutationactionopr}, while the first coupling does not.
In the following we will focus on the first coupling in (\ref{eqn:abeliancoupling}).

The theory couples to the backgrounds as: (terms with two $i$s and $j$s are summed over $i\geq j$, while terms with two $i$s are summed over $i$)
\begin{equation}
\frac{2\pi}{n_{ij}n_j}\int \tilde u^i\delta \tilde u^j X^{ij}+
\frac{2\pi}{n_{ij}}\int u^iu^jC_2^{ij}+\frac{2\pi}{n_i}\int u^i B_3^i~,
\end{equation}
where $\tilde u^i$ is a lift of $u^i$ to an integral one-cochain, and we first set $B_2^i=0$. 
In the absence of backgrounds $B_2^i$ the theory is well-defined, with the backgrounds obeying the standard cocycle conditions.
It is easy to check the dynamical fields are independent of the lift, so we can omit the tilde notation.

If we turn on non-trivial $B_2^i$, we need to couple $B_2^i$ to additional terms to make the dynamical fields $u^i$ independent of the lifts.
Changing the lift $\tilde u^i\rightarrow \tilde u^i+n_i c^i$ with integral one-cochain $c^i$ shifts the action by
\begin{equation*}
\frac{2\pi}{n_{ij}n_j}\int\left(n_ic^iB_2^j+n_jc^jB_2^i-n_jB_2^i\cup_1\delta c ^j \right)X^{ij}~,
\end{equation*}
where we used $\delta u^i=B_2^i$ mod $n_i$, $\delta X^{ij}=0$ mod $n_{ij}$ and (\ref{eqn:cochainrules}).
To cancel the dependence on the lift, we add additional couplings to $B_2^i$,
\begin{equation}\label{eqn:Abeliantheorycoupling}
\frac{2\pi}{n_{ij}n_j}\int \left(\tilde u^i\delta \tilde u^j-\tilde u^iB_2^j-\tilde u^j B_2^i+B_2^i\cup_1\delta \tilde u^j \right)
X^{ij}+
\frac{2\pi}{n_{ij}}\int u^iu^jC_2^{ij}+\frac{2\pi}{n_i}\int u^i B_3^i~,
\end{equation}
Since the theory (\ref{eqn:Abeliantheorycoupling}) does not depend on the lift of $u^i$, we will drop the tilde notation from now on.
We still need to make sure the theory is invariant under the $\mathbb{Z}_{n_i}$ gauge symmetry of $u^i$, which requires the dynamical fields to be independent of the bulk. 
The theory (\ref{eqn:Abeliantheorycoupling}) depends on the bulk by
\begin{equation*}
\frac{2\pi}{n_{ij}n_j}\int_{5d} \delta\left( \left(u^i\delta u^j-u^iB_2^j-u^j B_2^i+B_2^i\cup_1\delta u^j \right)
X^{ij}\right)
+\frac{2\pi}{n_{ij}}\int_{5d} \delta\left( u^iu^j C_2^{ij} \right)+\frac{2\pi}{n_i}\int\delta\left(  u^i B_3^i\right)~,
\end{equation*}
which can be simplified using (\ref{eqn:cochainrules}), $\delta u^i=B_2^i$ mod $n_i$, $\delta X^{ij}=0$ mod $n_{ij},$  and completing the square with $(\delta u^i-B_2^i)(\delta u^j-B_2^j)=0$ mod $n_i n_j$:\footnote{
The exact term $\frac{2\pi}{n_{ij}}\int_{5d}\delta\left((B_2^i\cup_1 u^j) C_2^{ij}\right)$ in the last line of (\ref{eqn:bulkAbeliangaugedepn}) can be cancelled by adding to (\ref{eqn:Abeliantheorycoupling}) the local counterterm $-\frac{2\pi}{n_{ij}}\int (B_2^i\cup_1 u^j) C_2^{ij}$ which does not depend on the lift of the dynamical fields.
}
\begin{align}\label{eqn:bulkAbeliangaugedepn}
&\sum_{i\geq j}\frac{2\pi}{n_{ij}n_j}\int_{5d}\left\{
\left(
u^i\delta B_2^j+u^j\delta B_2^i+\delta B_2^i\cup_1B_2^j-B_2^iB_2^j\right)X^{ij}
-\left(u^j B_2^i-B_2^i\cup_1 B_2^j \right)   \delta X^{ij}\right\}\cr
&
+\sum_{i>j}
\frac{2\pi}{n_{ij}}\int_{5d}\left\{
(u^j B_2^i-u^iB_2^j)C_2^{ij}+(u^iu^j-B_2^i\cup_1u^j)\delta C_2^{ij}\right\}
+\frac{2\pi}{n_{ij}}\int_{5d} \delta((B_2^i\cup_1 u^j) C_2^{ij})\cr
&+\sum_i\frac{2\pi}{n_i}\int_{5d}\left( B_2^iB_3^i-u^i\delta B_3^i\right)\cr
&=
\sum_i\frac{2\pi}{n_i}\int_{5d} 
	u^i\left(
	2{\delta B_2^i\over n_i}X^i-B_2^i{\delta X^i\over n_i}
	+\sum_{j>i}\left(
		{n_i\over n_{ij}}{\delta B_2^j\over n_j}X^{ji}-B_2^j{\delta X^{ji}\over n_{ij}}+{n_i\over n_{ij}} B_2^jC_2^{ji}\right)
	\right.\cr
	&\qquad\;\left.
	+\sum_{j<i}\left( 
		{n_i\over n_{ij}}{\delta B_2^j\over n_j}X^{ij}-{n_i\over n_{ij}} B_2^jC_2^{ij}\right)
	-\delta B_3^i\right)		\cr
&\qquad+\sum_{j>i}\frac{2\pi}{n_{ij}}\int_{5d}\left\{(u^iu^j-B_2^i\cup_1u^j)\delta C_2^{ij}+\delta\left((B_2^i\cup_1 u^j) C_2^{ij}\right)\right\}+S_\text{anom}~,
\end{align}
where $S_\text{anom}$ only depends on the background fields
\begin{align}\label{eqn:threegroupAbeliananom}
S_\text{anom}=
&\sum_i\frac{2\pi}{n_i}\int_{5d}
\left\{
B_2^i B_3^i-\frac{1}{n_i}\left( (B_2^iB_2^i-\delta B_2^i\cup_1 B_2^i)X^{i}-(B_2^i\cup_1 B_2^i)\delta X^{i}\right)\right\}
\cr
&-\sum_{i>j}\frac{2\pi}{n_{ij}n_j}\int_{5d}\left( (B_2^iB_2^j-\delta B_2^i\cup_1 B_2^j)X^{ij}-(B_2^i\cup_1 B_2^j)\delta X^{ij}\right)~.
\end{align}
Thus for the dynamical fields $u^i$ to live only in $(3+1)d$, the background fields must satisfy the following constraints:\begin{align}\label{eqn:constraintAbeliangaugethreegroup}
&\delta C_2^{ij}=0\text{ mod }n_{ij}\cr
&\delta B_3^i
=
2{\delta B_2^i\over n_i}X^i-B_2^i{\delta X^i\over n_i}
+\sum_{j>i}\left(
	{n_i\over n_{ij}}{\delta B_2^j\over n_j}X^{ji} + {n_i\over n_{ij}} B_2^jC_2^{ji}-B_2^j{\delta X^{ji}\over n_{ij}}\right)
\cr
&\qquad\qquad	+\sum_{j<i}\left( {n_i\over n_{ij}}{\delta B_2^j\over n_j}X^{ij}- {n_i\over n_{ij}} B_2^jC_2^{ij}\right)
\text{ mod }n_i~.
\end{align}
Such backgrounds describe a three-group symmetry. 
Consider the background gauge transformation
\begin{equation}
B_2^i\rightarrow B_2^i+\delta \lambda_1^i,\qquad C_2^{ij}\rightarrow C_2^{ij}+\delta \lambda_1^{ij}~.
\end{equation}
The backgrounds $B_3^i$ transform as
\begin{align}\label{eqn:threegroupgeneralgaugetransf}
&B_3^i\longrightarrow B_3^i
-\lambda_1^i{\delta X^i\over n_i}
+\sum_{j>i}\left(
	{n_i\over n_{ij}} (\lambda_1^jC_2^{ji}+B_2^j\lambda_1^{ji}+\lambda_1^j\delta \lambda_1^{ji})
	-\lambda_1^j{\delta X^{ji}\over n_{ij}} \right)
\cr
&\qquad\qquad	-\sum_{j<i}\left({n_i\over n_{ij}} (\lambda_1^jC_2^{ij}+B_2^j\lambda_1^{ij}+\lambda_1^j\delta \lambda_1^{ij})\right)
\text{ mod }n_i~.
\end{align}

The 't Hooft anomaly of the three-group symmetry is given by $S_\text{anom}$ in (\ref{eqn:threegroupAbeliananom}).

\subsubsection{Couplings}

From the constraints (\ref{eqn:constraintAbeliangaugethreegroup}) we can turn on the following couplings.
First, for each $\xi\in H^1(BG,H^3({\cal B},U(1))')$ we substitute $(X^{ij})=X^*\xi$.
Then we fix the backgrounds for the higher-form symmetries: $(B_2^i,C_2^{ij})=X^*\eta_2$ with $\eta_2\in H^2(BG,{\cal A})$, and $B_3^i=X^*\nu_3$ with $\nu_3\in C^3(BG,{\cal B})$, where the coupling parameters $(\eta_2,\nu_3)$ are constrained by (\ref{eqn:constraintAbeliangaugethreegroup}).
Namely,
\begin{align}
&\delta \nu_3^i=
2{\delta \eta_2^i\over n_i}\xi^i-\eta_2^i{\delta \xi^i\over n_i}
+\sum_{j>i}\left({n_i\over n_{ij}}{\delta \eta_2^j\over n_j}\xi^{ji}-\eta_2^j{\delta \xi^{ji}\over n_{ij}}+\eta_2^j \eta_2^{ji}\right)
\cr
&\qquad\qquad	+\sum_{j<i}\left( {n_i\over n_{ij}}{\delta \eta_2^j\over n_j}\xi^{ij}-\eta_2^j\eta_2^{ij}\right)
\text{ mod }n_i~,
\end{align}
where the superscripts are the components of each part in the coefficient groups (which are products of cyclic groups).
The couplings $(\eta_2,\nu_3)$ are subject to the gauge transformation (\ref{eqn:threegroupgeneralgaugetransf}) with the transformation parameters $(\lambda_1^i,\lambda_1^{ij})=X^*\lambda_1$ for $\lambda_1\in C^1(BG,{\cal A})$.

The 't Hooft anomalies of different couplings are given by substitution in (\ref{eqn:threegroupAbeliananom}),
\begin{align}\label{eqn:Abeliangaugeanomalycoupling}
S_\text{anom}=\int_{5d}X^*\omega_5,\quad
\omega_5=&\sum_i\frac{2\pi}{n_i}
\left\{
\eta_2^i \nu_3^i-\frac{1}{n_i}\left( (\eta_2^i\eta_2^i-\delta \eta_2^i\cup_1 \eta_2^i)\xi^{i}
	-(\eta_2^i\cup_1 \eta_2^i)\delta \xi^{i}\right)
\right\}
\cr
&-\sum_{i>j}\frac{2\pi}{n_{ij}n_j}\left( (\eta_2^i\eta_2^j-\delta \eta_2^i\cup_1 \eta_2^j)\xi^{ij}-(\eta_2^i\cup_1 \eta_2^j)\delta \xi^{ij}\right)~.
\end{align}
One can verify $\omega_5\in H^5(BG,U(1))$.
\subsection{Correlation functions}
\label{sec:correlationfunctionZn}

In this section we will compute the correlation functions in ${\cal B}=\prod_i\mathbb{Z}_{n_i}$ gauge theory with trivial Dijkgraaf-Witten action \cite{Dijkgraaf:1989pz}. In particular, we will compute the correlation functions of the symmetry surface defects and the codimension-one symmetry defects in $H^3({\cal B},U(1))'$. It is sufficient to discuss the case ${\cal B}=\mathbb{Z}_{n}\times\mathbb{Z}_{n'}$. 
The theory can be written in the continuous notation as: \cite{Maldacena:2001ss,Banks:2010zn,Kapustin:2014gua}
\begin{equation}\label{eqn:ZnZnpgaugetheorycont}
\frac{n}{2\pi}adb_2+\frac{n'}{2\pi}a'db_2'~,
\end{equation}
where $a,a'$ are $U(1)$ one-form gauge fields and $b_2,b_2'$ are $U(1)$ two-form gauge fields. The equation of motions constrain $a,b_2$ to have $\mathbb{Z}_n$ holonomy, and similarly $a',b_2'$ to have $\mathbb{Z}_{n'}$ holonomy.
There are symmetry surface defects $\oint aa'$, $\oint da/n$, $\oint da'/n'$, $\oint b_2$, and $\oint b_2'$.
The codimension-one symmetry defects in $H^3({\cal B},U(1))'=\mathbb{Z}_n\times\mathbb{Z}_{n'}\times\mathbb{Z}_{\gcd(n,n')}$ are generated by $\oint ada$, $\oint a'da'$ and $\oint ada'$.

First, we show there is $\mathbb{Z}_n$ braiding between $\oint_{\gamma} a$ and $\oint_\Sigma b_2$.
Since $\oint_\Sigma b_2=\int \delta(\Sigma^{\perp})b$, where $\delta(\Sigma^{\perp})$ is the two-form Poincar\'e dual to $\Sigma$,
 the equation of motion for $b_2$ is modified by the insertion to be $da=-(2\pi/n)\delta(\Sigma^\perp)$. 
Substituting this to $\oint_\gamma a$ gives the correlation function
\begin{equation}\label{eqn:Znbraidingcomputation}
\left\langle
\exp\left(i\oint_{\Sigma}b_2\right)\exp\left(i\oint_\gamma a\right)
\right\rangle
=
\exp\left(-{2\pi\over n}\text{Link}(\gamma,\Sigma)\right)~.
\end{equation}
Thus the lines $\oint a,\oint a'$ and surfaces $\oint b,\oint b'$ are non-trivial operators.

Consider the correlation functions with insertion of the ``fractional symmetry line operator'' $\exp(i\oint_\Sigma da/n)$. A similar computation shows that the only correlation function is the contact correlation function with the surface operator $\oint_{\Sigma'}b$:
\begin{equation}
\left\langle 
\exp\left( i\oint_\Sigma da/n\right) 
\exp\left( i\oint_{\Sigma'} b\right)\right\rangle=
\exp\left(-{2\pi i\over n^2}\#(\Sigma,\Sigma')\right)~,
\end{equation}
where $\#(\Sigma,\Sigma')$ is the intersection number of the two surfaces\footnote{
For $\gamma=\partial\Sigma$ the intersection number can be written as Link$(\gamma,\Sigma')$. However, since $\int da/n$ is not a genuine line operator but is a surface operator, the correlation function is a contact term.
}.
Thus the surface $\oint da/n$ is a redundant operator and should not be included in the list of non-trivial operators.

\begin{figure}[t]
  \centering
    \includegraphics[width=0.35\textwidth]{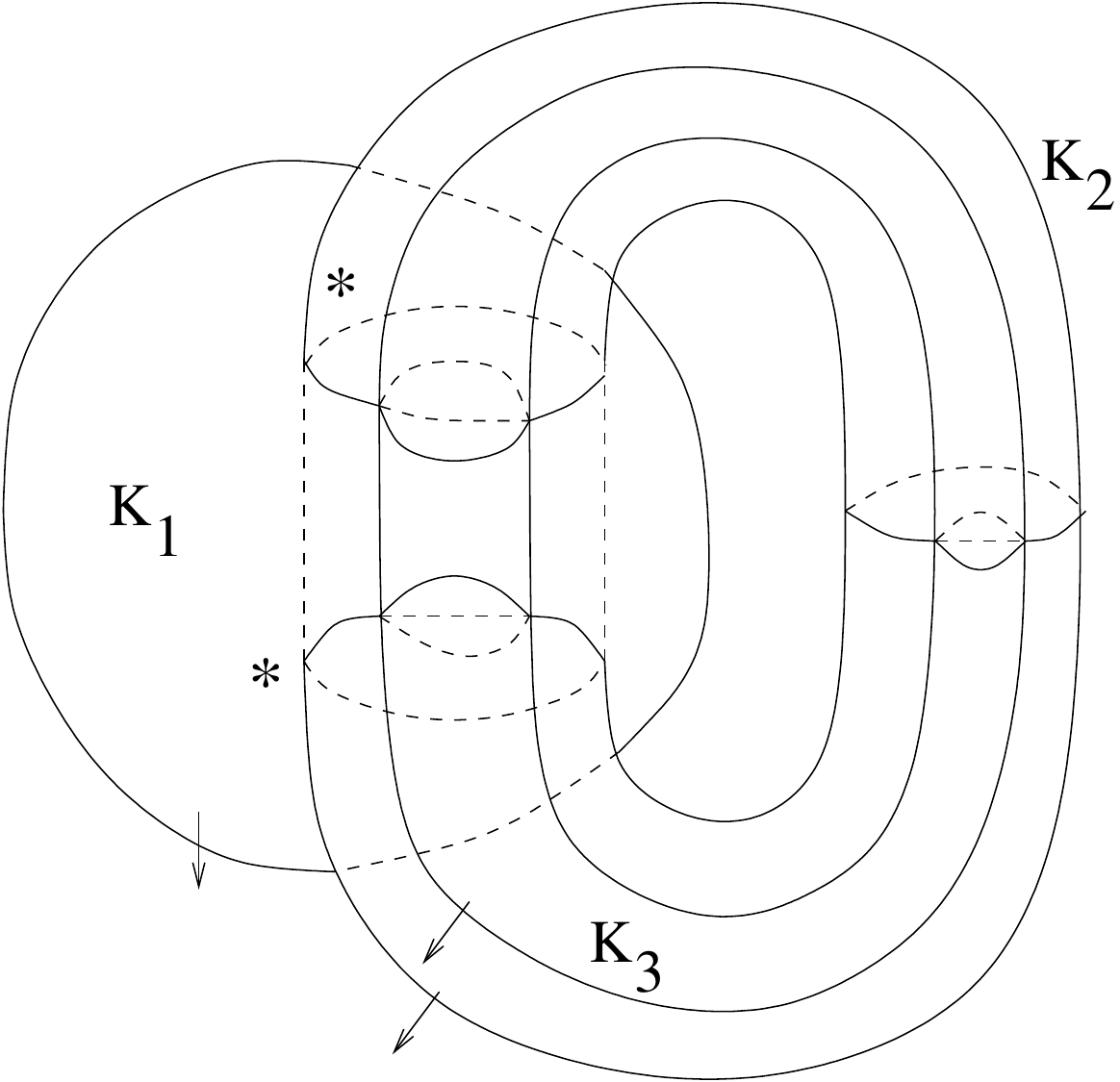}
      \caption{An example of a non-trivial triple linking number between surfaces $K_1,K_2,K_3$. 
      The figure is taken from \cite{Carter:2003TLK}}\label{fig:triplelink}
\end{figure}

Next, consider the correlation functions of the surface defect
\begin{equation}
{\cal W}_{aa'}=\exp\left( i{\text{lcm}(n,n')\over 2\pi}\int aa'\right)~,
\end{equation}
where $\text{lcm}(n,n')$ is the least common multiple of $n,n'$.
The operator is gauge invariant since $a,a'$ have $\mathbb{Z}_n$ and $\mathbb{Z}_{n'}$ holonomies, respectively.
This surface operator has trivial correlation function with all line operators.
We will compute the correlation function with two surface operators $\oint b,\oint b'$ at $\Sigma,\Sigma'$ and ${\cal W}_{aa'}$ at $\Sigma''$.
The equations of motion for $b,b'$ give $da=-(2\pi/n)\delta(\Sigma^\perp)$ and $da'=-(2\pi/n')\delta(\Sigma'^\perp)$.
On $S^4$ spacetime the equations can be solved by (modulo gauge transformations)
\begin{equation}
a=-\frac{2\pi}{n}\delta({\cal V}^\perp),\qquad a'=-\frac{2\pi}{n'}\delta({\cal V}'^\perp)~,
\end{equation}
where $\Sigma=\partial {\cal V}$, $\Sigma'={\cal V}'$, and we used $d\delta({\cal V^\perp})=\delta((\partial {\cal V})^\perp)$ for three-volume ${\cal V}$ from integration by parts\footnote{
We use the convention that the integration of $n$-form $\omega_n$ over $n$-dimensional locus $\Gamma_n$ gives the spacetime integral $\oint_{\Gamma_n}\omega_n=\int \delta(\Gamma_n^\perp)\omega_n$. Then using Stokes theorem and integration by parts one can show that $\delta((\partial\Gamma_n)^\perp)=(-1)^{D-n+1}d\delta(\Gamma_n^\perp)$ in $D$-dimensional spacetime.}.
Thus the correlation function is (where $\Sigma=\partial {\cal V}$, and similarly for ${\cal V}',{\cal V}''$)
\begin{align}\label{eqn:triplebraiding}
&\left\langle
\exp\left(i\oint_{\Sigma}b_2\right)\exp\left(i\oint_{\Sigma'}b_2'\right)
\exp\left(i{\text{lcm}(n,n')\over 2\pi}\oint_{\Sigma''} aa'\right)
\right\rangle\cr
&
=\exp\left( i{\text{lcm}(n,n')\over 2\pi}\int \delta(\Sigma''^\perp) aa'\right)
=\exp\left(
\frac{2\pi i}{\gcd(n,n')}\int \delta(\Sigma''^\perp)\wedge \delta({\cal V}^\perp)\wedge \delta({\cal V}'^\perp)\right)
\cr
&=
\exp\left(
\frac{2\pi i}{\gcd(n,n')}\text{Tlk}(\Sigma,\Sigma',\Sigma'')\right)~ ,
\end{align}
where Tlk is the triple linking number of three oriented surfaces \cite{Carter:2003TLK}. The triple linking number is an integer invariant of the surface-links, with the integral representation $\text{Tlk}(\Sigma_1,\Sigma_2,\Sigma_3)=\int \delta({\cal V}_1^\perp)\wedge\delta({\cal V}_2^\perp)\wedge \delta(\Sigma_3^\perp)$, where $\Sigma_i$ are closed oriented surfaces and $\partial{\cal V}_i=\Sigma_i$. An example of a surface-link that has non-trivial triple linking number is given in figure \ref{fig:triplelink}.

Therefore the surface operator ${\cal W}_{aa'}$ is a non-trivial operator, but nevertheless it has trivial braiding with every line operator.

As an exercise, one can use the correlation function (\ref{eqn:triplebraiding}) to check that the permutation 0-form symmetry (\ref{eqn:permutationsymma1a2a3}) preserves the correlation functions in $\mathbb{Z}_{n_1}\times\mathbb{Z}_{n_2}\times\mathbb{Z}_{n_3}$ gauge theory with zero Dijkgraaf-Witten action \cite{Dijkgraaf:1989pz}. The $\mathbb{Z}_{n_i}$ gauge field can be expressed as a pair of one-form and two-form gauge fields $a^i,b_2^i$ as in (\ref{eqn:ZnZnpgaugetheorycont}). Consider a different correlation function, between $\oint b_2^1,\oint b_2^2,\oint b_2^3$, which is trivial:
\begin{equation}\label{eqn:corretest}
\left\langle \prod_{i=1}^3 \exp\left(i\oint_{\Sigma_i}b_2^i\right)  \right\rangle=1~.
\end{equation}
Under the permutation 0-form symmetry (\ref{eqn:permutationsymma1a2a3}), 
\begin{equation}
\exp(i\oint b_2^1)\;\longrightarrow\; \exp\left(i\oint b_2^1+i\frac{n_2n_3}{2\pi\ell}\oint a^2 a^3\right),\quad \ell\equiv\gcd(n_1,n_2,n_3)~,
\end{equation}
and similarly for $\oint b_2^2,\oint b_2^3$. The correlation function (\ref{eqn:corretest}) between $\oint b_2^1,\oint b_2^2,\oint b_2^3$ becomes
\begin{align}
&\left\langle
\exp\left(i\oint_{\Sigma_1}b_2^1\right)
\exp\left(i\oint_{\Sigma_2}b_2^2\right)
\exp\left(i\oint_{\Sigma_3}b_2^3\right)
\right\rangle\;\longrightarrow\cr
&\left\langle 
\exp\left(i\oint_{\Sigma_1}\left( b_2^1+\frac{n_2n_3}{2\pi\ell}a^2a^3\right)\right)
\exp\left(i\oint_{\Sigma_2}\left( b_2^2+\frac{n_3n_1}{2\pi\ell}a^3a^1\right)\right)
\exp\left(i\oint_{\Sigma_3}\left( b_2^3+\frac{n_1n_2}{2\pi\ell}a^1a^2\right)\right)
  \right\rangle\cr
&=
\exp\left(
\frac{2\pi i}{\ell}\left(\text{Tlk}(\Sigma_1,\Sigma_2,\Sigma_3)
+\text{Tlk}(\Sigma_2,\Sigma_3,\Sigma_1)+\text{Tlk}(\Sigma_3,\Sigma_1,\Sigma_2)\right)
\right)=1~,
\end{align}
where the last equality used the following identity for the triple linking number: \cite{Carter:2003TLK}
\begin{equation}
\text{Tlk}(\Sigma_1,\Sigma_2,\Sigma_3)+\text{Tlk}(\Sigma_2,\Sigma_3,\Sigma_1)+\text{Tlk}(\Sigma_3,\Sigma_1,\Sigma_2)=0~,
\end{equation}
which can be derived from the integral representation of the triple linking number. Thus the correlation function (\ref{eqn:corretest}) is indeed preserved by the permutation 0-form symmetry (\ref{eqn:permutationsymma1a2a3}).

We remark that the non-trivial surface operator ${\cal W}_{aa'}$ is a counterexample to the claim in
 \cite{Lan:2018vjb} which states that in any bosonic $(3+1)d$ TQFT there is no non-trivial surface operator that has trivial braiding with all line operators. The argument there relies on condensing all particle excitations without changing the surface operators that have trivial braiding with all line operators. On the other hand, the surface operator ${\cal W}_{aa'}$ is defined by line operators (particles) wrapping one-cycles on the surface in spacetime, and thus condensing all particles makes such surface operators trivial (another way to see this is that such a surface operator no longer has a non-trivial correlation function (\ref{eqn:triplebraiding}) after the condensation and thus it is no longer a non-trivial operator).\footnote{
A similar example is the pure $U(1)$ gauge theory, which has symmetry surface operator $\oint F$ with $F$ the $U(1)$ field strength. It has trivial braiding with the electric Wilson lines (while non-trivial braiding with the magnetic 't Hooft lines).
On the other hand, condensing the basic electric particle changes the theory to be trivial at low energy, and it changes this non-trivial surface operator to be trivial, even though the surface operator has trivial braiding with the condensing particle.}

We would like to make some comparisons with a different triple-linking process that is discussed in \cite{Wang:2014xba,Jiang:2014ksa,Chen:2015gma,Tiwari:2016zru,Putrov:2016qdo}.
They discussed $\mathbb{Z}_{n_1}\times\mathbb{Z}_{n_2}$ or $\mathbb{Z}_{n_1}\times\mathbb{Z}_{n_2}\times\mathbb{Z}_{n_3}$ gauge theory with a non-trivial Dijkgraaf Witten action. In contrary, the gauge theories discussed here have zero Dijkgraaf-Witten action.
In addition, all of triple-linking processes discussed in \cite{Wang:2014xba,Jiang:2014ksa,Chen:2015gma,Tiwari:2016zru,Putrov:2016qdo} is between three surface operators where each of them has non-trivial braiding with some line operator. In contrary, in the triple-linking process (\ref{eqn:triplebraiding}) one of the surface operator $\oint aa'$ has trivial braiding with all line operators.

Consider now the correlation functions of codimension-one symmetry defects.
They have only contact correlation functions with non-local operators.

First, consider the codimension-one symmetry defects
\begin{equation}
{\cal U}_{aa'}=\exp\left({i \over 2\pi}\oint a da'\right)~.
\end{equation}
A similar computation as above shows that it has the following correlation function with two surface operator $\oint b,\oint b'$:
(where $\partial{\cal V}=\Sigma$)
\begin{equation}\label{eqn:correlUaap}
\left\langle\; {\cal U}_{aa'}({\cal V}'')\exp(i\oint_\Sigma b)\exp(i\oint_{\Sigma'} b')\;\right\rangle
=\exp\left(\frac{2\pi i}{nn'}\#({\cal V}'',{\cal V},\Sigma')\right)~.
\end{equation}

Similarly, one can compute the correlation function that involve ${\cal U}_a=\exp\left(\frac{i}{2\pi}\oint ada\right)$ with two surface operators of the same type, $\oint_{\Sigma} b$ and $\oint_{\Sigma'} b$. The result is the square of (\ref{eqn:correlUaap}) with $n=n'$.

\subsection{Selection rules depend on the couplings}
\label{sec:selectionrule}

Here we will use $\mathbb{Z}_n$ gauge theory to demonstrate how higher-form symmetries lead to selection rules for the observables \cite{Gaiotto:2014kfa}. 
For simplicity, we will consider the trivial coupling $\xi=0$ so $(\eta_2,\nu_3)$ are ordinary cocycles.

The $\mathbb{Z}_n$ gauge theory (\ref{eqn:Zngaugetheory}) has ${\cal A}=\mathbb{Z}_n$ one-form symmetry and ${\cal B}=\mathbb{Z}_n$ two-form symmetry, generated by the surface operator $V=\exp({2\pi i\over n}\oint v_2)$ and the line operator $U=\exp({2\pi i\over n}\oint u)$, respectively.
From their braiding correlation functions (\ref{eqn:Znbraidingcomputation}) we find that $U$ carries one-form charge $(-1)$, and $V$ carries two-form charge $(-1)$. 
The one-form and two-form symmetry transformations can be expressed as $u\mapsto u-\lambda_1$ with $\lambda_1\in Z^1(M;\mathbb{Z}_n)$, and $v_2\mapsto v_2-\lambda_2$ with $\lambda_2\in Z^2(M;\mathbb{Z}_n)$.

A line (or surface) operator with unit higher-form charge and supported on a cycle $\gamma$ (or $\Sigma$) transforms under the one-form (two-form) symmetry by the phase $\exp({2\pi i\over n}\oint_{\gamma}\lambda)$ (or $\exp({2\pi i\over n}\oint_{\Sigma}\lambda_2)$).
Thus if the cycle $\gamma$ or $\Sigma$ has a non-trivial homological class, the correlation functions of these non-local operators vanish unless the total one-form and total two-form charges both vanish.

Consider the spacetime $S^1\times M_3$ with $S^1$ being the time direction.
We wrap $k$ line operators of one-form charges $q_{\cal A}^{(i)}\in\mathbb{Z}_n$ on $S^1$. This is equivalent to inserting $k$ point defects (particle excitations) in space that extend over time, which modifies the Hilbert space on ${\cal M}_3$.
The Hilbert space is empty unless these point defects can fuse to the trivial defect.
In particular, the Hilbert space is necessarily empty unless
\begin{equation}\label{eqn:selectiononeform}
\sum_{i=1}^k q_{\cal A}^{(i)}=0\text{ mod }n~,
\end{equation}
which is consistent with the discussion above.

If there is a mixed 't Hooft anomaly between the one-form and two-form symmetries {\it i.e.} the symmetry line operators that generate the two-form symmetry are charged under the one-form symmetry, then inserting such symmetry line operators on $S^1$ contributes to the selection rule (\ref{eqn:selectiononeform}) by their one-form charges.
In the example of $\mathbb{Z}_n$ gauge theory, the symmetry line operator $Q_{\cal B}\oint u$ that generates the two-form symmetry element $Q_{\cal B}\in{\cal B}$ also carries the one-form charge $q_{\cal A}=-Q_{\cal B}\in\mathbb{Z}_n$.
Thus an insertion of such symmetry line defect on $S^1$ modifies the selection rule (\ref{eqn:selectiononeform}) into
\begin{equation}\label{eqn:selectionruleoneformQB}
\sum_{i=1}^k q_{\cal A}^{(i)}=Q_{\cal B}\text{ mod }n~,
\end{equation}

On the other hand, inserting a symmetry line operator at $\gamma$ that corresponds to $Q_{\cal B}\in {\cal B}$ is equivalent to activating a background for the (subgroup) two-form symmetry generated by the symmetry line operator. The background is given by Poincar\'e duality as $B_3=Q_{\cal B}\,\text{PD}(\gamma)$.
Thus on spacetime $S^1\times {\cal M}_3$ with $\gamma=S^1$, we have $Q_{\cal B}=\oint_{{\cal M}_3} B_3$.
The selection rule (\ref{eqn:selectionruleoneformQB}) becomes
\begin{equation}
\sum_{i=1}^k q^{(i)}_{\cal A}=\oint_{{\cal M}_3} B_3\text{ mod }n~,
\end{equation}

Therefore the couplings to the 0-form symmetry background $X$ by the fixed background $B_3=X^*\nu_3$ for $\nu_3\in H^3(BG,\mathbb{Z}_n)$ implies the selection rules
\begin{equation}\label{eqn:selectionnu}
\sum_{i=1}^k q_{\cal A}^{(i)}=\oint_{{\cal M}_3} X^*\nu_3\text{ mod }n~.
\end{equation}

Similarly, consider the case ${\cal M}_3={\cal M}_2\times S'^1$ (with a prime to distinguish it from the temporal circle). Wrapping $k'$ surface operators of two-form charges $q_{\cal B}^{(i)}$ on $S^1\times S'^1$ leads to the selection rules on the Hilbert space modified by line defects
\begin{equation}\label{eqn:selectioneta}
\sum_{i=1}^{k'} q^{(i)}_{\cal B}=\oint_{{\cal M}_2} X^*\eta_2\text{ mod }n~.
\end{equation}
Thus different couplings $(\eta_2,\nu_3)$ lead to different selection rules (\ref{eqn:selectionnu}),(\ref{eqn:selectioneta}).

In the general case, we can use the bilinear braiding between the generators of one-form and two-form symmetries to define the maps ${\cal M}_{\cal AB}:\;{\cal A}\rightarrow {\hat{\cal B}}\cong {\cal B}$ and ${\cal M}_{\cal BA}:\;{\cal B}\rightarrow {\hat{\cal A}}\cong {\cal A}$. Then the selection rules are given by (\ref{eqn:selectionnu}) and (\ref{eqn:selectioneta}) with the right hand sides  pushed forward by the maps ${\cal M}_{\cal BA}$ and ${\cal M}_{\cal AB}$, respectively.

\section{Application to adjoint QCD$_4$ with two flavors}
\label{sec:adjointQCD4}

In \cite{Cordova:2018acb,Bi:2018xvr} it is proposed as a possibility that the UV theory of $SU(2)$ gauge theory with one massless adjoint Dirac fermion ({\it i.e.} two Weyl fermions) flows to the IR theory of a single free massless Dirac fermion with decoupled $\mathbb{Z}_2$ gauge theory. 
Here we will discuss how the backgrounds of global symmetries couple to the UV and the IR theories.
We will use the result in section \ref{sec:Z2gaugetheoryunitary} for the $\mathbb{Z}_2$ gauge theory, and the method of coupling the IR theory to backgrounds of the UV symmetry \cite{Benini:2018reh}. For other proposals of the possible low energy behavior of adjoint QCD, see \cite{Cordova:2018acb} and the references therein.

The $SU(2)$ gauge theory has $\mathbb{Z}_2$ one-form symmetry that transforms the Wilson lines of half-integer $SU(2)$ isospins, and $\mathbb{Z}_8$ 0-form symmetry that transforms the adjoint fermion (it is the remnant of the classical $U(1)$ symmetry broken by a gauge-global Adler-Bell-Jackiw anomaly \cite{Adler:1969gk,Bell:1969ts}, under which the fermions have charge one)\footnote{
The $\mathbb{Z}_8$ symmetry is called an axial rotation on the fermions in \cite{Bi:2018xvr}.
}.
The theory can be defined on a non-spin manifold \cite{Cordova:2018acb}.
In the $SU(2)$ gauge theory, the symmetries have the 't Hooft anomaly: \cite{Cordova:2018acb}
\begin{equation}\label{eqn:anomalyZ2matching}
\pi\int_{5d}X_2 w_3+\frac{\pi}{2}\int_{5d}{\frak P}(X_2)X~,
\end{equation}
where $X_2$ is the background for $\mathbb{Z}_2$ one-form symmetry, and $X$ is the background for $\mathbb{Z}_8$ 0-form symmetry, and $w_3$ is the third Stiefel-Whitney class of the spacetime manifold.
Since the free Dirac fermion does not have a one-form symmetry, such an anomaly can only be matched by the low energy $\mathbb{Z}_2$ gauge theory.

\subsection{Broken 0-form symmetry}

In the proposal of \cite{Cordova:2018acb}, the $\mathbb{Z}_8$ 0-form symmetry is spontaneously broken to $\mathbb{Z}_4$ at low energies \cite{Anber:2018tcj} (which is based on the known physics of adjoint QCD on small $S^1$ \cite{Unsal:2007vu,Unsal:2007jx} as reviewed in \cite{Dunne:2016nmc}). Thus the theory has two vacua, exchanged by a $\mathbb{Z}_8$ transformation that is not in $\mathbb{Z}_4$.
The theory admits a domain wall interpolating between the two vacua. The anomaly for the $\mathbb{Z}_8$ elements that are not in $\mathbb{Z}_4$ induces an anomaly on the $(2+1)d$ domain wall as described by the SPT phase
\begin{equation}\label{eqn:domainwallanom}
\frac{\pi}{2}\int_{4d}{\frak P}(X_2)~.
\end{equation}
The anomaly on the domain wall (\ref{eqn:domainwallanom}) corresponds to $\mathbb{Z}_2$ one-form symmetry generated by a deconfined line operator of spin $\frac{1}{4}$ mod 1 on the wall \cite{Hsin:2018vcg}.
For instance, such an anomaly on the wall can be realized by $U(1)_2$ or the bosonic $(2+1)d$ $\mathbb{Z}_2$ gauge theory with non-trivial Dijkgraaf-Witten action \cite{Dijkgraaf:1989pz} {\it i.e.} the semion-antisemion theory.
Another possibility is $SU(2)_k$ Chern-Simons matter theory with $N_a$ massless Majorana fermions in the adjoint representation where the anomaly matching requires odd $(k+N_a)$. We remark that when $N_a=1,k=0$ this theory is conjectured to flow to a massless Majorana fermion (the Goldstino for broken ${\cal N}=1$ supersymmetry) and a decoupled $U(1)_2$ \cite{Gomis:2017ixy}.

The two vacua associated with the $\mathbb{Z}_8\rightarrow\mathbb{Z}_4$ symmetry breaking have $\mathbb{Z}_4$ 0-form symmetry, and thus we need to match the anomaly for the $\mathbb{Z}_4$ subgroup symmetry.
The anomaly is given by substituting $X=\iota(X_1)$ in (\ref{eqn:anomalyZ2matching}) with $\mathbb{Z}_4$ ordinary gauge field $X_1$, where $\iota:\mathbb{Z}_4\rightarrow \mathbb{Z}_8$ is the inclusion map. Concretely, $X=2\tilde X_1$ where tilde denotes a lift in $\mathbb{Z}_8$ (it is independent of the lift, since changing $\tilde X_1$ by a multiple of 4 leaves $X$ invariant mod 8). The anomaly is
\begin{equation}\label{eqn:anomalyZ2matchingZ4}
\pi\int_{5d}X_2 w_3+\pi\int_{5d}{\frak P}(X_2)X_1
=\pi\int_{5d}X_2 w_3+\pi\int_{5d}X_2 X_2 X_1
=\pi\int_{5d}X_2\left(w_3+w_2X_1\right)
~,
\end{equation}
where the last equality uses $X_2X_2X_1=Sq^2(X_2X_1)$ for $\mathbb{Z}_4$ cocycle $X_1$ and $\pi\int Sq^2(X_2X_1)=\pi\int w_2X_2X_1$.
Moreover, in this theory the $SU(2)$ fundamental Wilson line is believed to be confined (see {\it e.g.} \cite{Shifman:2013yca} and the references therein), and thus the UV one-form symmetry does not act on the deconfined Wilson line in the low energy $\mathbb{Z}_2$ gauge theory {\it i.e.} the UV one-form symmetry is unbroken.
This implies that the background $B_2$ of the $\mathbb{Z}_2$ one-form symmetry in the $\mathbb{Z}_2$ gauge theory (that transforms the deconfined $\mathbb{Z}_2$ Wilson line) cannot depend on the background $X_2$ of the UV one-form symmetry.
Then the constraint that the low energy theory reproduces the anomaly in the UV leaves the only one coupling of the $\mathbb{Z}_2$ gauge theory to the backgrounds:
$B_3=\text{Bock}(X_2)+X_2X_1$, $B_2=w_2$ and $\xi=0$, where $X_1$ is the background for $\mathbb{Z}_4$ 0-form symmetry. The $\mathbb{Z}_2$ gauge theory is
\begin{equation}\label{eqn:Z2gaugeproposalone}
\pi\int u\left(\text{Bock}(X_2)+X_2X_1\right)~,
\end{equation}
with dynamical $\mathbb{Z}_2$ gauge field $u$ that satisfies $\delta u=w_2$. This means that the $\mathbb{Z}_2$ Wilson line corresponds to a fermion particle.
It has the 't Hooft anomaly
\begin{equation}\label{eqn:anomalyordinary}
\pi\int_{5d} B_2B_3=\pi\int_{5d}w_2\text{Bock}(X_2)+\pi\int w_2X_2X_1
=\pi\int_{5d}X_2 w_3+\pi\int_{5d}X_2w_2X_1~,
\end{equation}
where we used $\pi\int_{5d}w_2\text{Bock}(X_2)=\pi\int_{5d}\text{Bock}(w_2)X_2=\pi\int_{5d}w_3X_2$ on closed orientable five-manifolds, and the relations for Steenrod squares.
Thus the theory has the anomaly that matches the anomaly (\ref{eqn:anomalyZ2matchingZ4}).

\subsubsection{Gauging the one-form symmetry}

Let us explore the consequence of the couplings in (\ref{eqn:Z2gaugeproposalone}) when gauging the $\mathbb{Z}_2$ one-form symmetry with dynamical two-form gauge field $X_2$. The microscopic theory becomes $SO(3)$ gauge theory with a Dirac fermion in the vector representation.
There is an option to add local counterterm for the two-form gauge field of the one-form symmetry
\begin{equation}
0,\quad {\pi\over 2}\int{\frak P}(X_2),\quad {2\pi\over 2}\int{\frak P}(X_2)=\pi\int X_2\cup w_2
,\quad {3\pi\over 2}\int{\frak P}(X_2)\quad \text{mod }2\pi\mathbb{Z}~,
\end{equation}
where we used $\pi\int X_2\cup X_2=\pi\int w_2\cup X_2$ on orientable manifold.
Since twice of ${\pi\over 2}\int{\frak P}(X_2)$ can be expressed as a mixed coupling with the spacetime manifold, we will consider the zero counterterm and ${\pi\over 2}\int{\frak P}(X_2)$, with the corresponding $SO(3)$ gauge theories denoted by $SO(3)_\pm$ following the notation in \cite{Aharony:2013hda}.
The two vacua in the $SU(2)$ gauge theory are related by the counterterm (\ref{eqn:domainwallanom}), and thus after gauging the one-form symmetry the two sides of the domain wall become distinct theories, which we denote by $SO(3)_+$ and $SO(3)_-$.

We proceed to describe the low energy physics on the two sides of the wall after gauging the one-form symmetry, which can be derived from (\ref{eqn:Z2gaugeproposalone}) as follows. First we turn off all the background fields (including putting the theory on a spin manifold with trivial $w_2$) except for $X_2$, which we then sum over in the path integral. We begin by considering the low energy theory for $SO(3)_+$.
Using integration by parts, $X_2$ couples to the $\mathbb{Z}_2$ gauge field linearly as $\pi\int \text{Bock}(u)X_2$, and thus integrating out $X_2$ imposes the constraint that $\text{Bock}(u)$ is trivial, which implies that the $\mathbb{Z}_2$ bundle is lifted to a $\mathbb{Z}_4$ bundle {\it i.e.} the gauge group is extended from $\mathbb{Z}_2$ to $\mathbb{Z}_4$.
Therefore we find the proposal in \cite{Cordova:2018acb} implies that the low energy theory dual to $SO(3)_+$ with a vector Dirac fermion is a free Dirac fermion together with a decoupled $\mathbb{Z}_4$ gauge theory with trivial Dijkgraaf-Witten action:
\begin{equation}
SO(3)_+\text{ with Dirac fermion }\Psi\text{ in }\mathbf{3}\;\longleftrightarrow\;
\text{Free Dirac fermion }\tilde\Psi+\mathbb{Z}_4\text{ gauge theory}~.
\end{equation}
We can also gauge the one-form symmetry and add the local counterterm $\pi\int {\frak P}(X_2)/2$. By completing the squares,
\begin{equation*}
\frac{\pi}{2}\int{\frak P}(X_2)+\pi\int X_2 \text{Bock}(u)
=\frac{\pi}{2}\int{\frak P}(X_2')-\frac{\pi}{2}\int {\frak P}(\text{Bock}(u)),\quad X_2'=X_2+\text{Bock}(u)~,
\end{equation*}
where the first term is a trivially gapped two-form $\mathbb{Z}_2$ gauge theory \cite{Kapustin:2014gua,Gaiotto:2014kfa,Hsin:2018vcg}, and the second term is also trivial since $H^4(\mathbb{Z}_2,U(1))=0$. Thus we find that the theory on the other side of the wall has the same low energy physics (up to classical local counterterms) as that before gauging the one-form symmetry:
\begin{equation}
SO(3)_-\text{ with Dirac fermion }\Psi\text{ in }\mathbf{3}\;\longleftrightarrow\;
\text{Free Dirac fermion }\tilde\Psi+\mathbb{Z}_2\text{ gauge theory}~.
\end{equation}
It is straightforward to investigate what becomes of the domain wall after gauging the one-form symmetry using the low energy theory descriptions following a similar discussion as in \cite{Hsin:2018vcg}. We will leave it to future work.

Now let us discuss the global symmetry after gauging the $\mathbb{Z}_2$ one-form symmetry.
There is dual $\mathbb{Z}_2$ one-form symmetry generated by $\exp(\pi i \oint X_2)$, with background denoted by $Y_2$ that couples as $\pi\int X_2 Y_2$. It is the magnetic one-form symmetry in $SO(3)$ gauge theory, which transforms the 't Hooft lines.
To cancel the mixed anomaly (\ref{eqn:anomalyZ2matchingZ4}), the background $Y_2$ must satisfy
\begin{equation}\label{eqn:2-groupgauging}
\delta Y_2=w_3+w_2X_1~.
\end{equation}
This means that the dual $\mathbb{Z}_2$ one-form symmetry participate in a two-group together with the Lorentz symmetry and the $\mathbb{Z}_4$ 0-form symmetry.
The two-group symmetry (\ref{eqn:2-groupgauging}) has different ``subgroup'' 0-form symmetries, which means that the theory can couple to only the 0-form symmetry background. For instance, $Y_2$ can be trivial if $w_2\neq 0,w_3=0$ and $X_1$ restricted to be a $\mathbb{Z}_2\subset \mathbb{Z}_4$ background. We emphasize that even when the manifold is non-spin, the $\mathbb{Z}_4$ 0-form symmetry is still a global symmetry, but it participates in a non-trivial two-group. This clarifies some statement in \cite{Bi:2018xvr}.

\subsection{Unbroken 0-form symmetry}
\label{sec:unbroken0-form}

Another proposal in \cite{Bi:2018xvr} suggests that the low energy theory preserves the full $\mathbb{Z}_8$ 0-form symmetry.
From (\ref{eqn:Z2gaugetheoryanomaly}) we find that to match the anomaly in the $SU(2)$ gauge theory, the $\mathbb{Z}_2$ gauge theory should couple to the $\mathbb{Z}_8$ 0-form symmetry background $X$ using non-trivial $\xi\in H^1(\mathbb{Z}_8,\mathbb{Z}_2)=\mathbb{Z}_2$, and $B_2=X_2$, $B_3=w_3$.
Substitution into (\ref{eqn:Z2gaugetheoryanomaly}) (with $X\rightarrow -X$) matches the anomaly (\ref{eqn:anomalyZ2matching}) in the $SU(2)$ gauge theory.
$B_3=w_3$ implies the theory has fermionic strings \cite{Thorngren:2014pza}.
The $\mathbb{Z}_2$ gauge theory coupled to the backgrounds with the anomaly (\ref{eqn:anomalyZ2matching}) is given by (\ref{eqn:Zngaugetheorycouplebackground}) with $X\rightarrow -X$:
\begin{equation}\label{eqn:Z2dualitypropop}
{\pi\over 2}\int\left( u\delta u +X_2\cup_1 \delta u\right)X+\pi\int uw_3~,
\end{equation}
where we used $\pi\int (u\delta u +X_2\cup_1 \delta u)X=\pi\int uX_2X$ and $X_2\cup_1 X_2=\text{Bock}(X_2)$.
In particular, $B_2=X_2$ implies the following: the one-form symmetry in the $SU(2)$ gauge theory that transforms the fundamental Wilson line corresponds to a one-form symmetry that acts on the deconfined $\mathbb{Z}_2$ Wilson line.
Thus within the couplings we discuss, for the low energy theory to preserve the $\mathbb{Z}_8$ 0-form symmetry, the $SU(2)$ fundamental Wilson line must be deconfined {\it i.e.} the one-form symmetry is spontanesouly broken (as opposed to the proposals in \cite{Anber:2018tcj,Cordova:2018acb}).
This is consistent with \cite{Cordova:2019bsd}, which proves that such UV 't Hooft anomaly cannot be realized by symmetry-preserving gapped theories.

\subsection{Goldstone mode of broken one-form symmetry}

In the scenario where the $\mathbb{Z}_2$ one-form symmetry is broken as discussed in section \ref{sec:unbroken0-form}, the one-form symmetry transforms the Wilson line in the low energy $\mathbb{Z}_2$ gauge theory. 
Thus the low energy $\mathbb{Z}_2$ gauge theory (\ref{eqn:Z2dualitypropop}) describes the Goldstone mode of the broken one-form symmetry. 
The action of the one-form symmetry on the line operators arises from the condition $\delta u=B_2$, which implies that under a one-form gauge transformation $B_2\rightarrow B_2+\delta\lambda_1$, $u$ transforms by $u\rightarrow u\rightarrow u+\lambda_1$. Therefore the one-form global symmetry (with $\lambda_1$ being a $\mathbb{Z}_2$ cocycle) transforms the deconfined $\mathbb{Z}_2$ Wilson line as
\begin{equation}\label{eqn:brokenhigherformacts}
\exp\left(\pi i\oint u\right)\;\longrightarrow\;\exp\left(\pi i\oint u\right)(-1)^{\oint \lambda_1}~.
\end{equation}

In this section we will show that by assuming the UV one-form symmetry is spontaneously broken and the $\mathbb{Z}_8$ 0-form symmetry is unbroken (as we argued above), one can derive the $\mathbb{Z}_2$ gauge theory (\ref{eqn:Z2dualitypropop}) for the corresponding Goldstone modes that matches the UV anomaly (\ref{eqn:anomalyZ2matching}).
We stress that the discussion here does not provide an argument that the one-form symmetry is broken if the $\mathbb{Z}_8$ 0-form symmetry is to be preserved.

We will discuss a more general setting with unbroken unitary 0-form symmetry and a collection of spontaneously broken higher-form symmetries with their background fields collectively denoted by $B$.
For simplicity, we will assume the higher-form symmetries do not participate in any higher-groups, and we will also assume they are finite groups\footnote{
For broken continuous higher-form symmetries the Goldstone mode is the photon (and its higher-form analogs) \cite{Gaiotto:2014kfa}.
}.
The 't Hooft anomaly corresponds to an SPT phase in the bulk, 
denoted by the partition function $Z^{\text SPT}_{M}[B]$ on the bulk manifold $M$.
For this argument we need to assume that the SPT phase with a trivial background for the broken higher-form symmetries is trivial, $Z_M^\text{SPT}[0]=1$. This is satisfied by the anomaly (\ref{eqn:anomalyZ2matching}).

On a manifold $M$ without a boundary, the SPT phase $Z^{\text SPT}_{M}[B]$ is invariant under a background gauge transformation of the higher-form symmetries $B\rightarrow B-\delta\lambda$ with cochains $\lambda$.
On the other hand, if the manifold $M$ has a boundary, the background gauge transformation gives a boundary variation by anomaly inflow:
\begin{equation}\label{eqn:boundaryvariation}
Z^{\text SPT}_{M}[B-\delta\lambda]=Z^{\text SPT}_{M}[B] Z_{\partial M}[B,\lambda]~.
\end{equation}
Equation (\ref{eqn:boundaryvariation}) is valid for any $\lambda$.

Next, we set $\lambda=u$ in (\ref{eqn:boundaryvariation}) with the constraint $\delta u=B$, and sum over $u$ on the boundary.
This means that $u$ is a dynamical field living on the boundary, which is consistent since $Z^\text{SPT}_M[0]$ has trivial bulk dependence.
We find that the following bulk-boundary system is well-defined:
\begin{equation}\label{eqn:Abelianboundarygaugetheory}
Z^{\text SPT}_{M}[B]\left( \sum_{u\text{ on }\partial M} Z_{\partial M}[B,u]\right)~.
\end{equation}
Namely, the boundary theory has the anomaly as described by the bulk SPT phase $Z^{\text SPT}_{M}[B]$.
In the absence of the background $B$, the constraint $\delta u=B$ implies that $u$ is a (higher-form) finite group Abelian gauge field that obeys the cocycle condition, and the higher-form symmetries transform the ``Wilson'' operator by $\oint u\rightarrow \oint u-\oint \lambda'$ with cocycles $\lambda'$ as in (\ref{eqn:brokenhigherformacts}).
Since the theory is deconfined, this implies that the higher-form symmetries are broken and $u$ corresponds to the Goldstone mode.
The boundary variation (\ref{eqn:Abelianboundarygaugetheory}) thus gives an Abelian (higher-form) gauge theory on the boundary $\partial M$ for the Goldstone mode $u$ that realizes the anomaly $Z^{\text SPT}_{M}[B]$\footnote{
See \cite{Kapustin:2014gua} for a similar discussion in $(3+1)d$ bulk where $B$ is replaced by a dynamical bulk gauge field.
See also \cite{Kobayashi:2019lep} for an alternative argument.
}. 
It is a TQFT that describes the spontaneously broken higher-form symmetries.
The construction is similar to the Wess-Zumino term that matches the 't Hooft anomaly of spontaneously broken continuous 0-form symmetries (see {\it e.g.} \cite{Wess:1971yu,Weinberg:1996kr}).

Let us return to the $SU(2)$ gauge theory with a massless adjoint Dirac fermion.
If the $\mathbb{Z}_8$ 0-form symmetry is not broken, and the UV $\mathbb{Z}_2$ one-form symmetry is broken, then the UV anomaly (\ref{eqn:anomalyZ2matching}) gives the action for the Goldstone mode by anomaly inflow as\footnote{
We used the identity ${\frak P}(X_2-\delta u)-{\frak P}(X_2)=\delta u\delta u-2\delta u X_2+\delta(X_2\cup_1\delta u)$.
}
\begin{equation}
\pi\int_{M_5}\left( \delta u w_3+\frac{1}{2}\left({\frak P}(X_2)-{\frak P}(X_2-\delta u)\right)X\right)
={\pi\over 2}\int_{M_4}\left( u\delta u +X_2\cup_1 \delta u\right)X+\pi\int_{M_4} uw_3~,
\end{equation}
where $M_4=\partial M_5$ is the spacetime manifold. This reproduces the $\mathbb{Z}_2$ gauge theory (\ref{eqn:Z2dualitypropop}).

\section{$U(1)$ gauge theory with time-reversal symmetry}\label{sec:U(1)timereversal}

In this section we first review the $U(1)\times U(1)$ one-form global symmetry in the pure $U(1)$ gauge theory \cite{Gaiotto:2014kfa}. Then we discuss different ways to couple the theory to the background of time-reversal symmetry (\ref{eqn:Tparticle}) 
(more precisely, the time-reversing Lorentz symmetry $O(4)$)\footnote{
We thank Anton Kapustin for suggesting this problem.}.

Since the two-form symmetry is trivial, the different couplings are classified by
\begin{equation}
H^2_{\cal T}(BO(4),U(1)\times U(1))~.
\end{equation}
The results are summarized in table \ref{tab:sixphases} for $\theta=0$ mod $2\pi$ and table \ref{tab:onephase} for $\theta=\pi$ mod $2\pi$, and they reproduce the classification in \cite{Wang:2016cto}.

The symmetry-enriched phases are combinations of the couplings and the SPT phases.
We find in total 22 non-anomalous symemtry-enriched phases for the time-reversal symmetric $U(1)$ gauge theory, in agreement with \cite{Wang:2016cto}.

In appendix \ref{sec:appendixUoneSO3} we will discuss the symmetry-enriched phases with additional unitary $SO(3)$ symmetry,  reproducing the classification in \cite{Zou:2017ppq}.

\subsection{One-form symmetry}

\subsubsection{$\theta=0$}

The $U(1)$ gauge theory in 4$d$ has $U(1)\times U(1)$ one-form global symmetry \cite{Gaiotto:2014kfa}, with currents
\begin{equation}
j^E = F,\quad j^M=\star F/2\pi,\qquad d\star j^E=d\star j^M=0~.
\end{equation}
The charged objects are the line operators, with electric and magnetic charges given by
\begin{equation}\label{eqn:generatoroneformuone}
q_\textbf{e} =\oint_{S^2}\star j^E,\qquad q_\textbf{m}=\oint_{S^2}\star j^M~,
\end{equation}
where $S^2$ surrounds the codimension-three line operators. They are the surface operators that generate the one-form symmetries \cite{Gaiotto:2014kfa}\footnote{
For a discussion of surface operators in gauge theories with continuous gauge groups, see \cite{Gukov:2006jk,Gukov:2008sn,Gukov:2014gja}.
}.
If we turn on background gauge field $B^E$ for the electric one-form symmetry, the quantization of the $U(1)$ gauge field strength $F$ is modified,
\begin{equation}\label{eqn:BE}
\oint \frac{F}{2\pi} = \oint \frac{B^E}{2\pi}\quad \text{mod }\mathbb{Z}~.
\end{equation}
This can also be imposed on-shell by adding to the action $B^E\star j^E$, but we will simply impose the modified quantization condition (\ref{eqn:BE}).

If we turn on a background gauge field $B^M$ for the magnetic one-form symmetry, the action has the additional coupling
\begin{equation}\label{eqn:BM}
\int B^M\star j^M = \int B^M\frac{F}{2\pi}~.
\end{equation}

Since the line operators are charged under the one-form symmetries, in the presence of the backgrounds $B^E,B^M$, the line operators are attached to additional surfaces.
For the line operator with electric and magnetic charges $(q_\textbf{e},q_\textbf{m})$ the additional surface is
\begin{equation}\label{eqn:surfacetheta0}
\int_\Sigma \left(q_\textbf{e}B^E+q_\textbf{m}B^M\right) ~,
\end{equation}
where $\partial \Sigma$ is the locus of the line operator.
Due to the angular momentum of the electromagnetic field, the dyon $(q_\textbf{e},q_\textbf{m})$ is a fermion for odd $q_\textbf{e}q_\textbf{m}$ if both $(q_\textbf{e},0)$ and $(0,q_\textbf{m})$ are bosons \cite{Goldhaber:1976dp}, and thus in the absence of the background $B^E,B^M$ the dyon $(q_\textbf{e},q_\textbf{m})$ is attached to the intrinsic surface
\begin{equation}\label{eqn:surfaceptheta0}
q_\textbf{e}q_\textbf{m}\pi\int_\Sigma w_2~.
\end{equation}
This means that the multiplication of projective representations of the Lorentz symmetry on the particles is not compatible with the fusion rules, which is analogous to the $(2+1)d$ symmetry-enriched phases with non-trivial $\kappa$ discussed in \cite{Barkeshli:2014cna}.

Thus the total surface is the sum of (\ref{eqn:surfacetheta0}) and (\ref{eqn:surfaceptheta0}).

Moreover, the one-form symmetry has a mixed anomaly. If we turn on both backgrounds $B^E$ and $B^M$, then the coupling (\ref{eqn:BM}) is not well-defined but has an anomaly described by the 5$d$ SPT
\begin{equation}\label{eqn:anom}
\int \frac{F}{2\pi}dB^M=\int \frac{B^E}{2\pi} dB^M~,
\end{equation}
where we used $\oint dB^M\in 2\pi\mathbb{Z}$.

The time-reversal symmetry (\ref{eqn:Tparticle}) does not commute with the electric one-form symmetry but commutes with the magnetic one-form symmetry. Thus it transforms the backgrounds by
\begin{equation}\label{eqn:Tbackground}
{\cal T}(B^E)=-B^E,\qquad {\cal T}(B^M)=B^M~.
\end{equation}
Note since time-reversal is an anti-unitary symmetry, there is a sign change comparing to the transformation of the charges.

\subsubsection{Non-zero $\theta$}

When we turn on a non-zero $\theta$ angle, the electric and magnetic charges are shifted due to the Witten effect \cite{Witten:1979ey}:
\begin{equation}
(q_\textbf{e},q_\textbf{m})=(n+\frac{\theta}{2\pi}m,m),\quad n,m\in\mathbb{Z}~.
\end{equation}
On the other hand, the surfaces that attach to the lines are not affected by tuning $\theta$.
We work in the basis where the $U(1)\times U(1)$ one-form symmetry couples to the quantized quantum number $(n,m)\in\mathbb{Z}\times\mathbb{Z}$.

Thus in the presence of the background $B^E,B^M$ the line $(q_\textbf{e},q_\textbf{m})$ with $q_\textbf{e}=n+\frac{\theta}{2\pi}m$, $q_\textbf{m}=m$ for integers $(n,m)$ is attached with the surface
\begin{equation}\label{eqn:surfacetheta}
(q_\textbf{e}-\frac{\theta}{2\pi}q_\textbf{m})\int_\Sigma B^E+q_\textbf{m}\int_\Sigma B^M+ (q_\textbf{e}-\frac{\theta}{2\pi}q_\textbf{m})q_\textbf{m}\pi\int_\Sigma w_2~,
\end{equation}
where the last term is from the angular momentum of the electromagnetism \cite{Wilczek:1981dr,Goldhaber:1988iw}.

For example, in the absence of the backgrounds $B^E,B^M=0$, a monopole $(q_\textbf{e},q_\textbf{m})=(0,1)$ is attached to the trivial surface at $\theta=0$, the surface $\pi\int w_2$ at $\theta=2\pi$, and again the trivial surface at $\theta=4\pi$. This is consistent with the statistics of the monopole with different $\theta$ angles \cite{Metlitski:2013uqa}.

\subsection{Coupling by one-form symmetry}\label{sec:bacgrounds}

The different phases can be described by different ways for the $U(1)$ gauge theory to couple to the background field of the Lorentz symmetry (or to define on a spacetime manifold).
This can be described by activating different backgrounds for the $U(1)\times U(1)$ one-form symmetry.

First, since shifting $\theta\rightarrow \theta+2\pi$ is equivalent to adding a magnetic background,
\begin{equation}
\theta=2\pi:\quad
\pi\int \frac{F}{2\pi}\frac{F}{2\pi}=\pi\int\frac{F}{2\pi}(w_2+w_1^2)=\int \frac{F}{2\pi}\Delta B^M,\quad \Delta B^M=\pi(w_2+w_1^2)~,
\end{equation}
it is thus sufficient to consider only the case $\theta=0$ or $\theta=\pi$.
If we turn on backgrounds for the electric and magnetic one-form symmetries $B^E,B^M$, one can show the $\theta=2\pi$ term is equivalent to $\theta=0$ with
\begin{equation}\label{eqn:shiftBMthetapi}
\Delta B^M = B^E+\pi w_2
\end{equation}
up to classical actions. The first term in the change of $B^M$ can be explained as from the shift in the spectrum of the line operators, while the second term comes from the angular momentum of the electromagnetic field.

\subsubsection{$\theta=0$}

The time-reversal symmetry acts on the one-form symmetry by (\ref{eqn:Tbackground}), which in Euclidean signature is
\begin{equation}
\tau:\quad (B^E,B^M)\quad\longrightarrow\quad (B'^E=-B^E,B'^M=B^M)~.
\end{equation}
Different couplings to the Lorentz symmetry correspond to different backgrounds for the ${\cal A}=U(1)_E\times U(1)_M$ electric and magnetic one-form symmetries, and they are classified by
\begin{equation}
H^2_\tau(BO(4),U(1)_E\times U(1)_M)=\mathbb{Z}_2^2\times\mathbb{Z}_2~.
\end{equation}
They correspond to the backgrounds
\begin{equation}
B^E=0,\pi w_2,\pi w_1^2,\pi (w_2+w_1^2),\quad B^M=0,\pi w_2~.
\end{equation}
The line operator with electric and magnetic charges $(q_\textbf{e},q_\textbf{m})$ is attached to the surface
\begin{equation}
\int \left( q_\textbf{e}B^E+q_\textbf{m}B^M +q_\textbf{e}q_\textbf{m}\pi\int w_2\right)~.
\end{equation}
When the surface is $\pi \int w_2$, the particle is in projective representation of the Lorentz symmetry and thus it is a fermion. 
It also satisfies $R^2=1$ for the Euclidean reflection $R$, and thus ${\cal T}^2=-1$ by Wick rotation. Namely, the particle is a Kramer doublet.
When the surface is $\pi\int w_1^2$, the particle is in the projective representation ${\cal T}^2=-1$ of the time-reversal symmetry. When the surface is $\pi\int (w_2+w_1^2)$, the particle is a fermion and satisfies ${\cal T}^2=1$.

Note $B^M=\pi w_1^2$ does not represent a non-trivial cocycle in $H^2_\tau(BO(4),U(1)_E\times U(1)_M)$.
One way to see this is that the time-reversal symmetry is not a symmetry of the wordline for particles with magnetic charges (\ref{eqn:Tparticle}), and thus we cannot interpret a surface $\pi\int w_1^2$ attached to the line with magnetic charges as the particle being in the projective representation of the time-reversal symmetry.
Thus we only need to consider $B^M=0,\pi w_2$.
Since the basic magnetic line operator is attached to the surface $\int B^M=0,\pi\int w_2$, this means that the monopole particle is a boson or fermion, depending on the choice. This is consistent with \cite{Wang:2016cto}.

Another way to show $B^M=\pi w_1^2$ does not represent a non-trivial coupling is that for $U(1)_\text{gauge}\rtimes \mathbb{Z}_2^{\cal T}$ the following coupling is a decoupled classical term 
\begin{equation}
\pi \int w_1^2\frac{F}{2\pi}=\pi \int w_1^2\left(\frac{F}{2\pi}-\frac{B^E}{2\pi}\right)+\pi\int w_1^2 {B^E\over 2\pi}~.
\end{equation}
To see this, note that for $B^E=0$ the coupling $\pi\int \frac{F}{2\pi}w_1^2$ is trivial (see {\it e.g.} \cite{Kapustin:2014gma}).
Now, different gauge bundles that satisfy (\ref{eqn:BE}) for $B^E=\pi w_2,\pi w_1^2,\pi(w_2,w_1^2)$ differ by twisted complex line bundles. Denote the integral lift of $B^E/\pi$ to be $F_c/ (2\pi)$, which is a classical field and by definition satisfies (\ref{eqn:BE}). Then $F-F_c$ is the field strength of a twisted complex line bundle ({\it i.e.} satisfies (\ref{eqn:BE}) with $B^E=0$), and thus
\begin{equation}\label{eqn:trivialcouplingw1square}
\int\frac{F}{2\pi}w_1^2 = \int \frac{F_c}{2\pi} w_1^2 + \int \frac{F-F_c}{2\pi}w_1^2=\int \frac{F_c}{2\pi} w_1^2\quad \text{mod }2\pi\mathbb{Z}~,
\end{equation}
which is a classical term that depends only on the background fields.

The possible electric background $B^E$ needs to be such that there is no anomaly (\ref{eqn:anom}). This leaves six choices as in table \ref{tab:sixphases}:
\begin{align}
&B^M=0,\quad B^E=0,\, \pi w_1^2,\, \pi w_2,\, \pi (w_2+w_1^2)\cr
&B^M=\pi w_2,\quad B^E=0,\pi w_1^2~.
\end{align}

Due to the modification of the quantization (\ref{eqn:BE}), the backgrounds $B^E=\pi w_2,\pi (w_2+w_1^2)$ modify the symmetry respectively into
\begin{equation}
\frac{Pin^+(4)_\text{global}\ltimes U(1)_\text{gauge}}{\mathbb{Z}_2},\qquad \frac{Pin^-(4)_\text{global}\ltimes U(1)_\text{gauge}}{\mathbb{Z}_2}~.
\end{equation}

It might be puzzling that while we start with a general unorientable four-manifold, the different couplings seem to require additional structures on the spacetime manifold\footnote{
Not every four-manifold admits a pin$^c$ or pin$^{\tilde c\pm}$ structure. For instance, $\mathbb{R}\mathbb{P}^2\times\mathbb{R}\mathbb{P}^2$ does not have a pin$^c$ or pin$^{\tilde c+}$ structure, while $\mathbb{R}\mathbb{P}^4$ does not have a  pin$^{\tilde c-}$ structure.
}.
The resolution is that these structures in general do not extend to the worldvolume of operator insertions. 
For instance, the background $B^E=\pi w_2$ inserts the generator $\int\star F$ at the Poincar\'e dual of $w_2$ \cite{Halperin:1972}, and if $w_2$ cannot be lifted to an integral class, the pin$^{\tilde c+}$ structure from the modification of quantization (\ref{eqn:BE}) does not extend to the worldvolume of the operator insertion.
The theory also does not depend on the choice of this structure, since different choices can be absorbed into the dynamical gauge field that is summed over in the path integral. 

So far we have only discussed non-anomalous couplings. The anomalous couplings are $B^E=\pi w_2,\pi (w_2+w_1^2)$, $B^M=\pi w_2$, 
and they have the anomaly $\pi\int w_2w_3$ from (\ref{eqn:anom}). The theories with these couplings is known to be anomalous \cite{Wang:2013zja,Thorngren:2014pza,Kravec:2014aza}, and they are called all-fermion electrodynamics, since the basic electric and magnetic particles are fermions. We will focus on the non-anomalous couplings.

\subsubsection{$\theta=\pi$}

Consider the time-reversal symmetry
\begin{equation}\label{eqn:transformthetapi}
\theta=\pi,\,(q_\textbf{e},q_\textbf{m})\quad\longrightarrow\quad \theta=\pi,\,(q_\textbf{e},-q_\textbf{m})~.
\end{equation}
For instance, the dyons $(\frac{1}{2},\pm 1)$ are related by time-reversal \cite{Metlitski:2015yqa,Wang:2016cto}.
Since the time-reversal symmetry changes $\theta$ to $-\theta$, to compensate for the change in $\theta$ the time-reversal symmetry acts differently on the one-form symmetry compared to how it acts when $\theta=0$.
To see this, note in the basis $(n,m)$ where $(q_\textbf{e},q_\textbf{m})=(n+\frac{\theta}{2\pi}m,m)$, the transformation takes $(n,m)$ to $(n',m')$ with
\begin{equation}
n'+\frac{\pi}{2\pi}m'=n+\frac{\pi}{2\pi}m,\quad m'=-m\quad \Rightarrow\quad n=n'+m',\quad m=-m'~.
\end{equation}
Since the background for one-form symmetry couples as $n\int B^E+m\int B^M$ (see (\ref{eqn:surfacetheta})), the background transforms as (in Euclidean signature, and up to possible gravitational terms)
\begin{equation}\label{eqn:transformoneformthetapi}
\tau:\quad (B^E,B^M)\quad \longrightarrow\quad (B'^E,B'^M)=(-B^E,B^M-B^E)~.
\end{equation}
It is easy to check that the symmetry has order two.
The possible couplings to the Lorentz symmetry are classified by
\begin{equation}
H^2_\tau(BO(4),U(1)_E\times U(1)_M)=\mathbb{Z}_2~,
\end{equation}
where the two elements correspond to $B^M=0,\pi w_2$.
On the other hand, the background $B^E$ is the same for all couplings. In the following we will determine its value.

The possible candidates for $B^E$ are $B^E=0,\pi w_1^2,\pi w_2,\pi (w_2+w_1^2)$. The relevant symmetries for these couplings are $Pin^{{\tilde c}+}(4)$, $Pin^{{\tilde c}-}(4)$ and (omitting the $SO(4)$ Lorentz group) $U(1)\rtimes \mathbb{Z}_2^{\cal T}$, $Pin^-(2)$.
We are looking for the SPT phase that has time-reversal symmetric $\theta=\pi$ term for the $U(1)$ gauge field when reduced on an orientable manifold, and it requires the $U(1)$ gauge field to be a spin$^c$ connection \cite{Seiberg:2016rsg}\footnote{
To see this, note $\theta=\pi$ for an ordinary $U(1)$ gauge field on an orientable manifold changes by $\theta=2\pi$ under time-reversal, which can be expressed as
\begin{equation}
S_{\theta=\pi}\longrightarrow S_{\theta=-\pi}=S_{\theta=\pi}+\pi\int \frac{F}{2\pi}w_2~.
\end{equation}
where $F$ is the field strength for ordinary $U(1)$ gauge field. The time-reversal symmetry is violated by the mixed term $\int F w_2/2$.
This mixed term can be removed (up to a gravitational term) by completing the square using ${F'\over 2\pi}={F\over 2\pi}+\frac{1}{2}\tilde w_2$ where $\tilde w_2$ is the lift of $w_2$ to an integral cocycle (this is possible for every orientable four-manifold). Thus $F'$ is the field strength of a spin$^c$ connection. One can further show the extra gravitational term in the transformaton can be cancelled by adding $\pi\int{\hat A}(R)$ \cite{Metlitski:2015yqa,Seiberg:2016rsg}.
}. 
This rules out the last two cases.
From the classification of SPT phases of $Pin^{{\tilde c}\pm}$ symmetry (see {\it e.g.} \cite{Wang:2014lca,Freed:2016rqq,Guo:2017xex}) we find that the $\theta=\pi$ term for the $U(1)$ gauge field can be lifted to an unorientable manifold only for
\begin{equation}\label{eqn:backgroundpiE}
B^E=\pi w_2~.
\end{equation}
More precisely, the $\theta=\pi$ term is given by $(-1)^{N_0}$ where $N_0$ is the mod 2 index of the Dirac operator with $U(1)$ charge one (see {\it e.g. }\cite{Metlitski:2015yqa}).
The background (\ref{eqn:backgroundpiE}) implies that the spacetime manifold in the Euclidean signature has a pin$^{\tilde c+}$ structure. Namely, the gauge and global symmetries combine into
\begin{equation}
Pin^{{\tilde c}+}(4)={Pin^+(4)_\text{global}\ltimes U(1)_\text{gauge}\over \mathbb{Z}_2}~, 
\end{equation}
where $(-1)^F$ is identified with the $\mathbb{Z}_2$ center of $U(1)$, and the Euclidean reflection does not commute with $U(1)$.

Another way to see the result is that the transformation (\ref{eqn:transformthetapi}) on the surface (\ref{eqn:surfacetheta}) implies that the transformation of backgrounds (\ref{eqn:transformoneformthetapi}) has the gravitational correction $B'^M=B^M-B^E+\pi w_2$ from the extra term $nm\pi\int w_2=n'm'\pi\int w_2+m'\pi\int w_2$ due to the angular momentum of the electromagnetic field. This is consistent with $B^E=\pi w_2$ enforced by time-reversal. (Note however, it cannot distinguish between $B^E=\pi w_2$ and $B^E=\pi (w_2+w_1^2)$ since a dyon is not time-reversal invariant).

\begin{table}[t]\centering
\begin{tabular}{|cc|c|}
\hline
Electric & Magnetic & Backgrounds\\ \hline
boson, ${\cal T}^2=1$ & boson & $B^E=B^M=0$\\ 
boson, ${\cal T}^2=-1$ & boson & $B^E=\pi w_1^2$, $B^M=0$\\ 
fermion, ${\cal T}^2=1$ & boson & $B^E=\pi (w_2+w_1^2)$, $B^M=0$\\ 
fermion, ${\cal T}^2=-1$ & boson & $B^E=\pi w_2$, $B^M=0$\\ 
boson, ${\cal T}^2=1$ & fermion & $B^E=0$, $B^M=\pi w_2$\\ 
boson, ${\cal T}^2=-1$ & fermion & $B^E=\pi w_1^2$, $B^M=\pi w_2$
\\ \hline
\end{tabular}
\caption{Six non-anomalous couplings of $U(1)$ gauge theory at $\theta=0$ mod $2\pi$ to the backgrounds of time-reversing Lorentz symmetry. The electric particle has electric and magnetic charges $(1,0)$, while the magnetic particle has charges $(0,1)$. The statistics and Kramer degeneracy of the particles are according to \cite{Wang:2016cto}. In this note we show the different couplings, or symmetry-enriched phases, correspond to different backgrounds for the electric and magnetic one-form symmetry as explained in section \ref{sec:bacgrounds}.}\label{tab:sixphases}
\end{table}

\begin{table}[t]\centering
\begin{tabular}{|cc|c|}
\hline
Electric & Magnetic & Backgrounds\\ \hline
fermion, ${\cal T}^2=-1$ & fermion & $B^E=\pi w_2$, $B^M=0$
\\\hline
\end{tabular}
\caption{One non-anomalous coupling of $U(1)$ gauge theory at $\theta=\pi$ mod $2\pi$ to the background of time-reversing Lorentz symmetry. The electric particle has charge $(1,0)$ (that corresponds to $(1,0)$ at $\theta=0$), while the magnetic particle has charge $(0,2)$ (that corresponds to the dyon $(-1,2)$ at $\theta=0$). 
We pick the basis of the one-form symmetry such that the backgrounds $B^E,B^M$ are properly quantized and couple to integer quantum numbers.
The statistics and Kramer degeneracy of the particles are according to \cite{Wang:2016cto}. In this note we show that the different symmetry-enriched phases correspond to different backgrounds for the electric and magnetic one-form symmetries.
}\label{tab:onephase}
\end{table}

Among the two couplings $B^M=0$ and $B^M=\pi w_2$, only the coupling $B^M=0$ has no anomaly (\ref{eqn:anom}).
In this non-anomalous phase the dyons $(q_\textbf{e},q_\textbf{m})=(\frac{1}{2},\pm 1)$ exchanged by time-reversal are attached to the trivial surface according to (\ref{eqn:surfacetheta}),
\begin{equation}
(\frac{1}{2} - \frac{\pi}{2\pi}(\pm 1))\int_\Sigma B^E + 0 + (\frac{1}{2} - \frac{\pi}{2\pi}(\pm 1))(\pm 1)\pi \int_\Sigma  w_2=0\text{ mod }2\pi\mathbb{Z}~,
\end{equation}
where we used $B^E=\pi w_2$.
Thus the two dyons are bosons. 
If $B^M=\pi w_2$, then the dyon would be a fermion, but this would result in an anomaly from the mixed anomaly (\ref{eqn:anom}) of the one-form symmetry. This is consistent with the conclusion in \cite{Wang:2016cto}.

The pure monopoles with $q_\textbf{e}=0=n+\frac{\theta}{2\pi}m$ have even magnetic charge $q_\textbf{m}=m$.
The basic magnetic monopole $(q_\textbf{e},q_\textbf{m})=(0,2)$ is attached to the following surface from (\ref{eqn:surfacetheta}),
\begin{equation}
(0- \frac{\pi}{2\pi}\cdot 2)\int_\Sigma B^E + 0 + (0- \frac{\pi}{2\pi}\cdot 2)\cdot 2\cdot\pi \int_\Sigma  w_2=-\int_\Sigma B^E=\pi\int_\Sigma w_2\text{ mod }2\pi\mathbb{Z}~.
\end{equation}
And thus the minimal magnetic monopole is a fermion. This is consistent with \cite{Wang:2016cto}.

Similarly, the basic electric particle $(q_\textbf{e},q_\textbf{m})=(1,0)$ is attached to the surface
\begin{equation}
(1- \frac{\pi}{2\pi}\cdot 0)\int_\Sigma B^E + 0 + (1- \frac{\pi}{2\pi}\cdot 0)\cdot 0\cdot\pi \int_\Sigma  w_2=\int_\Sigma B^E=\pi\int_\Sigma w_2\text{ mod }2\pi\mathbb{Z}~.
\end{equation}
Thus the minimal electric particle -- which is invariant under time-reversal symmetry-- is a fermion and a Kramer doublet. (Note this means it satisfies $R^2=1$ for Euclidean reflection $R$). This again agrees with \cite{Wang:2016cto}.

\subsection{Combined with SPT phases}\label{sec:U(1)SPTtotal}

Consider adding additional classical counterterm to the theory. Such counterterms correspond to the symmetry-protected topological phases for the time-reversal symmetry. There are 4 candidate SPT phases \cite{Kapustin:2014tfa}, generated by 
\begin{equation}\label{eqn:SPTU(1)}
\pi\int w_2^2,\quad \pi\int w_1^4~.
\end{equation}

In the symmetry-enriched phases with the couplings that activate the backgrounds
\begin{equation}\label{eqn:BEspecial}
B^E=\pi w_1^2,\pi (w_2+w_1^2)~,
\end{equation}
the following generator for the SPT phases (\ref{eqn:SPTU(1)}) becomes trivial
\begin{equation}
\pi\int w_1^4=\pi\int \frac{2F}{2\pi} w_1^2\in 2\pi\mathbb{Z}~,
\end{equation}
where we used (\ref{eqn:BE}) and that $2F$ is a properly quantized $U(1)$ gauge field (since (\ref{eqn:BEspecial}) takes value in the $\mathbb{Z}_2\subset U(1)$ subgroup). Thus there are only two SPT phases from the remaining non-trivial generator $\pi\int w_2^2$.
This is consistent with \cite{Wang:2016cto}, and the classification of the SPT phases for $Pin^{\tilde{c}\pm}$ symmetry (see {\it e.g.}\cite{Freed:2016rqq,Guo:2017xex}). 

Another way to understand the trivialization of the SPT phases due the coupling (\ref{eqn:BEspecial}) is from the mixed anomaly for the $U(1)\times U(1)$ one-form symmetry. In the presence of the background $B^E=\pi (w_2+w_1^2),\pi w_1^2$, the $U(1)$ magnetic one-form symmetry with parameter $\pi \tilde w_1$ (where tilde denotes a lift to integral one-cochain) produces the anomalous shift
\begin{equation}
\frac{1}{2\pi}\int B^E \pi \delta \tilde w_1 =\pi \int (w_2+w_1^2)w_1^2=\pi \int w_1^2 w_1^2=\pi\int w_1^4\quad \text{mod }2\pi\mathbb{Z}~.
\end{equation}
Thus this SPT can be absorbed by a magnetic one-form global symmetry transformation. Concretely, the SPT phase is equivalent to depositing the worldline SPT phase $\pi\oint w_1$ on the line operators charged under the $\mathbb{Z}_2\subset U(1)$ magnetic one-form symmetry.
This is an example of the trivialization of SPT phases by the dynamics discussed in section \ref{sec:wSPTgeneral}.

Therefore for $\theta=0$ there are $3\times 2+3\times 4=18$ non-anomalous phases, and together with the 4 non-anomalous phases with $\theta=\pi$ there are $22$ non-anomalous symmetry-enriched phases for $U(1)$ gauge theory with time-reversal symmetry. This reproduces the result in \cite{Wang:2016cto}.

We remark that for $\theta=\pi$ the theory necessarily has a gravitational term, no matter how different SPT phases (\ref{eqn:SPTU(1)}) are added.
To see this, note that $(-1)^{N_0}$ reduces to the following term when the manifold is orientable \cite{Atiyah:1963zz}
\begin{equation}
(-1)^{N_0}=\exp\left(\frac{\pi i}{2}\int \frac{F}{2\pi}\wedge\frac{F}{2\pi}+ \frac{i}{192\pi} \int \text{Tr }R\wedge R\right)~,
\end{equation}
where $R$ is the curvature 2-form, and the last term cannot be cancelled by the SPT phases (\ref{eqn:SPTU(1)}).

\section{SPT phases from gauging higher-form symmetries}
\label{sec:appendixSPT}

In this section we discuss obtaining SPT phases from gauging a higher-form symmetry in symmetry-enriched TQFT. In particular, we will show that the the Gu-Wen fermionic SPT phase \cite{Gu:2012ib} with $G$ symmetry in $(3+1)d$ can be obtained from gauging a two-form symmetry in bosonic $\mathbb{Z}_2$ gauge theory with fermionic particle and loop excitation with symmetry fractionalization $\rho_3\in H^3(G,\mathbb{Z}_2)$.

We can couple the theory to both the 0-form symmetry background $X$ and the two-form symmetry background $B_3$ by the shift 
\begin{equation}\label{eqn:shiftbg}
B_3\rightarrow B_3+X^*\nu_3~.
\end{equation}
If we turn off $B_3=0$ after the shift, then this is equivalent to coupling the theory to the background $X$ by a fixed two-form symmetry background (\ref{eqn:4dbackgrounds}).

Since the two-form symmetry does not have an anomaly on its own, we can gauge the two-form symmetry with a dynamical gauge field $b_3$.
In this appendix we will study how the theory obtained from gauging the two-form symmetry depends on the coupling $\nu_3$.

For simplicity, we consider the example of $\mathbb{Z}_2$ two-form symmetry.
We will also assume there is no three-group symmetry, $\delta B_3=0$. Thus $\nu_3\in H^3(BG,\mathbb{Z}_2)$.

There are two cases depending on whether the two-form symmetry has a mixed anomaly with other symmetry, which by
is $\int_{5d} B_2B_3$ with $B_2=X^*\eta_2$.

If the two-form symmetry does not have a mixed anomaly with the 0-form symmetry (trivial $\eta_2$), 
then $X^*\nu_3$ can be absorbed into the dynamical gauge field $b_3$ that is summed over in the path integral. 
Thus the resulting theory does not depend on $\nu_3$.

If the two-form symmetry has a mixed anomaly with the 0-form symmetry (non-trivial $\eta_2$), then the theory has the following bulk dependence
\begin{equation}\label{eqn:gauginganom}
\pi\int_{5d} (b_3+X^*\nu_3)X^*\eta_2~.
\end{equation}
To cancel the dependence of the dynamical field $b_3$ in the bulk, the 
background $Y$ of the dual $\mathbb{Z}_2$ 0-form symmetry that couples as $\pi\int b_3Y$ must satisfy
\begin{equation}\label{eqn:gaugingdualzero}
\delta Y= X^*\eta_2~.
\end{equation}
This implies that after gauging the two-form symmetry, the 0-form symmetry is extended by the dual $\mathbb{Z}_2$ 0-form symmetry to be a group extension determined by the cocycle $\eta_2$.

The constraint (\ref{eqn:gaugingdualzero}) also implies that the 't Hooft anomaly of the 0-form symmetry for different $\nu_3$ is trivialized.
This gives rise to different SPT phases of the 0-form symmetry.
The different SPT phases are described by $\nu_4\in C^4(BG,\mathbb{R}/\mathbb{Z})$ that cancels the anomaly of 0-form symmetry in (\ref{eqn:gauginganom})
\begin{equation}\label{eqn:gaugingnewSPT}
2\pi\int_{5d} X^*\delta \nu_4=\pi\int_{5d} X^* \nu_3\eta_2~.
\end{equation}

\subsection{Example: $\mathbb{Z}_2$ gauge theory bosonic shadow for Gu-Wen phase}

The $\mathbb{Z}_2$ gauge theory in $(3+1)d$ has one-form and two-form symmetry with a mixed anomaly described by $\pi\int_{5d} B_2B_3$.
Consider the symmetry-enriched phase by coupling the the theory to the background $X$ of the bosonic Lorentz symmetry (without time-reversal symmetry) and unitary 0-form symmetry $G$ using the coupling $\xi=0$, $\eta_2=w_2$ and $\nu_3=\rho_3\in H^3(BG,\mathbb{Z}_2)$.
Namely, we turn on the background gauge field
\begin{equation}
B_2=X^* w_2,\quad B_3=X^*\rho_3~.
\end{equation}
As discussed in Section \ref{sec:classification}, this means that the Wilson line in the $\mathbb{Z}_2$ gauge theory is a fermion and the magnetic string carries worldvolume 't Hooft anomaly of the $G$ symmetry specified by $\rho_3$.

We can gauge the two-form symmetry by shifting
\begin{equation}
    B_3=b_3+X^*\rho_3
\end{equation}
and summing over $b_3$.
Then after gauging the $\mathbb{Z}_2$ two-form symmetry, (\ref{eqn:gaugingdualzero}) implies that the Lorentz symmetry is extended to be $Spin(4)$ so the new theory depends on the spin structure, and the dual $\mathbb{Z}_2$ 0-form symmetry is identified with the $\mathbb{Z}_2$ fermion parity.
Thus gauging the $\mathbb{Z}_2$ two-form symmetry gives a fermionic SPT phase with internal $G$ symmetry that depends on $\rho_3$ 
by $\nu_4\in C^4(BG,\mathbb{R}\mathbb{Z})$ in (\ref{eqn:gaugingnewSPT}) and satisfies
\begin{equation}
\delta \nu_4=\frac{1}{2} Sq^2(\rho_3)\quad\text{mod }\mathbb{Z}~,
\end{equation}
where we used $\pi\int_{5d}X^*\rho_3w_2=\pi\int_{5d}X^*Sq^2(\rho_3)$ on orientable manifolds. 
This example is discussed in \cite{Kapustin:2014dxa,Kapustin:2017jrc}, and it reproduces the fermionic SPT phases studied by Gu and Wen in \cite{Gu:2012ib} using supercohomology.

\section{More Examples}\label{sec:moreexamples}

\subsection{$U(1)$ gauge theory with $\mathbb{Z}_2$ unitary symmetry}\label{sec:U(1)BC}

In this example we will discuss $U(1)$ gauge theory couples to the background of unitary $\mathbb{Z}_2$ 0-form symmetry.
Without the restriction of time-reversal symmetry, the $\theta$ parameter can be continuously tuned to zero, and thus we only need to consider the case $\theta=0$.

If the $\mathbb{Z}_2$ 0-form symmetry does not permute the line operators, then there is only one coupling, corresponding to $B^E=B^M=0$:
\begin{equation}
H^2(B\mathbb{Z}_2,U(1)\times U(1))=0~.
\end{equation}
Thus we will focus on the 0-form symmetry that permutes the line operators.

The theory has intrinsic unitary $\mathbb{Z}_2$ charge conjugation 0-form symmetry
\begin{equation}\label{eqn:Uonechargeconj}
{\cal C}:\quad (q_\textbf{e},q_\textbf{m})\longrightarrow (-q_\textbf{e},-q_\textbf{m})~.
\end{equation}
The gauge coupling can be continuously tuned to the self-dual point with $\theta=0$, where the electromagnetic duality ({\it e.g.} \cite{Montonen:1977sn,Witten:1995gf}) becomes a $\mathbb{Z}_4$ unitary 0-form symmetry
\begin{equation}\label{eqn:EMdualityperm}
{\cal D}:\quad (q_\textbf{e},q_\textbf{m})\longrightarrow (-q_\textbf{m},q_\textbf{e})~.
\end{equation}
The theory at $\theta=0$ also has the time-reversal symmetry ${\cal T}$ in (\ref{eqn:Tparticle}).
The symmetries satisfy the following algebraic relations:\footnote{
The electromagnetic duality also maps ${\cal T}$ to ${\cal CT}$ at $\theta=\pi$ \cite{Metlitski:2015yqa}, and thus the set of symmetry-enriched phases of $U(1)$ gauge theory with symmetry ${\cal CT}$ is equivalent to the set of symmetry-enriched phases with symmetry ${\cal T}$ discussed in section \ref{sec:U(1)timereversal} (see also \cite{Wang:2016cto}).
}
\begin{equation}
{\cal D}^2={\cal C},\quad
{\cal CD}={\cal DC},\quad {\cal CT}={\cal TC},\quad {\cal D}^{-1}{\cal T}{\cal D}={\cal CT}~.
\end{equation}

In this section we will focus on the unitary $\mathbb{Z}_2$ 0-form symmetry that acts by charge conjugation.

\subsubsection{The charge conjugation symmetry}

The charge conjugation symmetry (\ref{eqn:Uonechargeconj}) acts on the background gauge field of the one-form symmetry by
\begin{equation}\label{eqn:chargecongU(1)}
{\cal C}(B^E,B^M)=(-B^E,-B^M)~.
\end{equation}

We will focus on the $\mathbb{Z}_2\times \mathbb{Z}_2\subset U(1)\times U(1)$ subgroup one-form symmetry, with backgrounds $B_e,B_m$. They correspond to the special case $B^E=\pi B_e,B^M=\pi B_m$, where we normalize $\oint B_e,\oint B_m=0,1$ mod 2.
If we make $B_m$ dynamical, it forces the $U(1)\cong SO(2)$ gauge bundle to have even first Chern numbers, and the gauge group becomes its double cover, which is still isomorphic to $SO(2)$. 
The $\pi$-rotation in the center of the original gauge group corresponds to a $\pi/2$-rotation in its double covering, and thus the order of this element changes from 2 to 4 in the covering. On the other hand, the $\pi/2$-rotation does not commute with the charge conjugation, as it differs from a $(-\pi/2)$-rotation by a non-trivial $\pi$-rotation.
These observations imply the following mixed anomaly in the presence of background $B^{\cal C}$ of the charge conjugation symmetry (normalized as $\oint B^{\cal C}=0,1$ mod 2):
\begin{equation}\label{eqn:anomalysoBC}
\pi\int_{5d}\left( B_m \text{Bock}(B_e) + B_m B_e B^{\cal C}\right)~.
\end{equation}
In particular, if we promote $B_m$ to be dynamical, then we can introduce new background $B_2$ for the emergent one-form symmetry with the coupling $\pi\int B_m B_2$. 
To cancel the anomaly (\ref{eqn:anomalysoBC}), $B_2$ needs to satisfy
\footnote{For a general discussion about gauging a subgroup of an anomalous symmetry, see \cite{Tachikawa:2017gyf}.}
\begin{equation}\label{eqn:constraintBCB2}
\delta B_2 = \text{Bock}(B_e)+B_eB^{\cal C}~.
\end{equation}
If $B^{\cal C}=0$, this implies the one-form symmetry is extended from $\mathbb{Z}_2$ to $\mathbb{Z}_4$. This is consistent with the $\mathbb{Z}_2$ element in the center of the original gauge group turning into a $\mathbb{Z}_4$ element.
The background for the $\mathbb{Z}_4$ one-form symmetry can be constructed from $(B_2,B_e)$ as $B_2'=2 B_2-\tilde B_e$, where $\tilde B_e$ is a lift of $B_e$ to $\mathbb{Z}_4$ cochain.
Using (\ref{eqn:constraintBCB2}) one can verify $B_2'$ is independent of the lift, and it is closed in $\mathbb{Z}_4$.
In the presence of non-trivial $B^{\cal C}$, due to the action of charge conjugation on the $\mathbb{Z}_4$ one-form symmetry,
 the background for the one-form symmetry should be modified to be a twisted cocycle, in agreement with (\ref{eqn:constraintBCB2}).

\subsubsection{Couplings}

Next we will discuss different ways to couple the theory to the background gauge field of $\mathbb{Z}_2$ 0-form symmetry that acts as
the charge conjugation symmetry. There are 4 couplings from $H^2_{\cal C}(B\mathbb{Z}_2,U(1)\times U(1))=\mathbb{Z}_2^2$, corresponding to the background gauge fields for the one-form symmetry
\begin{equation}\label{eqn:couplingC}
B_e=0,(B^{\cal C})^2,\quad B_m=0,(B^{\cal C})^2~.
\end{equation}
These cocycles are non-trivial under the inclusion map $\mathbb{Z}_2\rightarrow U(1)$, since the charge conjugation symmetry acts on the one-form symmetry (\ref{eqn:chargecongU(1)}).
Further imposing the identification (\ref{eqn:identificationsymmetryset}) from the $\mathbb{Z}_4$ 0-form symmetry (\ref{eqn:EMdualityperm}) at the self-dual point, we find that among (\ref{eqn:couplingC}) there are three distinct couplings.
The coupling with $B_e=B_m=(B^{\cal C})^2$ is anomalous by (\ref{eqn:anomalysoBC}), in agreement with \cite{Zou:2017ppq}. The anomaly only receives a contribution from the last term in (\ref{eqn:anomalysoBC}) since $\text{Bock}((B^{\cal C})^2)$ is trivial, and the anomaly is given by
\begin{equation}
\pi \int_{5d} (B^{\cal C})^5~,
\end{equation}
which is the unique non-trivial $(4+1)d$ bosonic SPT phase for unitary $\mathbb{Z}_2$ symmetry, as classified by $H^5(\mathbb{Z}_2,U(1))=\mathbb{Z}_2$ \cite{Dijkgraaf:1989pz,Chen:2011pg}.

It is straightforward to generalize the discussion to a general unitary 0-form symmetry $G$ that acts as the charge conjugation symmetry by $\sigma_1\in \text{Hom}(G,\mathbb{Z}_2)$. The SPT phases that are trivialized by the dynamics can be obtained from the anomaly (\ref{eqn:anomalysoBC}). For instance, consider $\theta=0$ and the $\mathbb{Z}_2\times \mathbb{Z}_2^{\cal T}$ 0-form symmetry, where the time-reversal acts as before, and the unitary $\mathbb{Z}_2$ symmetry acts as charge conjugation. The coupling with $B_e=w_1^2$ trivializes the SPT phase $\pi\int \left(w_1^4+w_1^3B^{\cal C}\right)$ by the magnetic one-form global symmetry transformation with the parameter $\lambda_1=\pi\tilde w_1$.

\subsubsection{$O(2)$ gauge theory}
\label{sec:Otwogaugetheory}

We briefly discuss a gapless example with intrinsic three-group symmetry.
We start with $U(1)$ gauge theory and turn on the background gauge field $B^{\cal C}$ for the charge conjugation $\mathbb{Z}_2$ 0-form symmetry as in section \ref{sec:U(1)BC}.
Then we promote the gauge field to be dynamical, changing the theory into an $O(2)$ gauge theory. The theory is still a free theory.
The gauge bundle becomes an $O(2)$ bundle, with first Stiefel-Whtiney class $w_1(O(2))=B^{\cal C}$.

Since charge conjugation symmetry identifies the lines $(q_\textbf{e},q_\textbf{m})\sim (-q_\textbf{e},-q_\textbf{m})$, the one-form symmetry becomes $\mathbb{Z}_2\times\mathbb{Z}_2$, corresponding to the $\mathbb{Z}_2$ center of $O(2)$ and the $\mathbb{Z}_2$ magnetic one-form symmetry generated by $\oint w_2(O(2))$. The charged objects are the Wilson lines and the 't Hooft lines.

The theory has ${\cal B}=\mathbb{Z}_2$ two-form symmetry generated by the line $\oint w_1(O(2))$. The charged objects are the Alice strings (see {\it e.g.} \cite{Schwarz:1982ec,Bucher:1992bd,Vilenkin:2000jqa}), which end on the codimension-one operator that implements the $\mathbb{Z}_2$ charge conjugation transformation. 
Since the charge conjugation symmetry is gauged, such codimension-one operators are trivial operators, and the Alice strings can move freely.
When a particle with charge $(q_\textbf{e},q_\textbf{m})$ goes around the Alice string, it becomes the particle with charge ${\cal C}(q_\textbf{e},q_\textbf{m})=(-q_\textbf{e},-q_\textbf{m})$.

Denote the background three-form gauge field for the $\mathbb{Z}_2$ two-form symmetry by $B_3$, which couples to the theory as $\pi\int w_1(O(2))B_3$.
In order to cancel the mixed anomaly (\ref{eqn:anomalysoBC}), where $B^{\cal C}=w_1(O(2))$, the background $B_3$ must satisfy
\begin{equation}
\delta B_3 = B_e B_m~.
\end{equation}
Thus the theory has 3-group symmetry that combines the ${\cal B}=\mathbb{Z}_2$ two-form symmetry and the ${\cal A}=\mathbb{Z}_2\times\mathbb{Z}_2$ one-form symmetry, described by the backgrounds $B_3$ and $B_e,B_m$. A similar discussion that obtains 2-group symmetry by gauging a subgroup can be found in \cite{Tachikawa:2017gyf,Benini:2018reh}.

We remark that since the line $\oint w_1(O(2))$ does not transform under the electric or magnetic one-form symmetry, this is an example with non-trivial $\Xi$ in (\ref{eqn:constraintmodify}).

Next, we discuss $O(2)$ gauge theory with unitary 0-form symmetry $G$. We will consider the case where $G$ does not permute the non-local operators. It can couple by
\begin{equation}
\xi\in H^1(BG,H^1(\mathbb{Z}_2,U(1))')=H^1(BG,\mathbb{Z}_2)~,
\end{equation}
which corresponds to the codimension-one symmetry defect $\oint w_1(O(2))^3$.
The additional couplings are parametrized by $(\eta_2,\nu_3)\in H^2(BG,\mathbb{Z}_2\times\mathbb{Z}_2)\times C^3(BG,\mathbb{Z}_2)$ with $\eta_2=(\eta_2^e,\eta_2^m)$ for the $\mathbb{Z}_2$ electric and magnetic one-form symmetry. The constraint on $\delta\nu_3$ is
\begin{equation}
\delta \nu_3 =\eta_2^e\cup\eta_2^m~.
\end{equation}
The constraint gives the following equivalence relation: under $\eta_2^e\rightarrow \eta_2^e+\delta s_1^e$ and $\eta_2^m\rightarrow \eta_2^m+\delta s_1^m$ with $s_1^e,s_1^m\in C^1(BG,\mathbb{Z}_2)$, $\nu_3$ transforms as
\begin{equation}
\nu_3\sim \nu_3+s_1^e\eta_2^m+\eta_2^es_1^m+s_1^e\delta s_1^m+\delta s_2~,
\end{equation}
where $s_2\in C^2(BG,\mathbb{Z}_2)$ is induced by a two-form gauge transformation.

For instance, consider the coupling $\eta_2^e=0$, then $\eta_2^m\in H^2(BG,\mathbb{Z}_2)$ and $\nu_3\in H^3(BG,\mathbb{Z}_2)$ with additional equivalence relation $\nu_3\sim \nu_3+s_1^e\eta_2^m$ for $s_1^e\in H^1(BG,\mathbb{Z}_2)$.
If $G=\mathbb{Z}_2$, the equivalence relation implies that for non-trivial $\eta_2^m$ in $H^2(\mathbb{Z}_2,\mathbb{Z}_2)$, the coupling $\nu_3$ is trivial.

\subsection{$\mathbb{Z}_2$ gauge theory with time-reversal symmetry}\label{sec:Z2gaugetheory}

In this example we discuss $\mathbb{Z}_2$ gauge theory with time-reversal symmetry (more precisely, the time-reversing $O(4)$ Lorentz symmetry). Denote the basic line and surface operators by $U,V$, then consider the time-reversal symmetry acting on the line operator $U^{q_\textbf{e}}$ and surface operator $V^{q_\textbf{m}}$ as
\begin{equation}
{\cal T}(q_\textbf{e},q_\textbf{m})=(q_\textbf{e},-q_\textbf{m})=(q_\textbf{e},q_\textbf{m})\quad \text{mod }2~.
\end{equation}
Thus the time-reversal symmetry does not act on the one- and two-form symmetries. Similarly it does not act on the $\mathbb{Z}_2$ background gauge fields of the one- and two-form symmetries.

We will focus on the couplings with trivial $\xi$. Since the one-form and two-form symmetries are isomorphic to $\mathbb{Z}_2$,
the couplings to time-reversal symmetry are classified by
\begin{equation}
H^2(BO(4),\mathbb{Z}_2)\times H^3_\tau(BO(4),\mathbb{Z}_2)~.
\end{equation}
They correspond to different backgrounds $B_2^{(2)}, B_3^{(2)}$ for the $\mathbb{Z}_2$ one-form and two-form symmetries: (they are normalized as $\oint B_2^{(2)}, B_3^{(2)}=0,1$ mod 2)
\begin{equation}
B_2^{(2)}=p w_1^2+q w_2,\qquad B_3^{(2)}=p' w_1^3+q'w_3\qquad (\text{mod }2)~,
\end{equation}
where the integers $p,q,p',q'=0,1$ mod 2. Thus the couplings have a $\mathbb{Z}_2^4$ classification.
Some of the couplings are anomalous due to the mixed anomaly between the one-form and two-form symmetries
\begin{equation}
\pi\int_{5d}B_2^{(2)}B_3^{(2)}~,
\end{equation}
which comes from the braiding between the generators of the two higher-form symmetries.
There are 12 couplings that are non-anomalous, corresponding to the $\mathbb{Z}_2$ valued backgrounds
\begin{align}\label{eqn:Znphases12}
&B_2^{(2)}=0,w_1^2,\qquad B_3^{(2)}=0,w_3,w_1^3,w_3+w_1^3\cr
&B_2^{(2)}=w_2,w_2+w_1^2,\qquad B_3^{(2)}=0,w_1^3~.
\end{align}
In particular, the coupling with the backgrounds $B_2^{(2)}=w_2,B_3^{(2)}=w_3$ has the gravitational anomaly $\pi\int w_2w_3$ from (\ref{eqn:anom}).
In this case the basic line operator is attached to the surface $\pi\int B_2^{(2)}=\pi\int w_2$, and thus it is the worldline of a fermion. The basic surface operator is attached to the surface $\pi\int B_3^{(2)}=\pi \int w_3$, and so it describes a fermionic string excitation. The gravitaional anomaly for this coupling is also explained in \cite{Thorngren:2014pza}.

In the presence of a surface operator $V_\Sigma=\pi\oint_{\Sigma} v_2$ supported at $\Sigma$, the equation of motion for $v_2$ enforces $B_2^{(2)}=\delta u+\delta_\perp(\Sigma)$ i.e. $w_1^2$ is trivial away from the surface $\Sigma$.
In particular, the background $B_3=w_1^3$ is non-trivial and it can be detected by an insertion of surface operator. As another example, the background $B_2^{(2)}=w_2$ implies that $B_3^{(2)}=w_3=\text{Bock}(w_2)$ is trivial only away from $V_\Sigma$, and thus it differs from the case $B_3^{(2)}=0$.\footnote{This clarifies an imprecise statement in the literature \cite{Wen:2016cij}.}
In fact, the coupling with $B_2^{(2)}=w_2,B_3^{(2)}=0$ is non-anomalous, while that with $B_3^{(2)}=w_3$ has a gravitational anomaly, thus the two couplings cannot be continuously connected to each other.

\subsubsection{Combined with SPT phases}

We can further combine the couplings with SPT phases for the global symmetry. This amounts to adding a topological action for the background gauge field.
As discussed in section \ref{sec:wSPTgeneral}, this either produces a new phase, or merely redefines the observables by worldvolume SPT phases and does not lead to distinct phases similar to section \ref{sec:U(1)SPTtotal}.

In the case of $\mathbb{Z}_2$ gauge theory with time-reversal symmetry, the one-form and two-form symmetry has the mixed anomaly described by the $5d$ bulk term $\pi\int_{5d} B_2^{(2)}B_3^{(2)}$. By anomaly inflow to $4d$ this means that under the $\mathbb{Z}_2$ higher-form global symmetries with parameters given by the $\mathbb{Z}_2$ cocycles $\lambda_1,\lambda_2$, the theory changes by
\begin{equation}
\pi\int B_2^{(2)} \lambda_2+B_3^{(2)}\lambda_1~.
\end{equation}
The anomalous shift gives rise to an SPT phase that can be absorbed by a higher-form global symmetry transformation.
We will use the parameters $\lambda_1= w_1$ or $\lambda_2=w_2,w_2+w_1^2,w_1^2$.
For example, with $B_2^{(2)}=w_2$ and $\lambda_2=w_2$, this implies the SPT phase
\begin{equation}
\pi \int w_2^2
\end{equation}
can be absorbed by depositing the worldvolume SPT phase $\pi\oint \lambda_2=\pi\oint w_2$ one the surface charged under the $\mathbb{Z}_2$ two-form symmetry. 
The possible couplings are summarized as follows:
\begin{itemize}
\item When $B_2^{(2)}=w_1^2$, the transformation $(\lambda_1,\lambda_2)=(0,w_1^2)$ trivializes the SPT $w_1^4$. The SPT phase is equivalent to depositing the worldvolume SPT phase $\pi\oint w_1^2$ on the surface operator. 

\item When $B_2^{(2)}=w_2$, the transformation $(\lambda_1,\lambda_2)=(0,w_2)$ trivializes the SPT $w_2^2$.
The SPT phase is equivalent to depositing the worldvolume SPT phase $\pi\oint w_2$ on the surface operator.
    
\item When $B_2^{(2)}=w_1^2+w_2$, the transformations $(\lambda_1,\lambda_2)=(0,w_1^2)$, $(0,w_2)$, and $(0,w_1^2+w_2)$ trivialize the SPT phases $w_1^4$, $w_2^2$, and $w_1^4+w_2^2$, respectively.
These SPT phases are equivalent to depositing the worldvolume SPT phases $\pi\oint w_2,\pi \oint w_2,\pi\oint (w_2+w_1^2)$ respectively on the surface operator.

\item When $B_3^{(2)}=w_1^3$, the transformation $(\lambda_1,\lambda_2)=(w_1,0)$ trivializes the SPT $w_1^4$.
The SPT phase is equivalent to depositing the worldline SPT phase $\pi\oint w_1$ on the line operator.    
\end{itemize}

The non-anomalous symmetry-enriched phases in the $\mathbb{Z}_2$ gauge theory with time-reversal symmetry are summarized in table \ref{tab:Z2wT}. There are in total 25 non-anomalous symmetry-enriched phases in $\mathbb{Z}_2$ gauge theory with time-reversal symmetry
\begin{table}[h!]\centering
\begin{tabular}{ll|c}
$B_2^{(2)}$ & $B_3^{(2)}$ & non-trivial generators of SPT phases\\ \hline
$0$ & $0,w_3$ & $\pi\int w_2^2,\pi\int w_1^4$\\
$0$ & $w_1^3,w_3+w_1^3$ & $\pi\int w_2^2$\\
$w_2$ & $0$ & $\pi\int w_1^4$\\
$w_2$ & $w_1^3$ & 0\\
$w_2+w_1^2$ & $0,w_1^3$ & 0\\
$w_1^2$ & $0,w_1^3,w_3,w_3+w_1^3$ & $\pi\int w_2^2$\\ \hline
$w_2$ & $w_3$ & $\pi\int w_1^4$\\
$w_2$ & $w_3+w_1^3$ & 0\\
$w_2+w_1^2$ & $w_3,w_3+w_1^3$ & 0
\end{tabular}
\caption{The different couplings of $\mathbb{Z}_2$ gauge theory to the time-reversing Lorentz symmetry, and the non-trivial SPT phases that the couplings admit. The couplings are represented by different backgrounds $B_2^{(2)},B_3^{(2)}$.
The SPT phases for time-reversal symmetry can be generated by $\pi\int w_2^2$ and $\pi\int w_1^4$, but some of them can be trivialized by the coupling, and we list the non-trivial generators for each coupling.
The entries above the line in the middle are non-anomalous, while those below the line have the gravitational anomaly $\pi\int w_2w_3$.
In total there are 25 non-anomalous symmetry-enriched phases.}\label{tab:Z2wT}
\end{table}

\subsubsection{$\mathbb{Z}_2$ gauge theory from Higgsing the $U(1)$ theory}\label{sec:ZnHiggs}

The $U(1)$ gauge theory can be Higgsed to $\mathbb{Z}_2$ gauge theory by condensing a scalar of charge $2$. We add to \eqref{eqn:U1action} the scalar field $\Phi$ with Lagrangian
\begin{equation}
|D_{2a}\Phi|^2-V(|\Phi|)~,
\end{equation}
where the potential makes $\Phi$ condense with $\langle|\Phi|^2\rangle=\rho^2$.
Writing $\Phi=\rho e^{i\varphi}$ with periodic scalar $\varphi\sim \varphi+2\pi$, this amounts to adding
\begin{equation}
\rho^2(d\varphi-2a)\star (d\varphi-2a)~.
\end{equation}
At low energy compared to the vacuum expectation value of the Higgs field, $\rho\rightarrow \infty$ and thus the gauge field $a$ is fixed to be \cite{Banks:2010zn}
\begin{equation}\label{eqn:znhiggsing}
2a=d\varphi~.
\end{equation}
In particular, this means the $\mathbb{Z}_2$ cocycle $\oint u=\oint a/\pi$ can be lifted to an integral cocycle $d\varphi/(2\pi)$ when $\Phi$ condenses.

The condition on $a$ can also be realized by adding to \eqref{eqn:U1action} a Lagrangian multiplier
\begin{equation}\label{eqn:ztwo}
\frac{2}{2\pi}b_2da~,
\end{equation}
where $b_2$ is a $U(1)$ two-form gauge field. 
This action can be identified with (\ref{eqn:Zngaugetheory}) for $n=2$, $a=\pi u$, $b_2=\pi v_2$.
The theory has electric $\mathbb{Z}_2$ one-form symmetry and magnetic two-form $\mathbb{Z}_2$ symmetry, where the charged objects are the Wilson line $W=e^{i\oint a}$ and the vortex string $V=e^{i\oint b_2}$ associated with Higgsing the $U(1)$ gauge field. Note at low energies the vortex string squares to a trivial operator $V^2=e^{2i\oint b_2}=1$. Denote the corresponding background gauge fields for the higher-form symmetry by $B_2=\pi B_2^{(2)}, B_3=\pi B_3^{(2)}$; they are normalized as $\oint B_2,B_3\in \pi\mathbb{Z}$.

The gauge field $B_2$ couples to the theory as
\begin{equation}
\frac{2}{2\pi}\int b_2 B_2~.
\end{equation}
The field $b_2$ imposes the constraint $da-B_2=0$.
Alternatively, since gauging the electric one-form symmetry leads to a selection rule of the Wilson lines,
the background $B_2$ modifies the gauge field by
\begin{equation}\label{eqn:z2BE}
\oint\frac{da}{2\pi} = \oint \frac{B_2}{2\pi} \quad \text{mod }\mathbb{Z}~.
\end{equation}
Thus we can identify $B_2$ with the background for the UV electric one-form symmetry (restricted to $\mathbb{Z}_2$)
\begin{equation}\label{eqn:dictB2}
B_2=B^E~.
\end{equation}
Note the constraint (\ref{eqn:z2BE}) means the Higgs field is also charged under a background field related to $B^E$, so the condition (\ref{eqn:znhiggsing}) is modified by a background field to satisfy (\ref{eqn:z2BE}).

The background $B_3$ couples to the theory as
\begin{equation}\label{eqn:z2BM}
\frac{2}{2\pi}\int aB_3~.
\end{equation}
Equating the couplings for the background of the UV magnetic one-form symmetry (\ref{eqn:BM}) and the IR magnetic two-form symmetry (\ref{eqn:z2BM}),
\begin{equation}\label{eqn:matchuone}
\int\frac{da}{2\pi}B^M=\frac{2}{2\pi}\int a\frac{1}{2}dB^M=\frac{2}{2\pi}\int aB_3+(\text{classical action})~.
\end{equation}
Thus, using $\text{Bock}(x)=w_1 x$ for $x\in H^3(M_4,\mathbb{Z}_2)$, we find\footnote{
If we consider $U(1)_\text{gauge}\times\mathbb{Z}_2^{CT}$ instead of $U(1)_\text{gauge}\rtimes\mathbb{Z}_2^T$, then (\ref{eqn:matchuone}) has an additional non-trivial term $-\frac{2}{2\pi}\int d(a dB^M/2)$ that leads to $B_3=\text{Bock}(B^M)+w_1B^M$.
}
\begin{equation}\label{eqn:dictB3}
B_3 = \frac{1}{2}dB^M=\text{Bock}(B^M)~.
\end{equation}
For instance, if $B^M=\pi w_1^2$, then $B_3=0$, in agreement with the fact that it is a classical action (\ref{eqn:trivialcouplingw1square}) in the UV. Similarly, if $B^M=\pi w_2$, then the coupling is trivial only on a $pin^+$ manifold, and thus it corresponds to $B_3=\pi w_3=\pi\text{Bock}(w_2)$.

The dictionary (\ref{eqn:dictB2}), (\ref{eqn:dictB3}) implies the six couplings of the $U(1)$ gauge theory at $\theta=0$ to the time-reversing Lorentz symmetry become to the following couplings after Higgsing to $\mathbb{Z}_2$ gauge theory
\begin{align}\label{eqn:Z2fromU1couplings}
&B_2=0,\pi w_1^2,\qquad B_3=0,\pi w_3\cr
&B_2=\pi w_2,\pi (w_2+w_1^2),\qquad B_3=0~.
\end{align}
In particular, the Higgs mechanism does not give rise to symmetry-enriched phases with $B_3=\pi w_1^3,\pi (w_3+w_1^3)$.
Together with the SPT phases, we find that the 18 non-anomalous symmetry-enriched phases at $\theta=0$ after Higgsing become $15$ distinct phases.
In particular, the following symmetry-enriched phases in the $U(1)$ gauge theory at $\theta=0$ become equivalent (denoted by $\sim$) after Higgsing to the $\mathbb{Z}_2$ gauge theory
\begin{align}
(B^E=\pi w_2,B^M=0;\pi\int w_2^2)&\;\sim \;(B^E=\pi w_2,B^M=0;0)\cr
(B^E=\pi w_2,B^M=0;\pi\int (w_2^2+w_1^4))&\;\sim\; (B^E=\pi w_2,B^M=0;\pi \int w_1^4)\cr
(B^E=\pi(w_2+w_1^2),B^M=0;\pi \int w_2^2)&\;\sim\; (B^E=\pi(w_2+w_1^2),B^M=0;0)~,
\end{align}
where the notation is $(B^E,B^M;\text{SPT phase})$.

\section*{Acknowledgements}

We thank Zhen Bi, Xie Chen, Meng Cheng, Clay C{\'o}rdova, Anton Kapustin, Hotat Lam, Kantaro Ohmori, Shu-Heng Shao, Ryan Thorngren and Juven Wang for discussions. 
The work was supported by the U.S. Department of Energy, Office of Science, Office of High Energy Physics, under Award Number DE-SC0011632, and by the Simons Foundation through the Simons Investigator Award.
The work of P.-S.H. was performed in part at Aspen Center for Physics, which is supported by National Science Foundation grant PHY-1607611.

\appendix

\section{Cochains and higher cup products}
\label{sec:appendixmath}

In this appendix we summarize some facts about cochains and higher cup products. For more details, see {\it e.g.} \cite{hatcher2002algebraic} and the appendix A of \cite{Benini:2018reh}.

We triangulate the spacetime manifold $M$ with simplicies, where a $p$-simplex is the $p$-dimensional analogue of a triangle or tetrahedron (for $p=0$ it is a point, $p=1$ it is an edge, etc).
The $p$-simplices can be described by its vertices $(i_0,i_1,\cdots i_p)$ where we pick an ordering $i_0<i_1<\cdots i_p$.

A simplicial $p$-cochain $f\in C^p(G,{\cal A})$ is a function on $p$-simplices taking values in an Abelian group ${\cal A}$ (we use additive notation for Abelian groups). For simplicity, we will take ${\cal A}$ to be a field (an Abelian group endowed with two products: addition and multiplication).

The coboundary operation on the cochains $\delta:C^p(M,{\cal A})\rightarrow C^{p+1}(M,{\cal A})$ is defined by
\begin{equation}
(\delta f)(i_0,i_1,\cdots i_{p+1})=\sum_{j=0}^{p+1}(-1)^jf(i_0,\cdots \hat i_j,\cdots i_{p+1})
\end{equation}
where the hatted vertices are omitted. The coboundary operation is nilpotent $\delta^2=0$.
When a cochain $x$ satisfies $\delta x=0$, it is called a cocycle.

The cup product $\cup$ for $p$-cochain $f$ and $q$-cochain $g$ gives a $(p+q)$-cochain defined by
\begin{equation}
(f\cup g)(i_0\cdots i_{p+q})=f(i_0,\cdots i_p)g(i_{p}\cdots i_{p+q})~.
\end{equation}
It is associative but not commutative. In this note we will omit writing the cup products.
The higher cup product $f\cup_1 g$ is a $(p+q-1)$ cochain, defined by
\begin{equation}
(f\cup_1 g)(i_0\cdots i_{p+q-1})=\sum_{j=0}^{p-1}(-1)^{(p-j)(q+1)}f(i_0,\cdots i_ji_{j+q},\cdots i_{p+q-1})g(i_j,\cdots i_{j+q})~.
\end{equation}
It is not associative and not commutative.

We have the following relations for a $p$ cochain $f$ and $q$ cochain $g$:
\begin{align}\label{eqn:cochainrules}
&f\cup g=(-1)^{pq}g\cup f+(-1)^{p+q+1}\left[
\delta(f\cup_1 g)-\delta f\cup_1 g - (-1)^p f\cup_1 \delta g
\right]\cr
&\delta(f\cup g)=\delta f\cup g+(-1)^p f\cup \delta g\cr
&\delta\left(f\cup_1 g\right)
=\delta f\cup_1 g+(-1)^p f\cup_1 \delta g + (-1)^{p+q+1} f\cup g+(-1)^{pq+p+q} g\cup f~.
\end{align}

Similarly, if there is $G$ action on ${\cal A}$ given by $\rho:G\rightarrow \text{Aut}({\cal A})$, one can define a twisted coboundary operation that is nilpotent. Similarly the cup products $\cup,\cup_1$ can be modified. The rules (\ref{eqn:cochainrules}) are still true (with $\delta$ meaning the twisted coboundary operation).

When the coefficient group is ${\cal A}=\mathbb{Z}_2$, there are additional operations in the cohomology called the Steenrod squares.
For the purpose of this note we only need the operations $Sq^1$ and $Sq^2$. $Sq^i$ maps a $\mathbb{Z}_2$ $p$-cocycle to a $\mathbb{Z}_2$ $(p+i)$-cocycle.
The definitions of $Sq^1$  and $Sq^2$ acting on $\mathbb{Z}_2$ one-cocycle $x_1$ and two-cycle $x_2$ are
\begin{equation}\label{eqn:Steenrodsquarerule}
Sq^1(x_1)=x_1\cup x_1,\quad Sq^1(x_2)=x_2\cup_1 x_2,\quad Sq^2(x_1)=0,\quad Sq^2(x_2)=x_2\cup x_2~.
\end{equation}
In particular, $Sq^1$ acts on the cohomology the same way as the Bockstein homomorphism for the short exact sequence $1\rightarrow\mathbb{Z}_2\rightarrow\mathbb{Z}_4\rightarrow\mathbb{Z}_2\rightarrow 1$.

\section{Dimensional reduction of $\mathbb{Z}_n$ gauge theory with coupling $\xi$}\label{sec:ZnunitaryZn}

In this appendix we consider $\mathbb{Z}_n$ gauge theories coupled to the background $X$ of a unitary 0-form symmetry $G$ by a non-trivial $\xi\in H^1(BG,H^3({\cal B},U(1))')=H^1(BG,\mathbb{Z}_n)$. 
We will study the consequence of the coupling $\xi$ by dimensionally reduce the theory to $(2+1)d$.

In the continuous notation, the $\mathbb{Z}_n$ gauge theory coupled to $X$ by $\xi$ is described by
\begin{equation}\label{eqn:Zncontinuous}
\int_{4d}\left(\frac{n}{2\pi}b_2da+\frac{n}{(2\pi)^2}ada X^*\xi\right)~.
\end{equation}
where $a,X^*\xi$ are one-form gauge fields and $b_2$ is a two-form gauge fields.
The continuous fields $a,b_2$ are related to the discrete fields roughly by $e^{i\oint a}=e^{\frac{2\pi i}{n}\oint u}$ and $e^{i\oint b_2}=e^{\frac{2\pi i}{n}\oint v_2}$, and similarly for $X^*\xi$.

We place the theory on $S^1\times {\cal M}_3$ with small radius $\beta=1/T$ for the circle, and study the implication of the coupling for the effective $(2+1)d$ theory.
We turn on a non-trivial chemical potential for the 0-form symmetry background field:
\begin{equation}\label{eqn:dimredhoonomy}
\oint_{S^1} X^*\xi = {2\pi k\over n}~
\end{equation}
and trivial holonomy of $X$ for every cycle on ${\cal M}_3$.
The coupling $adaX^*\xi$ in (\ref{eqn:Zncontinuous}) reduces to the Chern-Simons term
\begin{equation}
\frac{k}{2\pi}\int_{3d} a d a~.
\end{equation}
The effective theory on ${\cal M}_3$ has one-form symmetry: part of it is from the original one-form symmetry in $(3+1)d$, while the other part is from the two-form symmetry\footnote{
The reduction of the $\mathbb{Z}_n$ gauge field also gives a decoupled $\mathbb{Z}_n$ 0-form in $(2+1)d$ that can be described by a periodic scalar $\phi\sim \phi+2\pi$ and the $U(1)$ two-form $b_2$ with the action $\frac{n}{2\pi}b_2d\phi$. 
This sector is bosonic, and it does not have an intrinsic one-form symmetry. It is decoupled form the Chern-Simons theory (\ref{eqn:dimredCS}), since we take the background $X$ to have trivial holonomy on cycles in ${\cal M}_3$, and thus we will omit it in the following discussion.
}. The $(2+1)d$ theory is
\begin{equation}\label{eqn:dimredCS}
\int_{3d}\left( \frac{k}{2\pi}ada+\frac{n}{2\pi}adb\right)
~,
\end{equation}
where $b$ is the $U(1)$ gauge field arising from the $b_2$ two-form gauge field in $(3+1)d$.
As a consistency check, this theory does not require a spin structure, just as the $(3+1)d$ theory.

The equations of motion for $a,b$ implies the line operators $U=e^{i\oint a},V=e^{i\oint b}$ obey the relations
\begin{equation}
U^{2k}V^n=1,\quad U^n=1~.
\end{equation}
In particular, $V$ has order $n^2/\ell$ with $\ell\equiv \gcd(n,2k)$. The theory has ${\cal A}_\text{3d}=\mathbb{Z}_\ell\times \mathbb{Z}_{n^2/\ell}$ one-form symmetry, generated by ${\cal W}_\ell=U^{2k/\ell}V^{n/\ell}$ and ${\cal W}_{n^2/\ell}=U^{-\beta}V^{\alpha}$ with integers $\alpha,\beta$ that satisfy $\alpha(2k/\ell)+\beta(n/\ell)=1$ (it is solvable since $\gcd(2k/\ell,n/\ell)=1$). 

Next, we will show the one-form symmetry in $(2+1)d$ can be reproduced from the three-group symmetry (\ref{eqn:3-group}) in $(3+1)d$.
The three-group symmetry (\ref{eqn:3-group}) reduces to
\begin{equation}\label{eqn:oneformsymmreduce}
\delta B_2'=2k\text{Bock}(B_2)~,
\end{equation}
where $B_2'$ is the two-form background field that comes from the reduction of $B_3$, and the coefficient $k$ comes from the holonomy of $X$ in (\ref{eqn:dimredhoonomy}) on the reduced circle.
This describes the two-form backgrounds of the group extension of $\mathbb{Z}_n$ by $\mathbb{Z}_n$ specified by the group cocycle $\omega(s_1,s_2)=\frac{2k}{n}\left(s_1+s_2-(s_1+s_2\;\text{mod }n)\right)$ where $s_1,s_2=0,1,\cdots n-1$. 
The group extension is $\mathbb{Z}_{n^2/\ell}\times\mathbb{Z}_{\ell}$. 
The backgrounds $B_2,B_2'$ with the constraint (\ref{eqn:oneformsymmreduce}) are equivalent to the following $\mathbb{Z}_{n^2/\ell}\times\mathbb{Z}_\ell$ two-cocycles:
\begin{equation}
B_2^{(n^2/\ell)}={n\over \ell}\tilde B_2'-{2k\over \ell}\tilde B_2\text{ mod }{n^2\over\ell},\qquad
B_2^{(\ell)}=\alpha \tilde B_2'+\beta \tilde B_2\text{ mod }\ell~,
\end{equation}
where $\alpha,\beta$ are integers satisfying $\alpha (2k/\ell)+\beta (n/\ell)=1$, and tilde denotes a lift to integral cochains.
The constraint (\ref{eqn:oneformsymmreduce}) implies that $\delta B_2^{(n^2/\ell)}=0$ mod $(n^2/\ell)$, $\delta B_2^{(\ell)}=0$ mod $\ell$ and $B_2^{(n^2/\ell)},B_2^{(\ell)}$ are independent of the lifts. 
The pair of two-cocycles $B_2^{(n^2/\ell)},B_2^{(\ell)}$ are the background fields of the $\mathbb{Z}_{n^2/\ell}\times\mathbb{Z}_\ell$ one-form symmetry.
Therefore the backgrounds $B_2,B_2'$ from dimensional reduction of the three-group symmetry describe the backgrounds of the correct one-form symmetry in $(2+1)d$.

For fixed background $X$, the Chern-Simons coupling $k$ depends on the coupling $\xi$ by the holonomy of $X^*\xi$ on the reduced circle (\ref{eqn:dimredhoonomy}).
The topological spin of the line operators depend on $k$: for the line $U^{q_\textbf{e}}V^{q_\textbf{m}}$ it is
\begin{equation}
h(q_\textbf{e},q_\textbf{m})=\left(
{ q_\textbf{e}q_\textbf{m}\over n}-\frac{kq_\textbf{m}^2}{n^2}\right)
\quad\text{mod }1~.
\end{equation}
Therefore different couplings $\xi$ can be distinguished from the statistics of the line operators in the $(2+1)d$ theory obtained by dimensional reduction.

\section{$U(1)$ gauge theory with $SO(3)\times \mathbb{Z}_2^{\cal T}$ symmetry}
\label{sec:appendixUoneSO3}

We will use our method to reproduce the classification in \cite{Zou:2017ppq} for the $U(1)$ gauge theory enriched by time-reversal and internal $SO(3)$ ordinary symmetry, where the time-reversal symmetry acts as in (\ref{eqn:Tparticle}), while the $SO(3)$ symmetry commutes with the gauge symmetry.

Consider first the case $\theta=0$ mod $2\pi$.
The couplings are classified by
\begin{equation}\label{eqn:uonesothreeanomalouscoupling}
H^2_\tau(B(O(4)\times SO(3)),U(1)_E\times U(1)_M)=\mathbb{Z}_2^4\times\mathbb{Z}_2^2~.
\end{equation}
These couplings correspond to turning on different backgrounds for the one-form symmetries.
The difference with section \ref{sec:U(1)timereversal} is that we can shift the backgrounds for the electric and magnetic one-form symmetries by $w_2(SO(3))$, the second Stiefel-Whitney class of the $SO(3)$ bundle.

Among the couplings some of them cannot be realized in a (3+1)$d$ system but must be on the surface of a $(4+1)d$ SPT phase specified by the anomaly (\ref{eqn:anom}).
There are 14 non-anomalous couplings listed in table \ref{tab:14phases}, in agreement with \cite{Zou:2017ppq}.

\begin{table}[t]\centering
\begin{tabular}{|l|l|l|}
\hline
Electric & Magnetic & Backgrounds\\ \hline
boson, ${\cal T}^2=1$, tensor & boson, tensor & $B^E=B^M=0$\\ 
boson, ${\cal T}^2=1$, spinor & boson, tensor & $B^E=\pi w_2(SO(3)), B^M=0$\\ 
boson, ${\cal T}^2=1$, tensor & boson, spinor & $B^E=0, B^M=\pi w_2(SO(3))$\\ 

boson, ${\cal T}^2=-1$, tensor & boson, tensor & $B^E=\pi w_1^2$, $B^M=0$\\ 
boson, ${\cal T}^2=-1$, spinor & boson, tensor & $B^E=\pi (w_1^2+w_2(SO(3)))$, $B^M=0$\\ 

fermion, ${\cal T}^2=1$, tensor & boson, tensor & $B^E=\pi (w_2+w_1^2)$, $B^M=0$\\ 
fermion, ${\cal T}^2=1$, spinor & boson, tensor & $B^E=\pi \left( w_2+w_1^2+w_2(SO(3))\right)$, $B^M=0$\\  

fermion, ${\cal T}^2=-1$, tensor & boson, tensor & $B^E=\pi w_2$, $B^M=0$\\ 
fermion, ${\cal T}^2=-1$, spinor & boson, tensor & $B^E=\pi \left( w_2+w_2(SO(3))\right)$, $B^M=0$\\  

boson, ${\cal T}^2=1$, tensor & fermion, tensor & $B^E=0$, $B^M=\pi w_2$\\ 
boson, ${\cal T}^2=1$, tensor & fermion, spinor & $B^E=0$, $B^M=\pi\left( w_2+w_2(SO(3))\right)$\\ 

boson, ${\cal T}^2=-1$, tensor & fermion, tensor & $B^E=\pi w_1^2$, $B^M=\pi w_2$\\

fermion, ${\cal T}^2=1$, spinor & boson, spinor & $B^E=\pi \left( w_2+w_1^2+w_2(SO(3))\right)$,\\
&& $B^M=\pi w_2(SO(3))$ \\ 

boson, ${\cal T}^2=-1$, spinor & fermion, spinor & $B^E=\pi\left( w_1^2+w_2(SO(3))\right)$, \\
& & $B^M=\pi\left( w_2+w_2(SO(3))\right)$\\
\hline
\end{tabular}
\caption{14 non-anomlaous couplings of $U(1)$ gauge theory at $\theta=0$ mod $2\pi$ to the backgrounds of $SO(3)\times \mathbb{Z}_2^{\cal T}$ symmetry. The electric particle has electric and magnetic charges $(1,0)$, while the magnetic particle has charges $(0,1)$. The statistics, Kramer degeneracy and $SO(3)$ projective representation (tensorial or spinorial representations, or equivalently integer or half-integer $SU(2)$ isospin) of the particles are according to \cite{Zou:2017ppq}. The different couplings correspond to different backgrounds for the electric and magnetic one-form symmetry.}\label{tab:14phases}
\end{table}

In particular, the following two couplings
\begin{align}\label{eqn:coupingso3response}
E_{f\frac{1}{2}}M_{b\frac{1}{2}}:\quad &B^E=\pi \left( w_2+w_1^2+w_2(SO(3))\right), B^M=\pi w_2(SO(3))\cr
E_{bT\frac{1}{2}}M_{f\frac{1}{2}}:\quad &B^E=\pi\left( w_1^2+w_2(SO(3))\right), B^M=\pi\left( w_2+w_2(SO(3))\right)~
\end{align}
are well-defined provided we add local counterterms involving the $SO(3)$ background gauge field. To see this, note the corresponding SPT phase is trivial and thus gives a	 local counterterm for the background gauge fields in $(3+1)d$:
\footnote{
Here we used several identities for Stiefel-Whitney classes, see {\it e.g.}\cite{milnor1974characteristic}: $v_i\cup x_{d-i}=Sq^i(x_{d-i})$ for any $x_{d-i}\in H^{d-i}(M_d,\mathbb{Z}_2)$ and $v_2=w_2+w_1^2$ is the second Wu class of the manifold. The Bockstein homomorphism for $\mathbb{Z}_2\rightarrow\mathbb{Z}_4\rightarrow\mathbb{Z}_2$ equals $Sq^1$ (see {\it e.g.} \cite{hatcher2002algebraic}). Wu's formula implies $Sq^1(w_2(SO(3)))=w_3(SO(3))$ and $Sq^2(w_3(SO(3)))=w_2(SO(3))w_3(SO(3))$ for any $SO(3)$ bundle.}
\begin{equation}
\frac{1}{2\pi}\int_{5d} B^E d B^M = 
\pi\int_{5d} Sq^2 w_3(SO(3))
+w_2(SO(3))w_3(SO(3))=0
\text{ mod }2\pi\mathbb{Z}~.
\end{equation}

Another way to see the couplings (\ref{eqn:coupingso3response}) can be made well-defined by a local counterterm of background gauge fields is that
the couplings are related to the couplings with the same $B^E$ but $B^M=0$ by $\theta=2\pi$ term (see equation (\ref{eqn:shiftBMthetapi})), up to a classical action of background gauge fields. This classical action contains a theta term of $SO(3)$ gauge field that can be described by ``$SO(3)_{1/4}$''. This reproduces the result in \cite{Zou:2017ppq}.

The other couplings in (\ref{eqn:uonesothreeanomalouscoupling}) are anomalous, and we can compute the corresponding 't Hooft anomalies using the anomaly for the one-form symmetry. The $5d$ SPT phases of the possible 't Hooft anomalies are generated by
\begin{equation}
S_\text{I}=\pi\int_{5d} w_2w_3,\quad
S_\text{II}=\pi\int_{5d} w_2 w_3(SO(3)),\quad
S_\text{III}=\pi\int_{5d} w_1^2 w_3(SO(3))~.
\end{equation}
The anomalies can all be realized in the theory, and they are called the class I, II and III anomalies in \cite{Zou:2017ppq}.

Next, we consider the case $\theta=\pi$ mod $2\pi$. The couplings are classified by
\begin{equation}
H^2_\tau(B(O(4)\times SO(3)),U(1)_E\times U(1)_M)=\mathbb{Z}_2^2~,
\end{equation}
where the difference with section \ref{sec:U(1)timereversal} is that the background for the magnetic symmetry can be shifted by $w_2(SO(3))$, and that gives total four (possibly anomalous) couplings.
From (\ref{eqn:surfacetheta}) the dyon $(q_\textbf{e},q_\textbf{m})=(\frac{1}{2},1)$ is attached to the surface $\int B^M$.
Thus for $B^M=0$ it has integer $SO(3)$ isospin, while for $B^M=\pi w_2(SO(3))$ it has half-integer $SO(3)$ isospin, in agreement with \cite{Zou:2017ppq}.
Due to the anomaly (\ref{eqn:anom}), only the coupling with $B^M=0$ is non-anomalous.
Thus to summarize, there are $14+1=15$ non-anomalous couplings to $SO(3)\times \mathbb{Z}_2^{\cal T}$ symmetry, in agreement with \cite{Zou:2017ppq}.

In addition, one can combine the couplings with the SPT phases for $SO(3)\times \mathbb{Z}_2^{\cal T}$ symmetry, which are classified by $\mathbb{Z}_2^4$ and can be generated by
\begin{equation}
\pi\int w_2^2,\quad\pi\int w_1^4,\quad \pi\int w_2(SO(3))w_1^2,\quad\pi\int w_2(SO(3))w_2~.
\end{equation}
In particular,  $\pi \int w_2(SO(3))^2=\pi\int w_2(SO(3))(w_2+w_1^2)$ is not an independent SPT phase.
In the presence of non-trivial coupling by the one-form symmetry, some of the SPT phases can be absorbed by a global magnetic one-form symmetry transformation with parameter $\lambda=\pi \tilde w_1$ that produces $\frac{1}{2\pi}\int B^E \delta(\pi \tilde w_1)$.
When $B^E\neq 0,\pi w_2$, this gives an SPT phase that is non-trivial on its own but is trivialized by coupling to the dynamics. 
In these cases there are only 3 independent generators for the SPT phases. 
Therefore in total there are $9\times 2^3+6\times 2^4=168$ non-anomalous phases, in agreement with \cite{Zou:2017ppq}.

\bibliographystyle{utphys}
\bibliography{biblio}{}

\end{document}